\documentclass[aip,twocolumn,10pt,letterpaper,bibnotes,floatfix,raggedbottom]{revtex4-1}
\usepackage{dcolumn}
\usepackage{bm}
\usepackage{graphicx}
\usepackage{sidecap}
\usepackage{amsmath}
\usepackage{amssymb}
\usepackage{physics}
\usepackage[protrusion=true,expansion]{microtype}
\usepackage{times}
\usepackage{multirow}
\usepackage{floatrow}
\usepackage{float}
\usepackage{hyperref}
\usepackage[dvipsnames]{xcolor}
\hypersetup{
    colorlinks=true,
    linkcolor=blue,
    filecolor=magenta,      
    urlcolor=cyan,
}
\floatstyle{plaintop}
\restylefloat{table}
\usepackage{dsfont} 

\begin{document}


\title{Dynamics of reconfigurable artificial spin ice: towards magnonic functional materials}

\author{Sebastian Gliga}
\email{sebastian.gliga@psi.ch}
\affiliation{Swiss Light Source, Paul Scherrer Institute, 5232 Villigen PSI, Switzerland}
\affiliation{School of Physics and Astronomy, University of Glasgow, Glasgow G12 8QQ, United Kingdom}

\author{Ezio Iacocca}
\affiliation{Department of Mathematics, Physics, and Electrical Engineering, Northumbria University, Newcastle upon Tyne NE1 8ST, United Kingdom}

\author{Olle G. Heinonen}
 \affiliation{\hbox{Materials Science Division, Argonne National Laboratory, Lemont, Illinois 60439, USA}}
 \affiliation{Center for Hierarchical Materials Design, Northwestern-Argonne Institute of Science and Engineering, Northwestern University, Evanston, Illinois 60201, USA}
 
 \begin{abstract}
Over the past few years, the study of magnetization dynamics in artificial spin ices has become a vibrant field of study. Artificial spin ices are ensembles of geometrically arranged, interacting magnetic nanoislands, which display frustration by design. These were initially created to mimic the behavior in rare earth pyrochlore materials and to study emergent behavior and frustration using two-dimensional magnetic measurement techniques. Recently, it has become clear that it is possible to create artificial spin ices, which can potentially be used as functional materials. In this Perspective, we review the resonant behavior of spin ices (which is in the GHz frequency range), focusing on their potential application as magnonic crystals. In magnonic crystals, spin waves are functionalized for logic applications by means of band structure engineering. While it has been established that artificial spin ices can possess rich mode spectra, the applicability of spin ices to create magnonic crystals hinges upon their reconfigurability. Consequently, we describe recent work aiming to develop techniques and create geometries allowing full reconfigurability of the spin ice magnetic state. We also discuss experimental, theoretical, and numerical methods for determining the spectral response of artificial spin ices, and give an outlook on new directions for reconfigurable spin ices.
 \end{abstract}
 \maketitle




\section{Introduction}

Artificial spin ices are superlattices composed of interacting magnetic nanoislands placed in a geometrical arrangement~\cite{Heyderman2013}. Originally, artificial spin ices were intended as macroscopic model systems mimicking the atomic frustration in rare earth pyrochlores~\cite{Wang2006}, with the advantage that their state could be directly measured using two-dimensional magnetic measurement techniques. 
Artificial spin ices were defined for crystallographic planes in the pyrochlores, leading to two fundamental arrangements: the square~\cite{Wang2006} and the kagome~\cite{Castelnovo2008} lattices.
Despite this dimensional reduction, artificial spin ices exhibit massively degenerate ground states~\cite{Heyderman2013} whose energy can be minimized by magnetic field-driven and thermal relaxation protocols~\cite{morgan2011,Farhan2013,Sklenar2018}. Building on these successes, artificial spin ices evolved into superlattices designed to explore geometric frustration, free from the crystallographic constraints of pyrochlore materials. Myriad of novel artificial spin ices emerged~\cite{Gilbert2014,Gilbert2016,Wang2016,Gliga2017,Sklenar2018,Luo2019,Saccone2019} featuring an interplay between frustration and topology~\cite{Nisoli2017}.

\begin{figure}[b]
\includegraphics[width=3in]{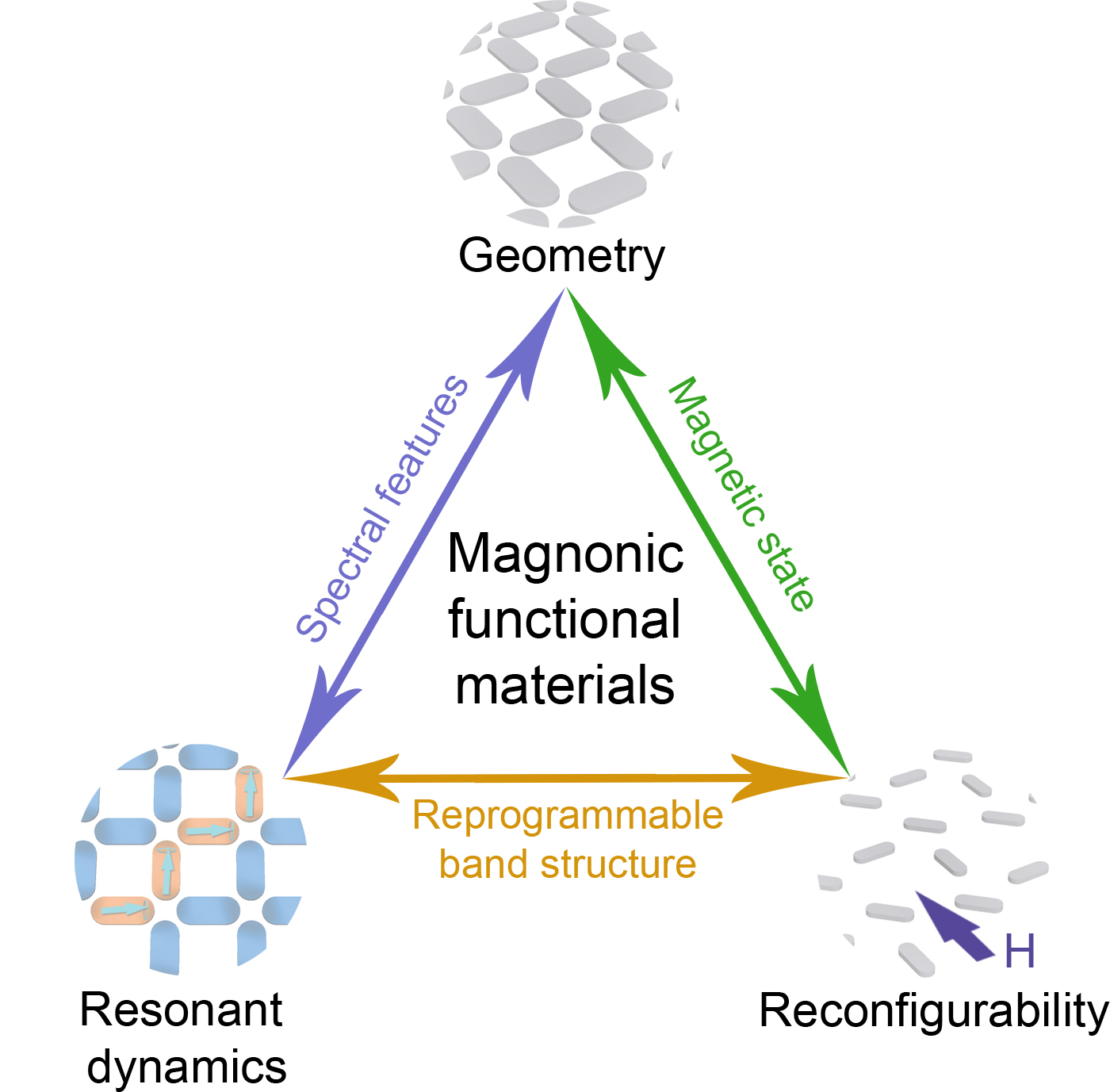}
\caption{ \label{fig:summary} Creating magnonic functional materials, such as magnonic crystals, based on artificial spin ices relies on the interplay of three main elements. The \emph{geometry} of the array determines the \emph{dynamics} of the magnetization as well as the \emph{reconfigurability} of its magnetic state (e.g. using external fields). The possibility of globally and locally reconfiguring the magnetic state is essential to achieving a reprogrammable band structure. 
}
\end{figure}

A subject of recent interest is the study of artificial spin ices in the context of magnetization dynamics. As superlattices, artificial spin ices are natural analogues of magnonic crystals~\cite{Nikitov2001,Kruglyak2010,Lenk2011,Demokritov2013,Chumak2015}, where spin waves are functionalized for logical applications by means of band structure engineering. 
Indeed, there is a growing interest in using spin waves (magnons) in information technology and computing\cite{chumak2019fundamentals}. This interest stems from the need for disruptive concepts requiring significantly lower energy consumption than traditional CMOS-based technology, in which information is processed using charge currents that dissipate significant power.
A key to achieving useful functionalities in magnonic crystals is the ability to reconfigure their magnetic state~\cite{Grundler2015}. The geometric frustration and degeneracy of ground states of artificial spin ices in principle make artificial spin ices strong candidates for reconfigurable magnonics.
In addition, the possibility of patterning virtually any planar geometry allows the definition of 
structures exhibiting both reconfigurable magnetic states and rich magnetization dynamics. These elements are essential for the creation of magnonic functional materials with a reprogrammable band structure, as illustrated in Fig.~\ref{fig:summary}.


In this Perspective article, we briefly review the progress made in the understanding and control of spin waves in  artificial spin ices. We survey methods to reconfigure artificial spin ices, theoretical models, experimental techniques, and salient advances in the study of magnetization dynamics in artificial spin ices. We discuss a number of outstanding challenges and perspectives for achieving reconfigurable artificial spin ices. 
Finally, we describe perspectives for the use of artificial spin ices for magnonics applications.
For an in-depth review of fabrication processes, recent developments in  ``connected'' artificial spin ices, and prospects for artificial spin ices as frustrated superlattices, we refer the reader to recent reviews in Refs.~\onlinecite{Lendinez2019,Skaervo2019}.

\section{Spin ice reconfigurability}
\label{sec:reconfigurable}

\begin{figure}[t]
\centering \includegraphics[width=2.8in]{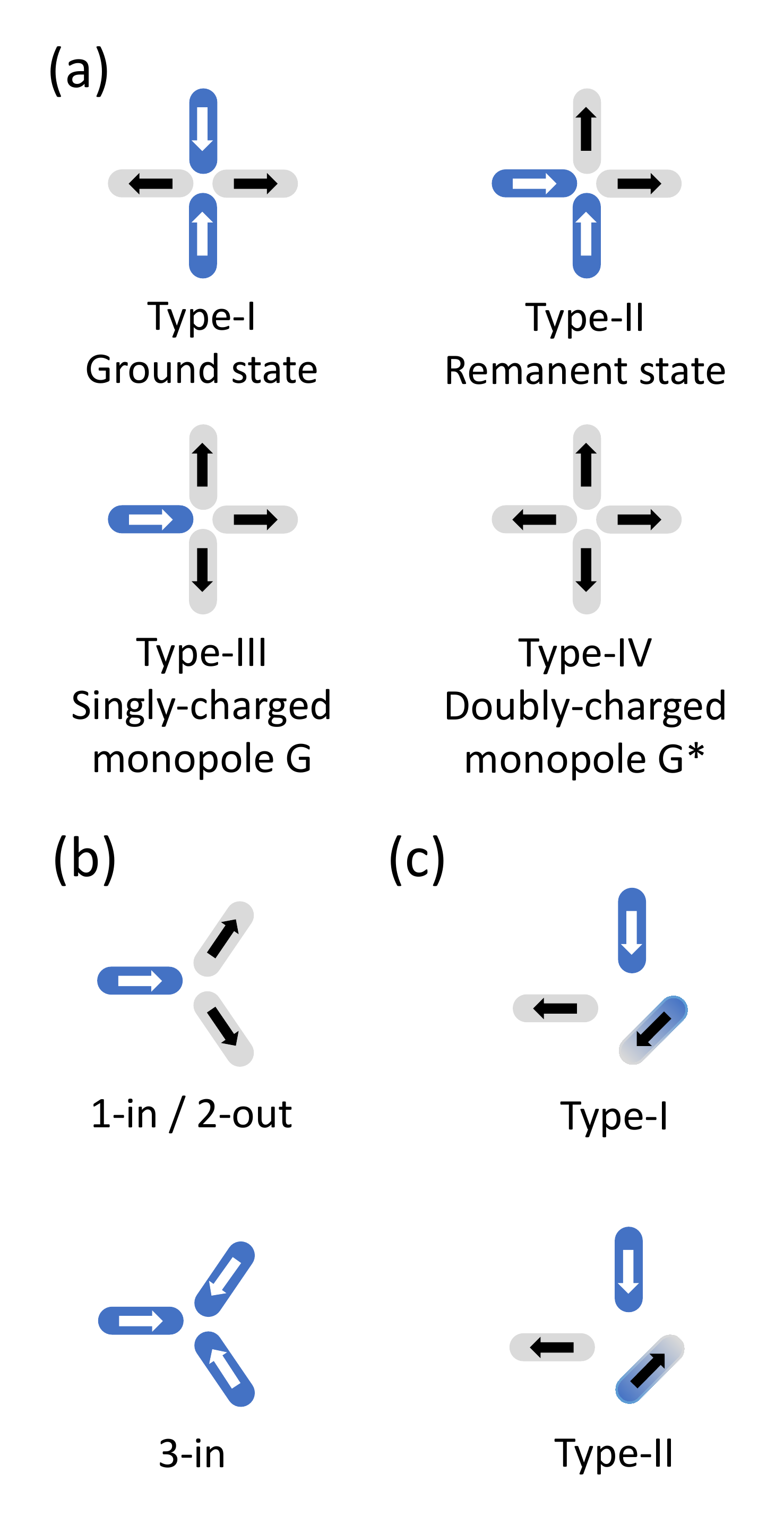}
\caption{ \label{fig:vertices} (a) Possible vertex configurations in square ice. Type-I is the ground state and has two possible degenerate configurations. Type-II is the remanent state, following saturation along one of the diagonals of the system and has four possible degenerate configurations, one for each diagonal. Both are charge-neutral and follow the \emph{ice rule}~\cite{Wang2006}. Types-III and IV are singly- and doubly-charged defects (monopoles). The notations $G$ and $G^*$ are from Ref.~\onlinecite{Gliga2013}. Type-III has eight possible degenerate states, while Type-IV has two. (b) Lowest-energy vertex configurations in kagome lattices. (c) Charge ice configurations equivalent to Type-I and Type-II configurations in the square ice (a). Each configuration can be toggled with almost 100\% yield by applying a field in the ($\bar{1}\bar{1}$) direction (Type-I) or (11) direction (Type-II). The equivalent position of the square ice nanoislands is shown by shaded nanoislands with dotted outline.
}
\end{figure}

%
The first reports on artificial spin ices were on square\cite{Wang2006} and kagome\cite{Castelnovo2008} ices. These systems are straight-forward to design as the unit cell contains a small number of elements. 
The relative simplicity of the lattices makes them attractive from the point of view of magnonic crystals. There is also considerable design freedom in that the nanoisland dimensions are decoupled from the lattice constant. This means that parameters, such as shape anisotropy and the magnetostatic coupling between elements, can be tuned independently to the extent allowed by lithographic limitations. Both lattices, however, suffer from the fact that their magnetic state is, in practice, often difficult to reconfigure. The square ice can be set in a well-defined remanent state by applying an external field along one of the array diagonals, resulting in a Type-II state, shown in Fig.~\ref{fig:vertices}(a). 
Moreover, the different nearest-neighbor distances between the four nanoislands at a vertex leads to non-equivalent interactions, the Type-I ground state is doubly-degenerate and can in principle be reached. 
However, the ground state is difficult to achieve using demagnetizing protocols\cite{Nisoli2007GS,morgan2011} and is more readily achieved during thermal relaxation as demonstrated in Ref.~\onlinecite{Farhan2013}, using very thin nanoislands (ca. 2 -- 3~nm thick). 
It is also possible to attain the square ice Type-I ground state by raising the temperature above the Curie temperature, $T_C$, of the magnetic elements and subsequently cooling down to room temperature~\cite{Porro2013,Zhang2013,Zhang2019_2}. However, due to the rather high Curie temperature of Permalloy (ca. 870 K), which is typically used in artificial spin ices, annealing may lead to interdiffusion and loss of magnetism. Alternatively, the Curie temperature of the system can be lowered by tuning the material composition, as demonstrated using FePd in Ref.~\onlinecite{Morley2018}.

While the kagome ice also has states with remanent magnetization that are relatively easy to obtain using external magnetic fields, in contrast to the square ice, the interactions between the three nanoislands at a vertex are degenerate, such that the global ground state could so far not be accessed, either through field-induced demagnetizing protocols or thermal relaxation~\cite{Anghinolfi2015}, but only through direct write of the magnetic state of each nanoelement\cite{Gartside2018}.
Other frustrated lattices, such as Shakti lattices~\cite{Morrison2013,Lao2018}, similarly, are not easily configured into their ground states.

Consequently, defining spin ice geometries in which the magnetic state is fully reconfigurable is essential for magnonic applications. In this context, a different spin ice geometry, the \emph{charge ice}, has recently been investigated by Wang {\em et al.}~\cite{Wang2016}. It consists in replacing specific nanoislands in the square ice with diagonally-oriented elements, while maintaining the locations of the magnetic charges (present at the extremities of the nanoislands) as in square ice. These geometric modifications result in great flexibility: they allow reconfiguring the entire lattice into eight distinct configurations with long-range order, using only an external uniform magnetic field applied at different angles. The equivalent Type-I and Type-II states are shown in Fig.~\ref{fig:vertices}(c).

The relative ease with which the charge ice can be reconfigured is clearly very attractive from the point of view of reconfigurable magnonic crystals. However, this reconfigurability comes at a price: the equivalence between the location of magnetic charges in the square and the charge ices imposes a condition on the relation between the length $\ell$ of the nanoislands and the center-to-center nanoisland separation $d$:
\begin{equation}
    d=\ell\left(1+\sqrt{2}\right).
\end{equation}
As a consequence, the magnetostatic interactions cannot be tuned independently of the nanoisland shape anisotropy, irrespective of nanoisland size. In particular, even if significantly reducing the nanoisland size, which is bound by lithographic limits, the nanoisland separation will remain relatively large and the magnetostatic interactions between nanoislands will be weak. As an example, for nanoislands of dimensions 35~nm$\times$100~nm, the nanoisland  separation is about 240~nm. Weak interactions between the nanoislands make magnon bands relatively flat with small or zero group velocities, and a weak dependence on the magnon spectrum on the magnetic configuration of the lattice. This is a serious impediment to using the charge ice as a reconfigurable magnonic crystal. We discuss possible solutions for enhancing inter-island coupling in Section\ref{sec:charge-ice}.

\section{Experimental techniques}
\label{sec:experimental-techniques}

So far, the most versatile techniques for measuring resonant dynamics in artificial spin ices have been broadband ferromagnetic resonance (FMR) and Brillouin light scattering (BLS) spectroscopy. The main advantage of FMR is its relative ease of use. However, it lacks spatial resolution,  and signal transmission is measured for large ensembles of nanoislands. Consequently, it does not allow, for example, the measurement of antisymmetric modes, in which oscillations of the magnetization with opposite phases cancel out and does not provide a spatial map of the magnetization dynamics. The magnetic structure needs to either be determined using other techniques such as magnetic force microscopy (MFM)~\cite{morgan2011,Zhang2013-crystallites} or transmission electron microscopy (TEM)~\cite{Phatak2011,Li2019_pin}, or through comparison with micromagnetic simulations. 
Microfocused BLS~\cite{SebastianFPhy}, on the other hand, typically allows for the measurement of the mode spectrum with a spatial resolution of a few hundreds of nanometers, allowing to identify, e.g. edge and bulk modes\cite{Li2016modes}. 

A number of other techniques can potentially be used for measuring magnetization dynamics in artificial spin ices and we expect that they will become more broadly used in the near future. In particular, X-ray based imaging exploiting the X-ray magnetic circular dichroism (XMCD)~\cite{Schuetz1987} effect has been employed to measure the evolution of the magnetic state during field-induced magnetization reversal~\cite{Mengotti2011} as well as during thermal relaxation~\cite{Arnalds2013,Farhan2013,Farhan2014,Gliga2017}. Time-resolved stroboscopic measurements taking advantage of the X-ray beam bunch structure currently achieve temporal resolutions below 100~ps\cite{Finizio2018} and allow imaging spin waves\cite{Wintz2016,Foerster2019,Albisetti20}. A pulsed or continuous wave excitation, phase-locked and time-delayed with respect to the photon bunches repetition frequency, is used to excite the magnetization precession. The response of the magnetization can thus be measured at different delay times~\cite{Boero2009}. So far, one of the main challenges of these types of dynamic measurements of spin ices has been achieving the spatial resolution required for detecting rather small variations of the magnetization on length scales of the order of a few tens of nanometers during resonant dynamics.

Recently, the stray field of an artificial spin ice was measured during magnetization reversal~\cite{Wyss2019} using a nanometer-sized superconducting quantum interference device (SQUID) fabricated on the apex of a sharp quartz tip and integrated into a scanning SQUID microscope~\cite{Finkler2010,Vasyukov2013,Vasyukov2018}. The lateral resolution, determined by the tip size, was sufficient to detect the magnetostatically-induced bending of the magnetization at the edges of individual nanoislands. Advantages of this technique include the possibility of combining sub-100~nm lateral resolution with field sensitivities of the order of a few tens of $\mathrm{nT}/\mathrm{Hz}^{1/2}$ and the possibility of measuring in external magnetic fields up to 1~T. In principle, suitable modifications of the electronics and of the detection scheme should allow performing stroboscopic measurements, given that the Josephson junction typically has a characteristic frequency in the GHz range. Alternatively, 
scanning techniques using nitrogen-vacancy (NV)-based sensing~\cite{Maletinsky2012,Rondin2012} have the advantage of working at room temperature and in ambient conditions.

Lateral resolutions of ca. 5~nm can in principle be obtained with aberration-corrected TEM~\cite{McVitie2015,Li2019_pin}. In this case, FMR measurements could be performed \emph{in-situ} using specially designed holders~\cite{Goncalves2017}, while at the same time having the possibility to image and control the magnetic state. 

\section{Theoretical formulation}

The magnetic configurations of ferromagnetic nanoislands in artificial spin ices and their collective dynamics are studied based on the Landau-Lifshitz-Gilbert (LLG) equation~\cite{Gilbert2004}
\begin{equation}
    \label{eq:LLG}
    \frac{\partial\mathbf{M}}{\partial t} = 
    -\gamma\mu_0\mathbf{M}\times\left[\mathbf{H}_\mathrm{eff}-\frac{
    \alpha}{\gamma \mu_0 M_s}\frac{d\mathbf{M}}{d t}\right],
\end{equation}
where $\gamma$ is the gyromagnetic ratio, $\mu_0$ is the vacuum permeability, $\mathbf{M}$ is the magnetization density vector, 
$M_s$ is the saturation magnetization density, and $\alpha$ is a phenomenological, dimensionless magnetic damping parameter. This form assumes that $\alpha\ll1$, as is typically the case for physical systems. For numerical implementations, it is more convenient to rewrite the equation in the form originally proposed by Landau and Lifshitz~\cite{Landau1953}:
\begin{equation}
    \label{eq:LL}
    \frac{\partial\mathbf{M}}{\partial t} = 
    -\frac{\gamma\mu_0}{1+\alpha^2}\mathbf{M}\times\left[\mathbf{H}_\mathrm{eff}+\frac{\alpha}{M_s}\mathbf{M}\times\mathbf{H}_\mathrm{eff}\right],
\end{equation}

The effective field $\mu_0\mathbf{H}_\mathrm{eff}=-\delta E/\delta\mathbf{M}$ parametrizes the relevant physical terms contained in the energy $E$. Typically, the effective field contributions used in artificial spin ices (and, indeed, in most micromagnetic simulations) are: 
\begin{equation}
    \label{eq:Heff}
    \mathbf{H}_\mathrm{eff}=\mathbf{H}_\mathrm{exc}+\mathbf{H}_\mathrm{ani}+\mathbf{H}_\mathrm{zee}+\mathbf{H}_\mathrm{dem},
\end{equation}
which includes exchange ($\mathbf{H}_\mathrm{exc}$), intrinsic anisotropy originating from crystalline spin-orbit coupling or from material structures such as layering, interfaces, or grain structures ~\cite{Stohr2006} ($\mathbf{H}_\mathrm{ani}$), an applied external field ($\mathbf{H}_\mathrm{zee}$), and nonlocal magnetostatic ({\em e.g.,} dipolar)  ($\mathbf{H}_\mathrm{dem}$) fields.

The exchange field can be written as $\mathbf{H}_\mathrm{exc}=M_s\lambda_\mathrm{exc}^2\mathbf{\Delta}\mathbf{m}$, where $\mathbf{m}=\mathbf{M}/M_s$ and $\mathbf{\Delta}$ is the Laplacian. The equation expresses the fact that the energy is minimized when the magnetization is  collinear within a characteristic length, the exchange length $\lambda_\mathrm{exc}$, which is typically on the order of $10$~nm for metallic ferromagnets~\cite{Stohr2006}. 

The magnetostatic field arises from volume charges $\propto\nabla\cdot\mathbf{M}$ and surface charges, due to the discontinuity of the normal component of the magnetization density at boundaries $\propto\mathbf{M}\cdot\mathbf{n}$, where $\mathbf{n}$ is an outward-pointing unit normal vector at an interface.
In artificial square ices, magnetostatic interactions provide the main coupling mechanism between the nanoislands. 

The LLG equation
~\eqref{eq:LLG} subject to the effective field~\eqref{eq:Heff} is, in general, a system of coupled nonlinear partial differential equations. Analytical solutions are typically found in cases where the magnetostatic field is simplified, e.g., in the thin film limit where it reduces to a local field~\cite{GarciaCervera1999}. It must be noted that the effect of magnetostatic field is fundamental to describe the profile and dispersion of long-wave spin waves in thin films, so-called magnetostatic waves~\cite{Stancil2009}. Such spin waves have been instrumental in magnonics research~\cite{Chumak2015}. 
Consequently, numerical techniques are often required in order to solve the LLG equation. 

\subsection{Micromagnetic simulations}
\label{sec:micromag}

Micromagnetic simulations~\cite{Brown1963b} are the most common and powerful tool for solving the LLG equation based on finite-difference or finite-element techniques, while taking magnetostatics into account.  
The system of equations is stiff, which means that time-integration is usually done using implicit time-steppers as small errors in explicit schemes can easily grow exponentially. A recent review of computational micromagnetics can be found in Ref.~\onlinecite{Abert2019}.

While artificial spin ices can be conceptually thought of as bar magnets~\cite{Castelnovo2008}, the magnetization within individual nanoislands is typically non-uniform~\cite{Gliga2015,Jungfleisch2016}. 
Thus, in order to simulate the properties of large arrays, 
a compromise has to be reached between spatial resolution and the spatial extent of the simulated domain. A common approach 
consists in reducing the artificial spin ice to a micromagnetic super-cell and imposing periodic boundary conditions, which mimic an infinite lattice. An accurate calculation of the static magnetic states can be obtained in this way ~\cite{Sklenar2018,Iacocca2017c}. The same principle can be naturally extended to determine resonances~\cite{Iacocca2016,Jungfleisch2016,Iacocca2017c,Iacocca2019,Dion2019}. An alternative approach relies on simulating small artificial spin ice lattices, which include as many unit cells as computational resources allow, without periodic boundary conditions. This approach was for example used to determine the effect of topological defects~\cite{Mengotti2009,Mengotti2011} on the dynamical spin wave spectrum of square ice~\cite{Gliga2013}. Reduced lattices have been also used to determine normal modes via the dynamical matrix method~\cite{Grimsditch2004} and to study in detail the magnetostatic interaction between nanoislands, e.g., in pairs of nanoislands~\cite{Dvornik2011}, frustrated vertices~\cite{Bang2019}, and pinwheel artificial spin ices~\cite{Paterson2019}.

Micromagnetic simulations have been also used to probe magnetic transport in ``connected'' artificial spin ices, discussed in Ref.~\onlinecite{Lendinez2019}.

\begin{figure*}[t]
 \centering 
\includegraphics[width=7.0in]{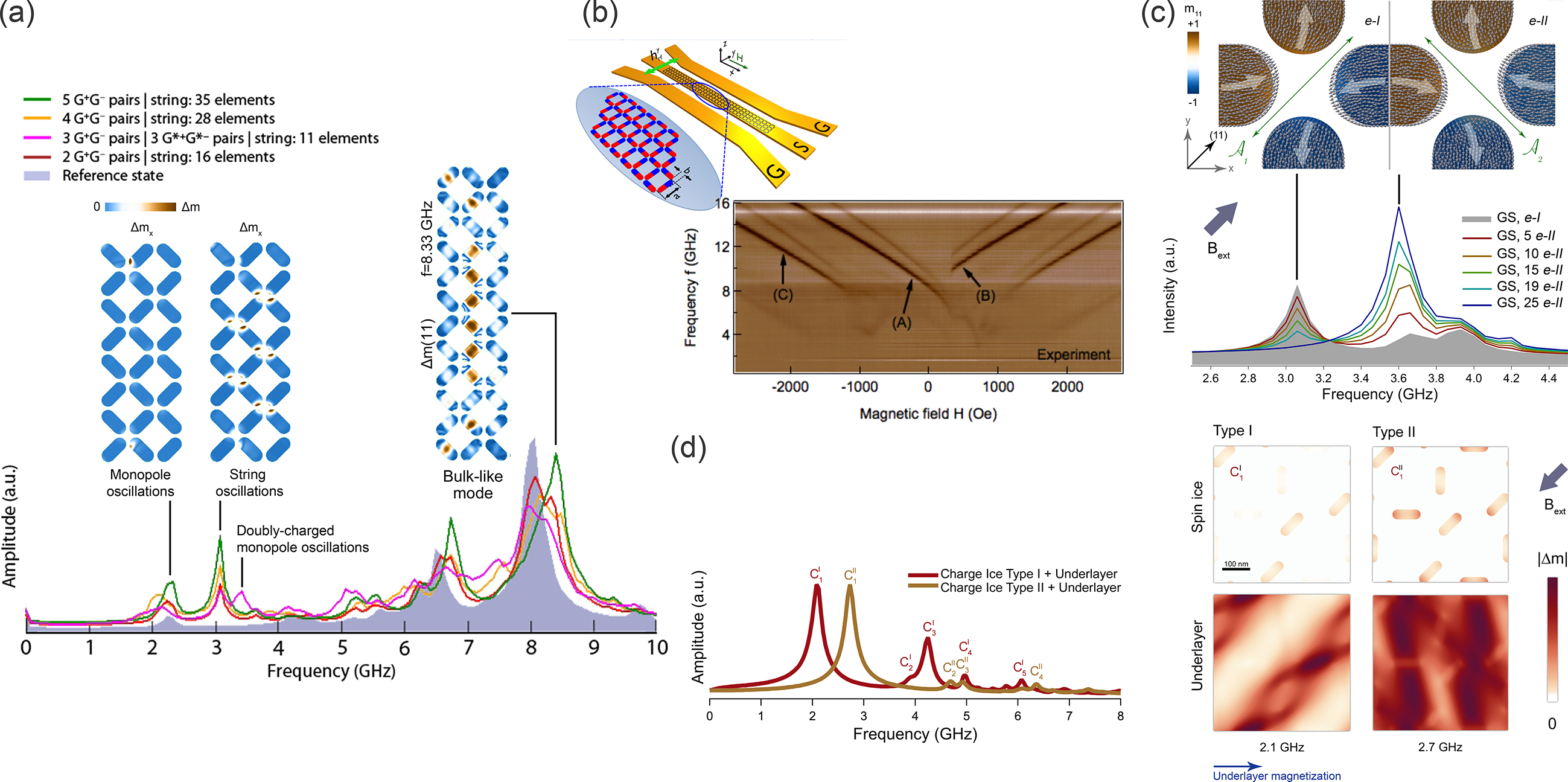}
\caption{ \label{fig:resonances} 
(a) Micromagnetic simulations of the evolution of the magnetization dynamics spectrum as a function of increasing string length (Type I vertices) and number of singly-charged (Type~III vertices) and doubly-charged (Type~IV vertices) monopole-antimonopole pairs compared to the remanent (reference) state (Type~II vertices). The bulk-like mode (shown at 8.33~GHz) is shifted with respect to the same mode in the reference state. 
Figure adapted and reprinted with permission from [\href{https://doi.org/10.1103/PhysRevLett.110.117205}{S. Gliga, A. K\'{a}kay, R. Hertel, O, Heinonen, Phys. Rev. Lett., \textbf{110}, 117205 (2013)}]. Copyright 2013 by the American Physical Society.
(b) Illustration of a spin ice array patterned on top of a coplanar waveguide for broadband ferromagnetic resonance measurements configuration. Measured ferromagnetic resonance response of a square ice. Reprinted figure with permission from [\href{https://doi.org/10.1103/PhysRevB.93.100401}{M. B. Jungfleisch, W. Zhang, E. Iacocca, J. Sklenar, J. Ding, W. Jiang, S. Zhang, J. E. Pearson, V. Novosad, J. B. Ketterson, O. Heinonen, and A. Hoffmann, Phys. Rev. B, \textbf{93}, 100401(R) (2016)}]. Copyright 2016 by the American Physical Society.
(c) Top: vertex symmetry breaking through bending of the magnetization at the edges of the nanoelements. Two possible degenerate configurations of the magnetization in strongly magnetostatically coupled nanoislands are considered (labeled \emph{e-I} and \emph{e-II}). Each have a single axis of magnetic
symmetry ($\mathcal{A}_1$ or $\mathcal{A}_2$), effectively  lowering the vertex symmetry, which would have two axes (both $\mathcal{A}_1$ and $\mathcal{A}_2$) in the absence of edge bending. Bottom: The ground state spectrum (gray) for a lattice only made of \emph{e-I}-type vertices displays a peak at 3 GHz. Introducing an increasing number of \emph{e-II} vertices in the system leads to a decrease of the original mode intensity and to the evolution of a peak at 3.6 GHz corresponding to the mode associated to the \emph{e-II} vertices. Reprinted figure with permission from [\href{https://doi.org/10.1103/PhysRevB.92.060413}{S. Gliga, A. K\'{a}kay, L. J. Heyderman, R. Hertel, and O. G. Heinonen, Phys. Rev. B, \textbf{92}, 060413(R) (2015)}]. Copyright 2015 by the American Physical Society.
(d) Simulated mode spectra of charge ice on an underlayer. While the interaction between nanoislands in this geometry is very weak, coupling it to a Permalloy magnetic underlayer leads to the very different spectra for Type I and Type II configurations. The spatial distribution of the lowest frequency Type I and Type II modes (shown on the right) reveals underlayer modes, which define spin wave channels. Figure adapted from Ref.~\onlinecite{Iacocca2019}.
}
\end{figure*}

\subsection{Semi-analytical models}

The computational cost of micromagnetic simulations can be greatly reduced by analytical methods. While simple band structures can be found exactly in some cases, e.g. the well-known Bloch waves for free electrons in a periodic potential~\cite{Balkanski2000}, computing realistic band structures typically necessitate 
semi-analytical models. 

Semi-analytical models require two key simplifications. First, the exchange field is considered to be negligible, assuming that the magnetization, $\mathbf{M}$, is approximately uniform within each nanoisland. Second, the magnetostatic field is treated as a dipole-dipole interaction between nanoislands with an effective field acting on nanoisland $i$ of the form:
\begin{equation}
    \label{eq:dipole}
    \mathbf{H}_{d}^i=\frac{V}{4\pi}\sum_j\left[\frac{3\mathbf{r}_{i,j}(\mathbf{M}_j\cdot\mathbf{r}_{i,j})}{|\mathbf{r}_{i,j}|^5}-\frac{\mathbf{M}_j}{|\mathbf{r}_{i,j}|^3}\right].
\end{equation}
The sum is performed over the whole lattice, $V$ is the volume of the magnetic element, and $\mathbf{r}_{i,j}$ is the distance between two magnetic elements $i$ and $j$. Note that care has to be taken when summing the long-range dipolar interactions over the entire lattice to ensure correct convergence. In Eq.~(\ref{eq:dipole}), the magnetic elements can represent an entire nanoisland or subdivisions of a nanoisland. With this approximation, and assuming conservative dynamics ($\alpha=0$), the LLG equation reduces to the Larmor torque equation expressed as 
a set of vector, coupled, ordinary differential equations that are much simpler to treat analytically. This approach has been used in the context of magnetic nanodots~\cite{Bondarenko2010,Verba2012,Lisenkov2016}, Fe-intruded yttrium iron garnet (YIG)~\cite{Shindou2013}, and ``decorated'' honeycomb lattices~\cite{Shindou2013b}.

A model that takes into account both the dipolar field and the edge canting of the magnetization in nanoislands was proposed in Ref.~\onlinecite{Iacocca2016}. To this effect, a tight-binding-like approach is used to calculate the effective field acting on the artificial spin ice unit cell. Key to this process is the reduction of the long-range dipole-dipole field into intra-unit-cell and inter-unit-cell components. The latter can be pre-computed to desired numerical accuracy as detailed in the Appendix of Ref.~\onlinecite{Iacocca2016}.

\section{Resonant magnetization dynamics}
\label{sec:resonant_mag_dyn}

\subsection{Square ice}
The magnetization dynamics in square ice was first investigated in Ref.~\onlinecite{Gliga2013}, in particular the influence of topological defects on the resonant dynamics of the square ice. Such defects occur when the magnetization at a vertex is not in a two-in/two-out state (so-called \emph{ice rule}, corresponding to Type-I and Type-II vertices), resulting for example in vertex states where the magnetization is in a one-in/three-out state (`monopoles' or Type-III vertices) or a even four-in state (doubly charged monopoles or Type-IV vertices), shown in Fig.~\ref{fig:vertices}(a). In the pyrochlore compounds, such defects have been found to display behavior similar to that of Dirac monopoles~\cite{Castelnovo2008}. These emergent monopoles occur in pairs (e.g., monopole-antimonopole) connected by a string, along which the magnetization is reversed with respect to a reference state. While it was known that these defects affect the equilibrium behavior and the magnetization reversal in spin ices~\cite{morgan2011,Phatak2011,Budrikis2012,Westphalen2008}, Gliga \emph{et al.}\cite{Gliga2013} found that each type of topological defect as well as the strings of reversed magnets connecting these defects display distinct and localized features, both spatially as well as in frequency, as summarized in Fig.~\ref{fig:resonances}(a). These features, in the GHz frequency range, thus act as fingerprints for each type of defect. 

The resonant dynamics of long-range ordered square artificial spin ices were first investigated experimentally by Jungfleish {\em et al.}~\cite{Jungfleisch2016}. Using broadband FMR spectroscopy, they found a number of modes within a range between 4~GHz and 16~GHz as a function of in-plane magnetic field, shown in  Fig.~\ref{fig:resonances}(b). Some of the lower-frequency modes disappeared or exhibited complex hysteretic behavior at low fields. Detailed comparison with micromagnetic simulations showed that the hysteresis in the mode spectrum was related to the magnetization configuration of particular nanoislands, resulting from the applied field history. 
The experimentally measured resonance spectroscopy was  quantitatively described by a semi-analytical model for the Type-II configuration.

More recently, Ghosh {\em et al.}~\cite{Ghosh2019} also experimentally investigated the resonant modes of square ices  using FMR spectroscopy and the evolution of those modes as a function of nanoisland thicknesses. By comparing their experimental results with micromagnetic modeling, they could identify bulk-like as well as (symmetric) edge modes in the spectra. They also identified local configurations during magnetization reversal as well as topological defects, as predicted in Ref.~\onlinecite{Gliga2013}. The resonant modes in square ice have been spatially mapped by Li {\em et al.}~\cite{Li2016} using micro-BLS. Recently, anti-spin ice systems consisting in thin films with geometrically-placed holes instead of nanoislands (analogous to antidots), have also been studied using BLS. These structures have been found to support  frequency-dependent spin wave confinement in regions between holes~\cite{Mamica2018}.

Beyond macrostates, it has also been found that seemingly small variations in the magnetic state of individual elements could equally affect the magnetization dynamics. In particular, due to magnetostatics, in elements above a certain thickness (ca. 5--10~nm in Permalloy, depending on lateral dimensions and thickness) the magnetization state changes from an `onion' state (mostly uniform) to \emph{C} or \emph{S} states in which the magnetization bends at the extremities of the element~\cite{Rougemaille2013}. These changes affect the magnetic symmetry of the vertices and the torques in the presence of an applied field, as shown in Fig.~\ref{fig:resonances}(c), resulting in distinct mode spectra for \emph{C} and \emph{S} configurations~\cite{Gliga2015}. Additionally, the presence of such internal degrees of freedom affects the thermal evolution of the system, giving rise to edge \emph{melting}, in which the magnetization stochastically switches between the \emph{C} and \emph{S} states. This behavior is reflected in the mode spectrum of the thermal magnetization dynamics in the form of 1/$f$-type flicker noise at low frequencies~\cite{Gliga2015}. 

\begin{figure}[t]
    \centering
    \includegraphics[width=3.3in]{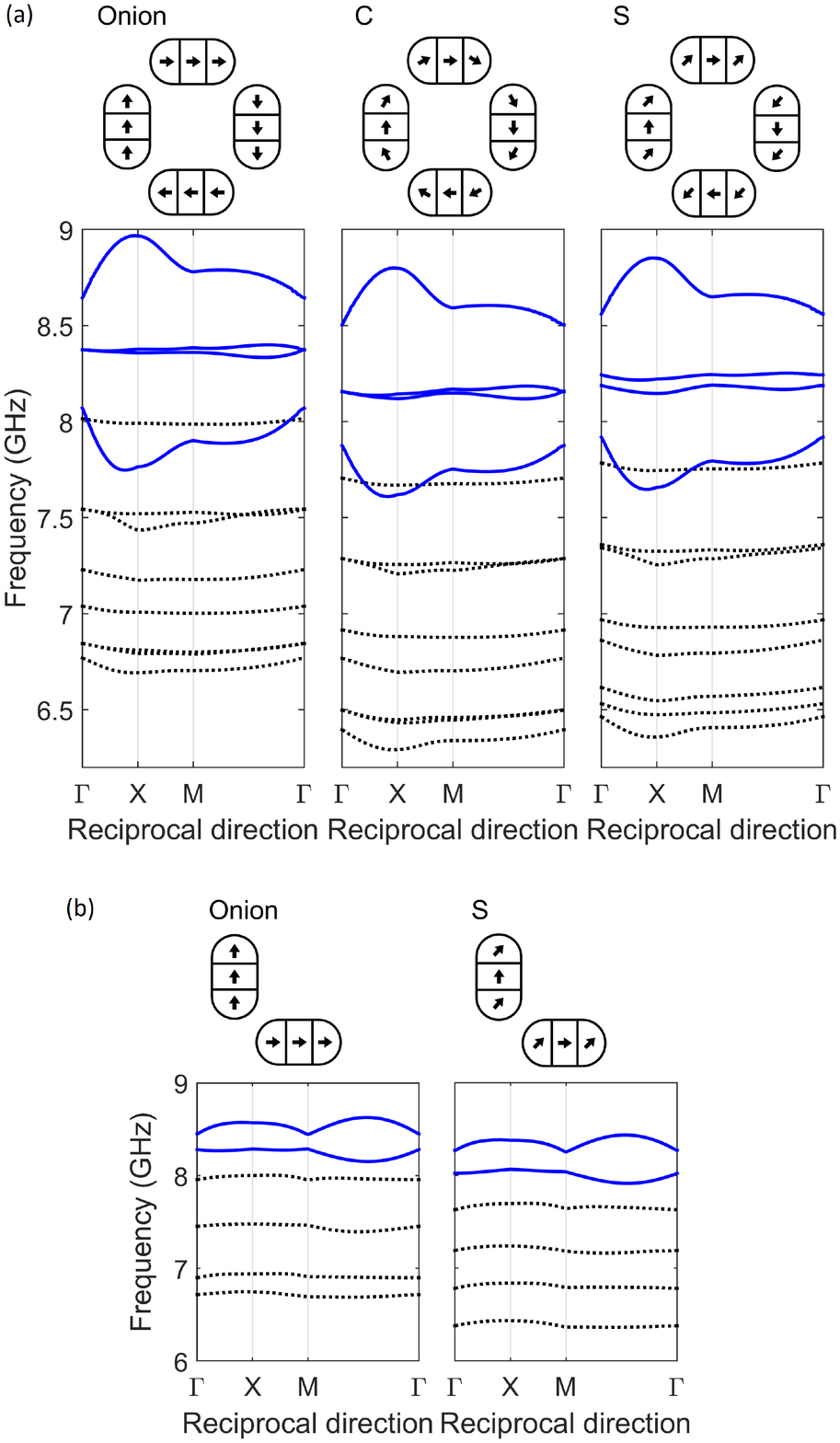}
    \caption{(a) Magnon band dispersions for a square ice in the Type-I state. The bulk and the edge modes are respectively depicted by solid blue and black dashed lines. (b) Magnon band dispersions for a square ice in the Type-II state. The bulk and edge modes are respectively depicted by solid blue and black dashed dashed lines. Reprinted figure with permission from [\href{https://doi.org/10.1103/PhysRevB.93.134420}{E. Iacocca, S. Gliga, R. L. Stamps, and O. Heinonen, Phys. Rev. B, \textbf{93}, 134420 (2016)}]. Copyright 2016 by the American Physical Society.}
    \label{fig:magnons}
\end{figure}

\subsection{Kagome ice}

The magnetization dynamics of kagome artificial spin ices were investigated experimentally by Dion {\em et al.}~\cite{Dion2019} and based on micromagnetic simulations by Arroo \emph {et al.}~\cite{Arroo2018sculpting}. In Ref.~\cite{Dion2019}, the patterned shape of the nanoislands, and thus the shape anisotropy, was altered in the three sublattices, such that the three-fold rotational symmetry was broken. Different resonant modes and responses could then be obtained by aligning an external magnetic field along the three inequivalent directions. Moreover, because of the different coercive fields of the nanoislands in the three sublattices, a variety of  microstates could be obtained through the application of a magnetic field in different directions. These microstates were shown to have different spin wave spectra using micromagnetic modeling. In Ref.~\cite{Arroo2018sculpting}, the mode spectrum in kagome ice was investigated, demonstrating that the magnetic microstate influences the spin-wave spectra. In addition to mode shifting, the uniform mode can be strongly be enhanced or suppressed, thus allowing its activation and deactivation. 
The magnetization dynamics in a connected kagome lattice was studied by Bhat, Watanabe, Baumgaertl, and Grundler~\cite{Bhat2019}. Despite being connected, they showed that topological defect configurations gave rise to different, distinguishable dynamics, and that Dirac strings connecting two topological defects induced pronounced modifications in magnon frequencies.

\subsection{Charge ice}
\label{sec:charge-ice}

The charge ice geometry, introduced in Section~\ref{sec:reconfigurable}, is based on square ice, and its global magnetic state is fully reconfigurable by applying magnetic fields at successive angles, in a well-defined order. The downside of this system is that the geometric constraints in building the lattice lead to largely spaced and, consequently, weakly coupled elements. Addition of an exchange biased magnetic underlayer~\cite{Iacocca2019} has been shown to lead to interactions between the spin ice and the magnetic film that significantly modify the modes of the spin ice, in particular suppressing the amplitude of the bulk modes and leading to significant differences in the mode spectra of Type-I and Type-II states (spectra in Fig.~\ref{fig:resonances}d). In addition, coupling to the uniaxially anisotropic thin film gives rise to modes in the underlayer, two of which are shown in Fig.~\ref{fig:resonances}d, and can act as spin wave channels, similar to those present in antidot magnonic crystals~\cite{Sklenar2013,Mamica2018}. 

\subsection{Magnonic band structure}

 The magnon bands in square ices was investigated numerically by Iacocca, \emph{et al.}~\cite{Iacocca2016}. Square ice in  Type-I and Type-II states was considered and the effect of the magnetic state as well as of the edge bending of the magnetization on the magnon dispersions was determined. 
Figures~\ref{fig:magnons}(a) and~\ref{fig:magnons}(b) depict the magnon dispersion for the two configurations with the possible edge bending states. Clearly, the magnon bands in the Type-I and Type-II configurations are very different. The Type-I configuration has four bulk bands representative of the four nanoislands in the unit-cell, while the Type-II configuration with only two nanoislands in the unit-cell has two bulk bands. Second, the Type-II bands, in particular the bulk bands, are much flatter than the Type-I dispersive bulk bands with a much smaller forbidden gap than between the Type-I dispersive bulk bands. Beyond introducing a semi-analytical model for calculating the band structure, that work clearly demonstrated that square spin ices can be viewed as reconfigurable magnonic lattices. Toggling between magnetic states results in fundamentally different magnonic band structures.

An interesting, and presently relevant, question arising in the context of band structures is whether magnonic bands can be designed to be topologically non-trivial, and perhaps even toggled between topologically trivial and non-trivial band structures. In general, systems with topologically non-trivial band structures will necessarily exhibit edge modes at the interface with a topologically trivial structure. Perhaps the best known example are time-reversal-invariant topological insulators, which exhibit edge states in the gap. In the case of a topological insulator, the edge states have a specific spin-momentum locking so that, in the absence of impurities that break time-reversal invariance, the edge states propagate without scattering. In general, the topological edge states exhibit some kind of chirality coupling the propagation vector to some internal degree of freedom.

Shindou and co-workers~\cite{Shindou2013,Shindou2013b} theoretically examined periodic magnetic structures that can exhibit topologically non-trivial magnon bands. While the structures they considered are technically not artificial spin ices, they are nevertheless interesting to discuss in the context of magnonic lattices. The first considered  structure~\cite{Shindou2013} was a yttrium iron garnet (YIG) thin film in which a periodic rectangular array of circular holes 
was introduced with lattice constants $(a_x,a_y)$. The holes were filled with Fe. The key to obtaining topologically non-trivial magnon bands is that the dipolar interaction between the Fe cylinders can act analogously to the spin-orbit interaction and cause a gapped band inversion when the Fe cylinders are in a periodic array (as opposed to in a continuous film). When the unit-cell size $\lambda=\sqrt{a_xa_y}$ becomes larger than a typical magnetic exchange length, a gap opens up and the lowest magnon band acquires a non-trivial topology. Reducing $\lambda$ takes the system to a topological transition where the band gap closes and the two lowest magnon bands form Dirac cones at the band closing points. Shindou and co-workers also demonstrated that when the system is gapped with a non-trivial topology, chiral edge states form which propagate unidirectionally.

A second structure considered in Ref.~\onlinecite{Shindou2013b} was a two-dimensional periodic array of square or honeycomb lattices of ferromagnetic particles. Here, the particles were assumed to be small enough such that each one could be treated as a single macrospin. In the presence of an external magnetic field perpendicular to the plane of the arrays, these arrays could admit magnon bands with non-trivial topologies with chiral edge modes.

An extension of the semi-analytical model in Ref.~\onlinecite{Iacocca2016} was used to investigate a square ice on top of a heavy metal thin film, such as Pt~\cite{Iacocca2017c}. It is well known that Py or other ferromagnetic thin films deposited on a heavy-metal spin-orbit scatterer, such as Pt, Pd, Ta, or W, leads to an interfacial Dzyaloshinskii-Moriya interaction (DMI)\cite{fert1980role,fert1990magnetic,thiaville2012dynamics,yanes2013exchange}. The DMI allows for a chiral magnetization structure\cite{heide2008dzyaloshinskii,Bode2007DMI}. The principle demonstrated in Ref.~\onlinecite{Iacocca2017c} was that the presence of DMI can lead to a topologically non-trivial magnon band structure in the square ice. Indeed, upon increasing the DMI strength, a band inversion can occur between the two lowest magnon bands forming Dirac cones with non-trivial topologies.

\section{Perspectives}
\label{sec:perspectives}

\subsection{Reconfigurability}

As indicated in Fig.~\ref{fig:summary}, one of the main challenges in using spin ices as reconfigurable magnonic lattices is finding a geometry whose magnetic state is easily reconfigurable and exhibits sufficiently strong magnetostatic interactions to generate rich dynamics. 
The square and kagome ices, along with modified geometries such as Shakti lattices can easily be reconfigured to a limited number of long-range ordered remanent states, using external fields. However, generally a large number of states, including the ground state, are difficult to access. Here, we discuss, solutions and perspectives to overcome this situation, including using local probes provided by scanning probe microscopy, spin transfer torques and lattice geometries.  

Recently, Gartside {\em et al.}~\cite{Gartside2016,Gartside2018} used a high-moment magnetic force microscopy (MFM) tip to controllably reverse the magnetization in selected spin ice nanoislands. This tip acts as a monopole source when in close proximity to the magnetic nanostructures. As the tip is moved perpendicular to the long axis of the nanoisland, a domain wall pair is nucleated. As the tip completes its motion across the nanoisland, the domain walls move apart towards the opposite ends of the nanoisland, leaving a reversed magnetization behind. They consequently named this technique \emph{topological defect-driven magnetic writing}. 
Reference~\onlinecite{Gartside2018} further demonstrated that it is possible to design configurations both in connected and magnetostatically coupled kagome spin ices, including the kagome ground state as well as more exotic out-of-equilibrium states. An MFM tip has also been used in Ref.\cite{Wang2016} to locally define the magnetic state of charge ice as well as in Ref.\cite{Lehmann2019} to define the magnetic state of individual elements in order to achieve different ferroto-toroidic states in an artificial spin ice structure forming an artificial magneto-toroidal crystal. 
While this technique is versatile, it also is rather slow as it hinges on mechanical motion of the tip. 
Other means of generating stray fields to nucleate domain walls can be envisioned, which do not rely on mechanical motion. As an example, Gartside {\em et al}. ``envisage a system comprising a three-dimensional network of nanowires whereby current-controlled domain walls replace the magnetic force microscope tip, greatly enhancing flexibility, throughput and integration with existing
technologies". 
However, such schemes also present limits: 
as the size of the nanoislands shrinks close to the domain-wall width, the reversal of nanoislands becomes coherent (Stoner-Wohlfarth switching) rather domain-wall driven. An external field, whether generated by domain walls in nanowires or by other means, will have to be large enough to overcome the coercive field of the nanoislands, which can typically be larger than the field required to nucleate domain walls. 

Another means of reconfiguring the magnetic state consists in using spin transfer torque (STT) to switch magnetic nanostructures and nanoislands. This effect is equally used in  commercially available STT magnetic random access memories (STT-MRAMs)\cite{Apalkov2013}. However, designing and realizing an artificial spin ice in which individual nanoislands are switched using STT requires complex deposition and patterning techniques. This can certainly be overcome in principle as for STT-MRAMs, but it would make reconfigurable artificial spin ices considerably more expensive and, perhaps, technologically out of range for academic laboratories. Another issue would be that stray fields from polarizing layers in STT devices would have to be mitigated. This can also in principle be done using, e.g. synthetic antiferromagnets as polarizing layers, but again would generate increasing complexity and cost.

A different avenue to reconfigure artificial spin ices is to explore new lattices that may allow for simpler protocols. As discussed in Sec.~\ref{sec:charge-ice}, charge ice allows for reconfiguration using a saturating field along different directions at the price of reduced coupling between the nanoislands~\cite{Wang2016}. In contrast, it has been shown that strongly coupled nanoislands can support long-range order according to a well-defined phase diagram. Sklenar \textit{et al.}~\cite{Sklenar2018} investigated a quadrupole lattice that could support both ferro-quadrupolar and antiferro-quadrupolar long-range order. Long-range ordered states were observed after annealing without any applied field and a full phase diagram was computed by Monte Carlo simulations, establishing clear field and temperature transitions for the long-range ordered state, paramagnetic state, and their coexistence. 

Another recently investigated geometry consists of a chiral pattern, obtained by rotating the elements in each vertex by 45$^{\circ}$. This \emph{chiral ice} has been found to exhibit ratchet behavior during thermal relaxation~\cite{Gliga2017}. Indeed, following saturation, the net vertex magnetization rotates in a single direction (e.g. clockwise) during thermal relaxation at room temperature. Thus, while the magnetization dynamics is locally stochastic, globally it unfolds in a well-defined direction. The final magnetic state can be defined by using a weak bias field. While this allows a certain degree of reconfigurability, which may lead to the creation of functional materials, it also requires very thin magnets (ca. 2--3 nm thick) and it is not clear at present to which extent the resonant dynamics is interesting. Li \emph{et al.}~\cite{Li2019_pin} have used thicker nanoislands in the same pinwheel geometry and showed that it is possible to obtain well-defined configurations during field-induced magnetization reversal. These thicker elements equally exhibit edge bending of the magnetization~\cite{Paterson2019,Wyss2019}, as well as chiral dynamics at the vertex level~\cite{Wyss2019} and might possess rich spectral features. One of the main advantages of this geometry is that the ground state is ferromagnetic~\cite{Macedo2018} and can trivially be obtained through saturation in an external field.  In addition, it is in principle possible to engineer physical defects in the lattice in order to achieve local control of the magnetic structure. In Ref.~\onlinecite{Drisko2016}, physical defects  such as a missing nanoisland accompanied by lattice distortion were introduced in square ice. Such defects have been found to lead to the formation of domain walls across ground state regions. In these systems, the spin ice cannot support continuous ground-state ordering, demonstrating that a single physical defect can alter the topology of the system, thus providing a possible path for tuning the magnetic ordering.

In a kagome lattice, Chopdekar \emph{et al.}~\cite{Chopdekar2013} tailored the shape anisotropy of specific nanoislands, and thus their switching fields, to achieve desired states with near perfect reliability. 

Finally, we note that while magnonic response is confined to the GHz regime, artificial spin ice dynamics over a large range of frequencies still remains unexplored, ranging from the quasi-static measurements of hysteretic behavior or thermal relaxation (a few Hz) to a few GHz. For example, it has been proposed that using microwave assisted switching, it should be possible to facilitate the nucleation of certain types of defects in connected kagome ices~\cite{Bhat2016}, and AC fields could also be used for clocking atificical spin ice-based logic architectures~\cite{Arava2018,Jensen2018}. 


\subsection{Intrinsic damping}

Another challenge in realizing magnonic crystals with spin ice is due to the relatively large damping in ferromagnetic transition metals, such as Ni, Co, Fe, and their common inter-metallic alloys. This is a result of the spin-orbit interactions, which cause dephasing of the magnetization dynamics (see, for example, Skadsem {\em et al.}~\cite{Skadsem2007}).  For example, in Permalloy the value of the damping constant $\alpha$  is of about 0.008~\cite{Liu2007PRL}. While this is small enough for a number of studies of magnetization dynamics, such as vortex motion in Permalloy disks, it still is sufficiently large that linear spin wave packets only propagate over distances of the order of a few hundred nanometers~\cite{Madami2011} before being damped out, and magnetization dynamics is damped out within 5~ns\cite{Liu2007PRL}. This obviously limits applications such as logic devices, in which wave packet propagation is desirable over large distances, for example between features as well as to gate the propagation of the wave packets. An obvious possibility to extend the propagation length of spin waves is to use materials with smaller intrinsic damping. Relatively recently, it was discovered that the intermetallic CoFe alloy can have a very small intrinsic damping~\cite{Schoen2016} of about $10^{-4}$. This occurs at a Co-concentration of about 25\%, when the density-of-states at the Fermi energy has a sharp minimum which limits the scattering of electrons. This is of great interest as CoFe alloys can be deposited using a range of different techniques, including sputtering. Additionally, there is a long history of using CoFe alloys in academic and industrial research, as well as in industrial applications. Another advantage of CoFe alloys is their large magnetic moment: their polarization $\mu_0 M_s$ is over 2.0~T at this Co concentration. 
It is also important to keep in mind that at 25\% Co, the alloy structure is bcc rather than fcc, with much larger magnetostriction and magnetocrystalline anisotropy than the fcc alloys.

Another material of interest is YIG, with a very low dimensionless damping of about $10^{-4}$. However, YIG is not easily grown in thin films and it is difficult to pattern. Recent advances in pulsed laser deposition have demonstrated YIG thin films of thickness down to 3.4~nm\cite{Mendil2019}. However, the saturation polarization is smaller, $\mu_0M_s\approx150$~mT than its bulk value of about $180$~mT at room temperature for relatively thick films (90~nm). The saturation magnetization density decreases with in thin films, and decreases especially rapidly below 10~nm. Other possibilities are Heusler alloys\cite{Felser2015}, especially half-metallic L2$_1$ Heusler alloys such as Co$_2$MnAl or Co$_2$MnSi\cite{Jourdan2014}, which can exhibit very small damping. 

We note that very low damping has recently been measured in magnetically soft epitaxial spinel NiZnAl ferrite thin films, which also exhibit strong magnetoelastic coupling\cite{Emori2017}. Such materials potentially allow the development of spin--mechanical devices, opening a new route for realizing reconfigurable magnonic crystals.

\subsection{Coupling schemes}

A number of interesting directions of research involve hybrid structures combining artificial spin ices with different systems. In Section~\ref{sec:charge-ice}, we have described a heterostructure in which the charge ice was coupled to a soft magnetic thin underlayer as a means of increasing and modifying the interaction between the nanoislands. Other types of heterostructures can equally be used to define specific behavior and functionalities. Wang {\em et al.}\cite{Wang2018} placed a charge ice on top of a Type-II superconducting thin film. The magnetic charges in the spin ice gave rise to stray fields that introduced a vortex lattice in the superconducting thin film. Using the ability to controllably reconfigure the charge ice, the state of the vortex lattice could be toggled between geometrically frustrated, highly degenerate vortex lattices, and non-frustrated ones. Different charge ice configurations gave rise to distinct transport properties, and Ref.~\onlinecite{Wang2018} demonstrated that this heterostructure can be used to realize reprogrammable superconducting electronic devices. Golovchanskiy et al.\cite{golovchanskiy2019} also built a hybrid metamaterial structure of Nb and Py thin films. In their work, they showed that the diamagnetic response of the superconductor affected the magnon dispersion, and furthermore they could distinguish between the Meissner state of the superconductor (in which all magnetic flux is expelled) and the mixed state, in which penetrating flux gives rise to a vortex lattice.
One can also envisage patterning artificial spin ices on top of Type-I superconductors. In these, the magnetic stray fields from the spin ice will not form vortices due to field penetration, as in Type-II superconductors. Instead, as long as the fields are below the critical field $H_c$, the superconductor will act as a perfect magnetic mirror. This should radically change the magnetostatic interactions between the nanoislands as well as within the nanoislands, as shown by Golovchanskiy et al.\cite{golovchanskiy2019}. In addition to reconfiguring the magnetic state of the artificial spin ice, one could also drastically alter the magnetostatic interactions simply by changing the temperature of the superconductor, above or below the critical temperature.

Other exotic materials, in particular materials with topologically non-trivial electronic properties, exhibit interesting interactions with magnetic materials and magnetic fields which can in principle be exploited. For example, topological insulators~\cite{Hasan2010,Bansil2016,kane2005z,kane2005,bernevig2006quantum,fu2007topological,moore2007topological} with time-reversal symmetry have gapless topological surface states, which cannot backscatter due to spin-momentum locking and are (in principle) dissipationless. Spin-momentum locking can be also exploited in spintronics applications: for example, the topological surface states in the topological insulator Bi$_2$Se$_3$\cite{xia2009observation,zhang2009topological} were used to switch the magnetization~\cite{Li2019} of the insulating ferromagnet BaFe$_{12}$O$_{19}$. The switching efficiency at low temperatures was much higher than in Pt/BaFe$_{12}$O$_{19}$ heterostructures using spin orbit torques. In topological insulators such as Bi$_2$Se$_3$, the topological surface states are protected by time-reversal symmetry, and magnetic impurities that break this symmetry generally suppress or even destroy the topological surface states. But the question of how topological insulators and their topological surface states interact with lattices of magnetic charges is an interesting one. A further question is whether the topological surface states and their charge and spin transport properties can be manipulated using reconfigurable artificial spin ices. Other topological materials include Dirac and Weyl semimetals\cite{Armitage2018}. Dirac semimetals\cite{wang2012dirac,liu2014stable,young2012dirac} are gapless, obey time-reversal and inversion symmetry, and have a number of Dirac cones that are doubly degenerate. Weyl semimetals\cite{wan2011topological,xu2011chern,lv2015experimental} can be obtained from Dirac semimetals by breaking either inversion or time-reversal symmetry; breaking either symmetry lifts the degeneracy of the Dirac cones, and each Dirac point separates into two non-degenerate Weyl nodes. Both Dirac and Weyl semimetals also admit surface states, Fermi arc states, which have specific spin-momentum lockings. While the Dirac Fermi arc states are protected by symmetry, the Weyl Fermi arc states are topologically protected and are more robust than the Dirac Fermi arc states. Weyl semimetals with broken time-reversal symmetry are in general magnetic and so respond to magnetic fields. This opens exciting directions for developing functional materials by combining magnetic Weyl semimetals with different materials. For example, heterostructures of magnetic Weyls exhibit an unusual inverse Edelstein effect, which converts a pure spin current to a charge current\cite{Zhang2019}; the unusual properties, such as a pronounced anisotropy, originate in the topological properties of the electronic states. It is thus likely that reconfigurable artificial spin ices could be combined with Dirac or Weyl semimetals to affect and manipulate the spin charge transport properties, especially of the Fermi arc states. Such heterostructures could possibly also be used to mediate interactions and entanglement of quantum states, thereby enabling new avenues for devices in quantum computing and quantum sensing.


\subsection{Three-dimensional structures}
\label{sec:3D}
\begin{figure}[t]
\centering \includegraphics[width=3.3in]{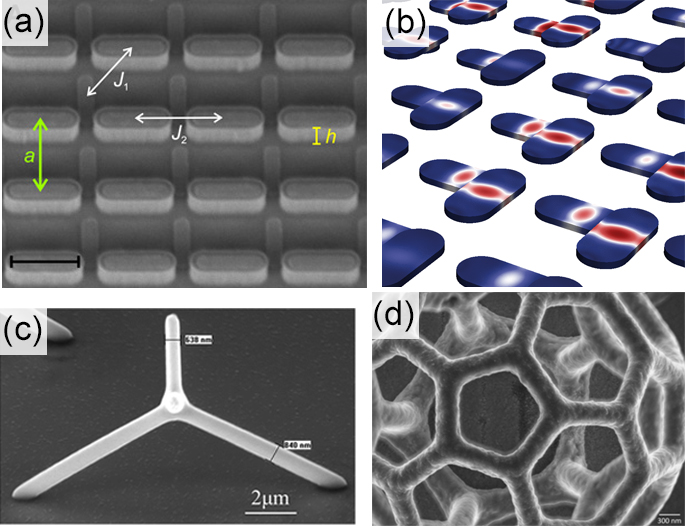}
\caption{ \label{fig:3D} (a) Schematic view of 3D square ice, with a height offset between the two nanoisland sublattices. $J_1$ and $J_2$ respectively represent the interactions between neighboring and opposite elements~\cite{Perrin2016}. 
Figure reproduced from [\href{https://advances.sciencemag.org/content/5/2/eaav6380}{A. Farhan, M. Saccone, C. F. Petersen, S. Dhuey, R. V. Chopdekar, Y.-L. Huang, N. Kent, Z. Chen, M. J. Alava, T. Lippert, A. Scholl, S. van Dijken, Sci. Adv. 5, eaav6380 (2019)}] under Creative Commons Attribution License. 
(b) Micromagnetic simulations of resonant modes in 3D square ice ground state~\cite{}. (c) Cobalt tetrapod structures fabricated using two-photon lithography. [\href{https://doi.org/10.1039/C7NR07843A}{S. Sahoo, S. Mondal, G. Williams, A. May, S. Ladak, A. Barman, Nanoscale 10, 9981 (2018)}] - Published by The Royal Society of Chemistry.
(d) SEM image of a Ni plated mesoscopic 'buckyball' structure fabricated by two-photon lithography. The structure is 5 $\mu$m in diameter. 
Reprinted from \href{https://doi.org/10.1016/j.mattod.2019.05.001}{Materials Today 26, 100, S. Gliga, G. Seniutinas, A. Weber, Ch. David, Architectural structures open new dimensions in magnetism: Magnetic buckyballs (2019)}, with permission from Elsevier.}
\end{figure}

Recently, the possibility of creating three-dimensional (3D) structures using using e-beam lithography~\cite{Kirchner2015}, two-photon lithography\cite{Maruo97} as well as focused electron beam induced deposition~\cite{Keller2018,Fowlkes2018} has opened radically new possibilities for defining artificial spin systems. The use of the third dimension allows increased configurability and optimization of the magnetostatic interaction between different elements of the system. 
A recent example is the creation of 3D square ice, in which the two sublattices are separated by a height offset~\cite{Perrin2016}, shown in Fig.~\ref{fig:3D}a. Such structures have permitted the realization of systems with extensive degeneracy and unbound monopoles~\cite{Farhan2019}, analogous to those found in the rare earth pyrochlore compounds~\cite{Castelnovo2008}. 
Indeed, while the magnetic moments in atomic spin ices are located at the vertices of a tetrahedral lattice, the artificial square ice is obtained by projecting these moments onto a plane, leading to unequal interactions between the four nanoislands. The height offset can be chosen such that it restores the equivalence of the magnetostatic interactions. In terms of magnetization dynamics, we expect  such systems to offer further possibilities for tailoring the mode spectrum and the band structure in artificial spin ices, not only exploiting topological defects, but equally the offset in the third dimension as simulated in Fig.~\ref{fig:3D}(b). Concurrently, more complex elementary structures have been created, which consist of connected bars and allow for the study of magnetic frustration in three dimensions. For example, the dynamics of `tetrapods' such as in Fig.~\ref{fig:3D}(c) has been investigated using the magneto-optical Kerr effect (MOKE), revealing that it is possible to measure localized oscillations of the magnetization in such structures. Larger and more complicated geometries, such as mesoscopic 'buckyballs'  (Fig.~\ref{fig:3D}(d)) have recently been fabricated\cite{Donnelly2015,Gliga2019MT} and their structure measured using X-ray ptychographic tomography\cite{Donnelly2015}. We expect that these proofs-of-concept will allow the development of novel materials combining specific mechanical and magnetic properties to create structures with reconfigurable functions. At the same time, dedicated techniques are necessary to measure the properties of such structures. Presently, tomography techniques are being actively developed, which allow probing 3D magnetic structures with X-rays\cite{Streubel2015,Donnelly2017,Witte2020}, neutrons~\cite{Manke2010} or electrons\cite{Phatak2014,Biziere2019}. 
Very recently, stroboscopic time-resolved measurements of  magnetization dynamics in 3D have been demonstrated based on magnetic laminography\cite{Donnelly2020trl}.
Beyond X-rays, ferromagnetic resonance measurements of 3D objects have been made possible by employing microresonator loops~\cite{Lenz2019}. Ultimately, the combination of such techniques will enable the full structural and magnetic characterization of 3D structures. 

\subsection{Modelling}

\begin{figure}[t]
\centering \includegraphics[width=3.3in]{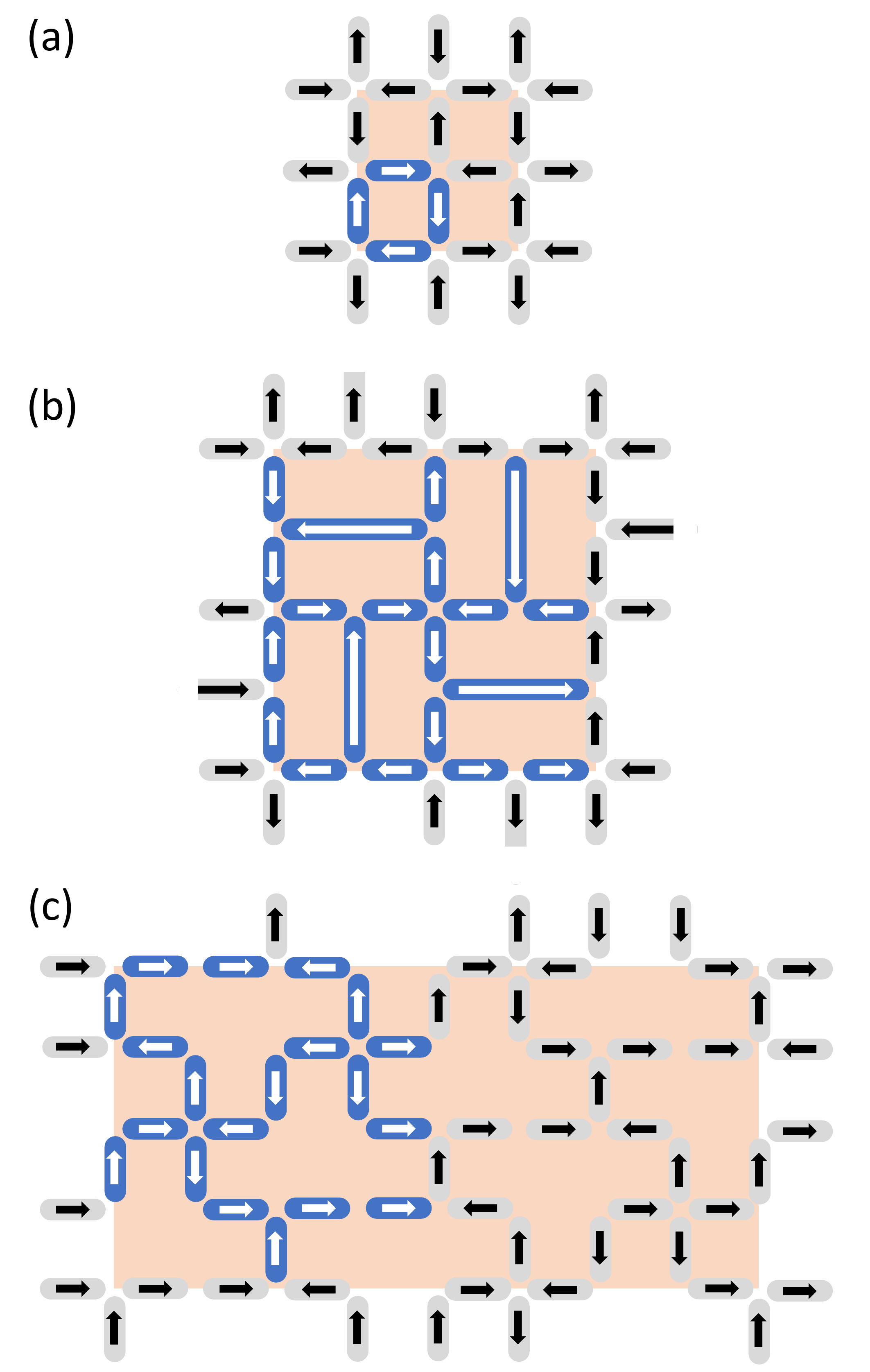}
\caption{ \label{fig:micropersp} Unit-cell versus super-cell configurations in selected states of (a) square, (b) Shakti, and (c) Tetris ice for numerical modeling using periodic boundary conditions with orthogonal translations. The unit-cell is indicated by blue-colored nanoislands. The orange-shaded area indicates the micromagnetic super-cell. }
\end{figure}
The plethora of experimental possibilities directly impacts the theoretical and numerical modelling of artificial spin ices. The interplay between short-range and long-range interactions discussed in section~\ref{sec:micromag} imposes serious constraints on the feasibility of implementing numerical models with predictive power. The first issue to address is the growing size of the artificial spin ice unit-cell, which also increases the full size of the micromagnetic domain and may also increase the size of the micromagnetic super-cell when imposing periodic boundary conditions. For example, a Type-I square ice has a unit cell consisting of four nanoislands and a micromagnetic super-cell consisting of eight nanoislands (four whole and eight split by the periodic boundary conditions). We illustrate this in in Fig.~\ref{fig:micropersp}(a) where the nanoislands in the unit cell are colored blue and the micromagnetic super-cell is indicated by the orange-shaded area. The reason why the number of nanoislands in the unit-cell and the super-cell do not coincide is that typical implementations of periodic boundary conditions require orthogonal translation vectors for the super-cell. Under such implementations, and using the same color-scheme in panels (b) and (c), we identify the unit-cell of a Shakti lattice~\cite{Gilbert2014} and a Tetris lattice~\cite{Gilbert2016}, both in a long-range ordered state. We note that the magnetic states depicted in Fig.~\ref{fig:micropersp}(b) and (c) do not correspond to the ground states identified, e.g., in Ref.~\onlinecite{Morrison2013} (see Figs. 8 and 10), but instead are illustrations of long-range ordered states with relatively simple unit-cells. The Shakti lattice in such a configuration has 20 nanoislands in both the unit-cell and the super-cell (12 whole and 16 split by the peridodic boundary conditions). The Tetris lattice in the configuration shown has 20 nanoislands in the unit-cell and 40 in the super-cell (32 whole and 16 split by the peridodic boundary conditions). These examples show that dynamic simulation of Shakti and Tetris lattices could increase the memory allocation fourfold and eightfold, respectively, compared to a Type-I square ice. In typical finite difference approaches~\cite{Vansteenkiste2014} where the entire domain is discretized, such an increase in memory allocation will necessarily impose a lower bound on the cell-size (due to the finite amount of RAM), resulting in a poor spectral resolution of the modes. This suggests that at least some novel artificial spin ices may preferentially be modelled by finite element methods~\cite{Gliga2013}, where non-magnetic regions can be managed more efficiently to compute magnetostatic fields~\cite{Abert2019}, in particular in combination with a boundary element method~\cite{Schrefl2003}. In addition, one can envision dedicated schemes including skewed periodic boundary conditions to allow for arbitrary translation vectors and thus equalize the super-cell to the unit-cell.

Another intriguing direction consists in using artificial intelligence methods to accelerate or complement micromagnetic modeling. Machine learning methods have recently been used to describe Stoner-Wohlfarth switching in single-domain particles\cite{miao2019}. This study employed supervised learning, in which machine learning models (random forest, support vector machine, and deep neural networks) were first trained on a large number of modeled examples of particles with different damping, anisotropy fields, external field strengths and directions, and the switching behavior predicted by the surrogate models were then validated against other modeling data sets.  In another work\cite{exl2019}, convolutional neural networks were used to construct a surrogate model for the time-stepping predictor in micromagnetic modeling of the LLG equation. The authors relied on dimensional reduction methods (principal component analysis) to reduce dimensions of the non-linear time-stepping problem. Unsupervised learning was then used to train a convolutional neural network which provided an estimator for time-stepping. The model was then applied to the micromagnetic benchmark problems 1 and 2 (see $\mu$MAG micromagnetic modeling activity group: http://www.ctcms.nist.gov/~rdm/mumag.org.html). There exist several potential directions for employing artificial intelligence methods to the dynamics of artificial spin ices. One possibility is to replace the micromagnetic description of the internal dynamics of nanoislands by surrogate machine learning models. These can be constructed, for example, by supervised training of deep neural networks using modeled dynamics of a single nanoisland as training data. The gain would be dimensional reduction by eliminating all but a small number of internal degrees of freedom of the nanoislands, and the remaining problem would be that of the larger-scale inter-island interactions coupled to the surrogate models. Another direction would be to replace the long-range inter-nanoisland interactions with a machine-learning based surrogate model, thus reducing the problem to that of individual nanoislands coupled to an effective field given by a surrogate model.

Another limitation of micromagnetic simulations is that dynamic excitations are computed for collective modes ($k=0$) and periodic boundary conditions allow for the excitation of even wavevectors harmonically proportional to the super-cell size. Such short wavelenghts are essentially irrelevant for the artificial spin ice band structure defined within the first Brillouin zone. Under the constraint that increasing the super-cell size to contain many unit-cells is an unfeasible approach, novel schemes to compute Bloch waves\cite{Rychly2015MC} must be found. A possible solution is to use semi-analytical models, such as in Ref.~\onlinecite{Iacocca2016}, where a tight-binding-like approach was used to collapse the $k$-dependent effective field into the unit-cell by invoking Bloch's theorem. The solution is then obtained by solving an eigenvalue problem. For such an approach to be viable, it would be important to determine irreducible micromagnetic super-cells to avoid aliasing. Implementing this type of micromagnetic simulation would constitute a hybrid approach where the detailed micromagnetic structure informs the computation of an effective field acting on the super-cell from which the eigenmodes can be calculated. Admittedly, the computational efficiency of such a hybrid approach would be low, but it may be feasible to implement in situations requiring accuracy at the nanometer level.

To explore the Brilliouin zones in artificial spin ices, the semi-analytical model proposed in Ref.~\onlinecite{Iacocca2016} is an attractive method. Computation of the Brillouin Zones is achieved through an involved determination of the matrix elements in the eigenvalue problem and results in a coarse spatial mode resolution compared to micromagnetic simulations. Because the magnetization vector is recast in terms of a Holstein-Primakoff transformation~\cite{Slavin2009}, i.e. a complex conjugate pair of dynamic variables, the resulting eigenvalue problem is of size $2NM$, where $N$ is the number of considered macrospins in each nanoisland and $M$ is the number of nanoislands per unit-cell. Therefore, the growing complexity of artificial spin ices will lead to both complicated expressions for the matrix elements and dense matrices. However, this would be a one-time exercise, suggesting the possibility of developing a matrix library as a function of artificial spin ice geometry and magnetic state.

\begin{figure}[t]
\centering \includegraphics[width=3.5in]{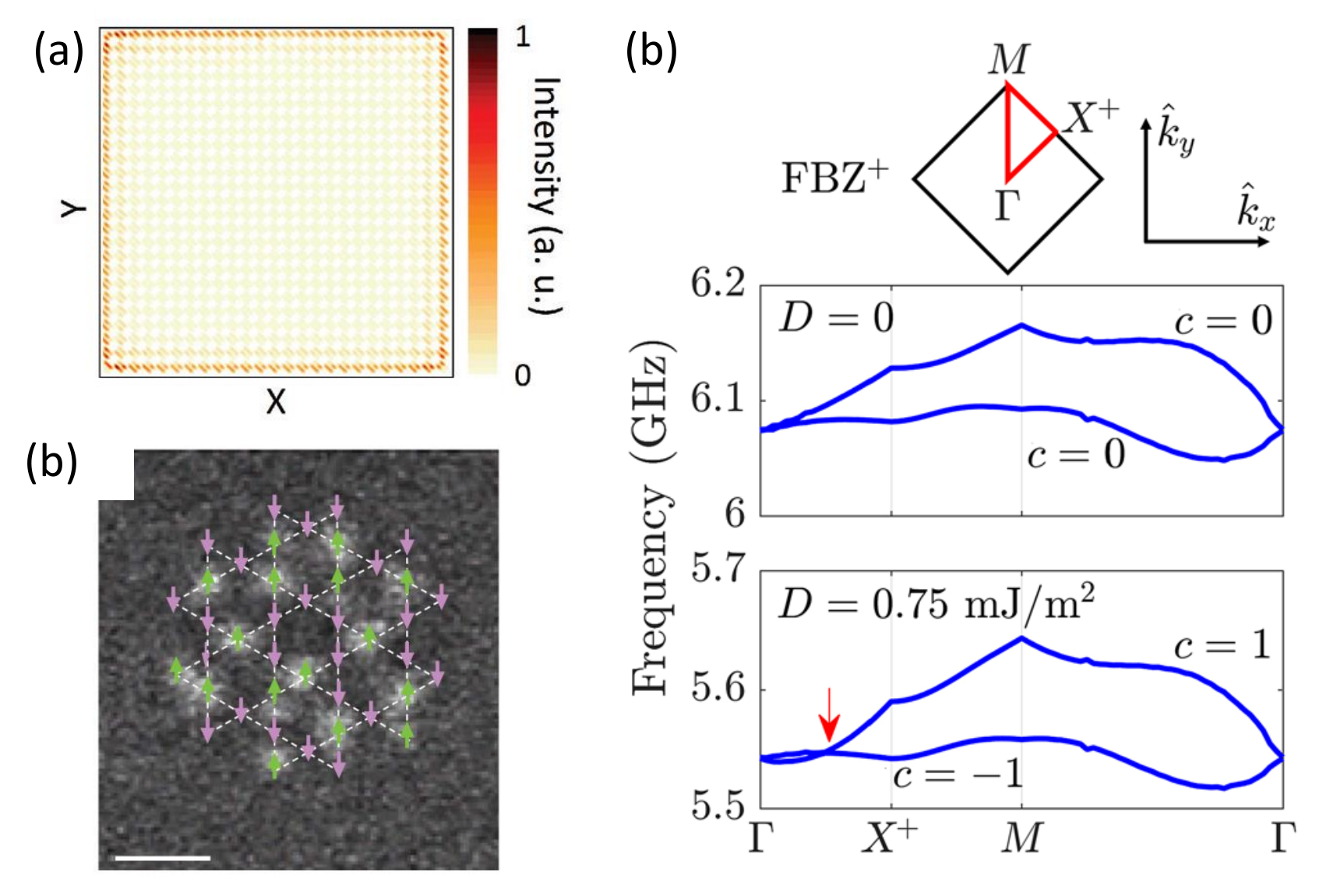}
\caption{ \label{fig:topologicalmagnons} (a) Spin-wave edge modes simulated in a $25\times25$ unit-cells of a decorated honeycomb lattice. Reprinted figure with permission from [\href{https://doi.org/10.1103/PhysRevB.87.174402}{R. Shindou, J.-I. Ohe, R. Matsumoto, S. Murakami, and E. Saitoh, Phys. Rev. B, \textbf{87}, 174402 (2013)}]. Copyright 2013 by the American Physical Society. (b) Spin wave band-diagram in the first Brillouin zone of a square ice patterned on top of of a heavy metal incuding interfacial DMI with parameter $D$. Reprinted figure with permission from [\href{https://doi.org/10.1103/PhysRevApplied.8.034015}{E. Iacocca and O. Heinonen, Phys. Rev. Applied, \textbf{8}, 034015 (2017)}]. Copyright 2017 by the American Physical Society. (c)  kagome artificial spin ice realized by chiral nanoislands where the DMI of the underlayer tilts the magnetization of the nanoislands' edges. From [\href{https://doi.org/10.1126/science.aau7913}{Z. Luo, T. Phuong Dao, A. Hrabec, J. Vijayakumar, A. Kleibert, M. Baumgartner, E. Kirk, J. Cui, T. Savchenko, G. Krishnaswamy, L. J. Heyderman, and P. Gambardella, Science \textbf{363}, 1435-1439 (2019)}]. Reprinted with permission from AAAS. }
\end{figure}

The predictive power of semi-analytical models is useful to explore the features of the first Brillouin zone. An interesting 
application is to determine the onset of topologically protected bands that can give rise to edge modes akin to surface conduction in topological insulators. Such ``topological magnons'' have been explored so far in two different contexts. One is the use of periodic lattices with broken symmetry~\cite{Shindou2013,Shindou2013b}, as discussed in section~\ref{sec:resonant_mag_dyn}. Numerical demonstration of edge modes in a ``decorated'' honeycomb lattice~\cite{Shindou2013b}is shown in Fig.~\ref{fig:topologicalmagnons}(a). Similar ideas were explored in artificial spin ices in Ref.~\onlinecite{Iacocca2017c}, where a square ice coupled to a heavy metal substrate was modelled to investigate chiral effects induced by the interfacial Dyzaloshinskii-Moriya interaction (DMI) on the spin ice. The DMI parameter $D$ was used to toggle the onset of topological bands. As shown in Fig.~\ref{fig:topologicalmagnons}(b), a non-zero $D$ parameter gives rise to a Dirac point indicated by a red arrow, and the concomitant change in the band's Chern number, $c$. The associated chirality in reciprocal space should lead to band inversion in the surface states, although no direct computation of this case has been presented. The use of DMI to induce chirality in ferromagnetic nanoislands has recently been  demonstrated experimentally~\cite{Luo2019}. So far, only the static magnetization has been tuned to realize arbitrary structures including a kagome lattice depicted in Fig.~\ref{fig:topologicalmagnons}(c). The dynamic behavior of such structures remains to be studied. It would indeed be interesting to excite spin waves in such systems and explore their topology as well as reconfigurability. A second approach to topological magnons has been to consider chirality on the atomic level. This is typically obtained in pyrochlore spin ices~\cite{Zhang2013} and in honeycombs lattices~\cite{Chisnell2015}. Recent studies have further shown the possiblity of toggling the chirality of edge modes by tuning the DMI to exchange interaction ratio~\cite{Wang2018_pra} or by inducing DMI by time-dependent interactions, or Floquet engineering~\cite{Owerre2017,Wang2018_pra}. The recently investigated kagome ice with nanoislands with varying anisotropies~\cite{Dion2019} could be a starting point to explore the onset of topological bands by pattering the structure on a heavy metal or by studying the next-nearest-neighbors interactions between the nanoislands.

Other type of artificial spin ices can be deliberately designed to take advantage of broken long-range order. The initial work by Gliga et al.~\onlinecite{Gliga2013} explored precisely how topological defects in a square ice would impact the collective dispersion of waves. Similarly, it would be interesting to consider structures exhibiting collective, topological frustration~\cite{Nisoli2017}, and not only vertex frustration. This would allow exploring possible links between geometrical topology and the topological character of excited spin waves. Finally, quasi-crystals,  i.e. ordered lattices without a well-defined spatial periodicity, are other types of structures, which have been explored in the context of magnonics~\cite{Rychly2015,Lisiecki2019} and in artificial spin ices~\cite{Bhat2013,Shi2018}. A predictive theoretical or numerical model of these artificial spin ices will no doubt be challenging to implement. Experiments will likely drive the need to develop numerically efficient techniques.

\subsection{Devices, logic, and information processing}

In magnonics\cite{Kruglyak2010,Lenk2011,Demokritov2013,Chumak2015}, spin waves are utilized as carriers of information for technologically relevant functions. One direction of research consists in exploring the design of logic gates and multiplexers based on spin-wave interferometry\cite{Hertel2004,schneider2008realization,vogt2014realization} and the directional propagation of waves\cite{Vogt2010,Zakeri2010,Jamali2013,Otalora2016}. In particular, this has led to the demonstration of NAND gates\cite{Hertel2004} as well as a three-terminal YIG-based magnonic transistor\cite{chumak2014magnon}. 
Another direction of research consists in the creation of periodically patterned, magnetostatically coupled nanostructures with feature sizes of the order of the magnon wavelength in which spin waves can propagate. These allow creating devices with tailored band structures, defining frequency ranges for which spin wave propagation is allowed and forbidden band gaps\cite{Nikitov2001}. 
Such structures, called magnonic crystals, in analogy to photonic crystals, potentially provide the basis for a wide range of magnon spintronic devices, including waveguides, signal filters, phase shifters and signal processing elements operating the GHz frequency range\cite{Melkov2006,Lee2009WG}. Magnon-based applications are therefore compatible with the speeds of current CMOS-based circuits and, if antiferromagnetic magnons can be harnessed, this could even be pushed to the THz range.
Furthermore, the band structure can be reprogrammed by exploiting different possible magnetic states in the unit cell of the periodic lattice\cite{Chumak2009,Topp2010,Tacchi2010}.

So far, artificial spin ices have shown potential for magnonic applications: they possess rich mode spectra, tunable band structures, which can be engineered to have non-trivial topologies in the presence of DMI and, when coupled to an underlayer, can in principle support spin wave propagation along defined channels. Going forward, these characteristics are essential for the development and demonstration of relevant building blocks for information transmission and detection, gating and switching, or logic devices. Generally, we expect such functionalities to be achieved in two ways. 
The first consists in tailoring the band structure through the geometry and magnetic state of the artificial spin ices to obtain desired functionalities required e.g. for spin wave filters, phase filters or to define spin wave propagation channels\cite{Iacocca2019}. Achieving such functionalities requires the possibility of reconfiguring the global state of the artificial spin ice, such as by means of an external field. For practical devices, however, a more promising route is the use of voltage-controlled anisotropy\cite{maruyama2009,amiri2012}, which allows for magnetization switching using applied electric voltages. This would be much easier to engineer, as the building blocks for integration already exists, and it would also lead to smaller power consumption than the generation of magnetic fields.
The second way consists in exploiting topological defects (e.g. `monopoles') and the strings of reversed nanoislands connecting them in order to define regions with specific functions, such as spin wave interference. This hinges upon the possibility of locally reconfiguring the magnetic state of the artificial spin ices using techniques, such as tip-induced reversal\cite{Gartside2018}, local contacts for the injection of spin-polarized currents, or applied voltages through the use of voltage-controlled anisotropy. Such schemes will also have to allow for large-scale processing, which is in principle rather an engineering than a fundamental materials science problem -- indeed, modern commercial spin transfer torque random access memories, for example, use processing technology compatible with CMOS technology. (For a recent review, see, for example,  Ref.~\onlinecite{hanyu2019spin}).

\begin{figure}[t]
\centering \includegraphics[width=3.3in]{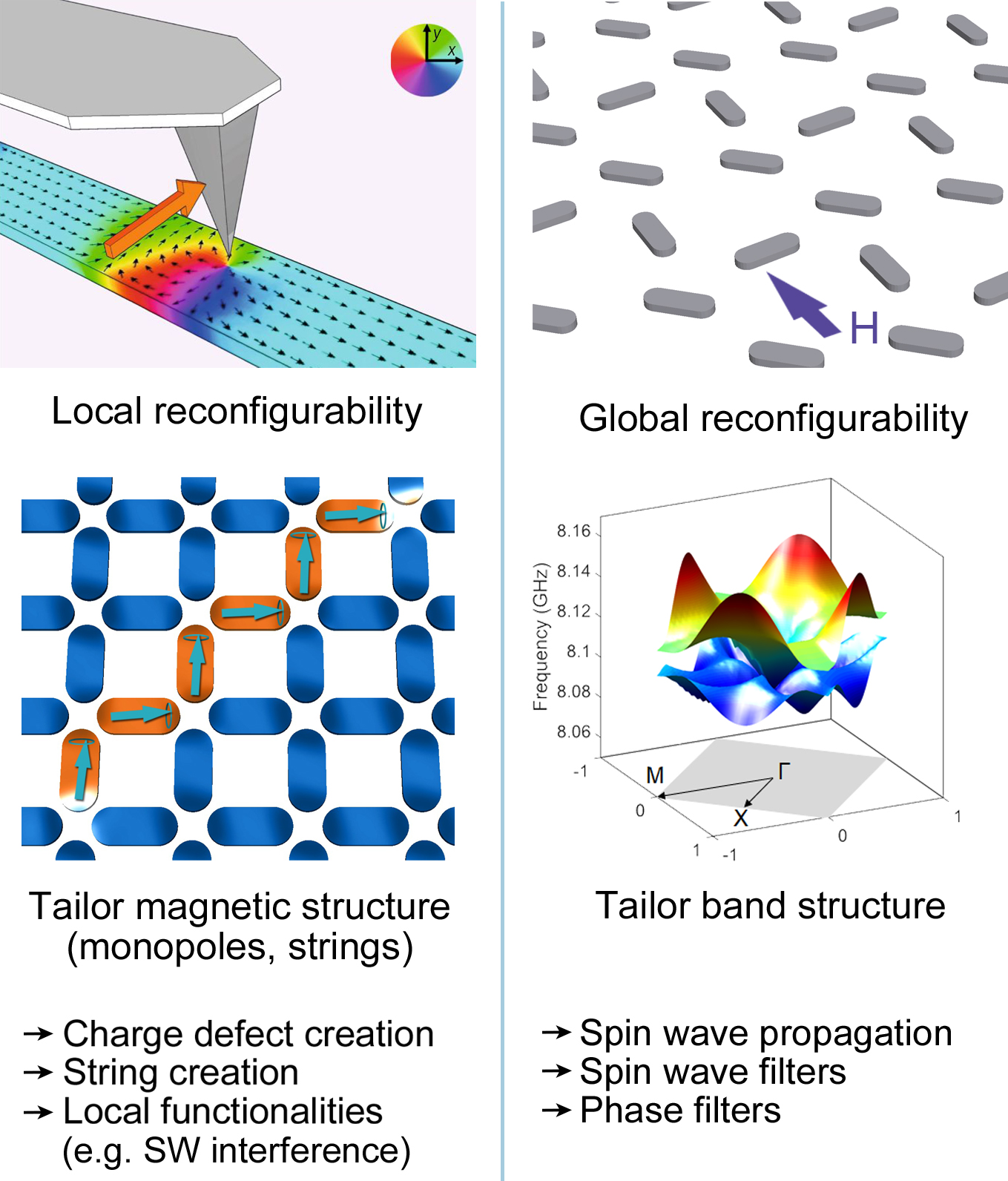}
\caption{ \label{fig:perspectives} Possible routes for creating reconfigurable magnonic devices based on artificial spin ices. 
(Left) Local reconfigurability, here illustrated by using an MFM tip , enables `writing' specific configurations, such as strings and charge defects 
(Reprinted by permission from Springer Nature: \href{https://doi.org/10.1038/s41565-017-0002-1}{[Nat. Nanotechnol. 13, 53--58, Realization of ground state in artificial kagome spin ice via topological defect-driven magnetic writing, J. C. Gartside, D. M. Arroo, D. M. Burn, V. L. Bemmer, A. Moskalenko, L. F. Cohen, W. R. Branford (2018)]}). 
These can be useful to create well-defined regions for spin wave interference and create e.g. logical gates. (Right) Global reconfigurability is achievable in specific geometries such as the charge ice, allowing the definition of on-demand band structures. These can lead to the creation of magnonic crystals to control spin wave propagation, e.g. through spin wave or phase filters.}
\end{figure}

Magnons can be generated in a number of different ways but the most likely path compatible with miniaturization and large-scale processing is using spin-torque oscillators (STOs). STOs can generate magnetic oscillations with frequencies between $\sim1$~GHz to over tens of GHz that directly couple to the magnonic medium. Using low-damping materials, such as YIG, magnons can propagate macroscopic distances without significant attenuation\cite{chumak2019fundamentals}. 
Detection of magnons should ideally be performed in a way that is compatible with large-scale integration, i.e. based on giant magnetoresistance or tunneling magnetoresistance, in which read-out elements are deposited on the magnetic medium. 
Artificial spin ices as reconfigurable magnonic lattices combine several functions that are necessary: they can be reconfigured to switch or multiplex information flow, and are not limited to information flow in only one direction. The nanoislands themselves can also be directly integrated with STOs for generation and detection as well as switching for local or global reconfiguration. Finally, the nanoislands can also serve as non-volatile memory element directly integrated into a magnonics-based IT and computing system. 

\section*{Acknowledgments}
S.G. is thankful to Armin Kleibert and Martino Poggio  for discussions. 
O.G.H. was funded by the Department of Energy, Office of Science, Basic Energy Sciences Materials Sciences and Engineering Division.
S.G. acknowledges funding by the European Union's Horizon 2020 research and innovation program under the Marie Sk\l{}odowska-Curie grant agreement No.~708674 as well as funding from the Swiss National Science Foundation. 
 The authors also are grateful for the computing resource support provided on Blues and Bebop, high-performance computing clusters operated by the Laboratory Computing Resource Center at Argonne National Laboratory,
Use of the Center for Nanoscale Materials, an Office of Science user facility, was supported by the US Department of Energy, Office of Science, Office of Basic Energy Sciences, under Contract No. DE-AC02-06CH11357.
%

\bibliographystyle{apsrev4-1}
\bibliography{References} 

\begin{thebibliography}{175}%
\makeatletter
\providecommand \@ifxundefined [1]{%
 \@ifx{#1\undefined}
}%
\providecommand \@ifnum [1]{%
 \ifnum #1\expandafter \@firstoftwo
 \else \expandafter \@secondoftwo
 \fi
}%
\providecommand \@ifx [1]{%
 \ifx #1\expandafter \@firstoftwo
 \else \expandafter \@secondoftwo
 \fi
}%
\providecommand \natexlab [1]{#1}%
\providecommand \enquote  [1]{``#1''}%
\providecommand \bibnamefont  [1]{#1}%
\providecommand \bibfnamefont [1]{#1}%
\providecommand \citenamefont [1]{#1}%
\providecommand \href@noop [0]{\@secondoftwo}%
\providecommand \href [0]{\begingroup \@sanitize@url \@href}%
\providecommand \@href[1]{\@@startlink{#1}\@@href}%
\providecommand \@@href[1]{\endgroup#1\@@endlink}%
\providecommand \@sanitize@url [0]{\catcode `\\12\catcode `\$12\catcode
  `\&12\catcode `\#12\catcode `\^12\catcode `\_12\catcode `\%12\relax}%
\providecommand \@@startlink[1]{}%
\providecommand \@@endlink[0]{}%
\providecommand \url  [0]{\begingroup\@sanitize@url \@url }%
\providecommand \@url [1]{\endgroup\@href {#1}{\urlprefix }}%
\providecommand \urlprefix  [0]{URL }%
\providecommand \Eprint [0]{\href }%
\providecommand \doibase [0]{http://dx.doi.org/}%
\providecommand \selectlanguage [0]{\@gobble}%
\providecommand \bibinfo  [0]{\@secondoftwo}%
\providecommand \bibfield  [0]{\@secondoftwo}%
\providecommand \translation [1]{[#1]}%
\providecommand \BibitemOpen [0]{}%
\providecommand \bibitemStop [0]{}%
\providecommand \bibitemNoStop [0]{.\EOS\space}%
\providecommand \EOS [0]{\spacefactor3000\relax}%
\providecommand \BibitemShut  [1]{\csname bibitem#1\endcsname}%
\let\auto@bib@innerbib\@empty
\bibitem [{\citenamefont {Heyderman}\ and\ \citenamefont
  {Stamps}(2013)}]{Heyderman2013}%
  \BibitemOpen
  \bibfield  {author} {\bibinfo {author} {\bibfnamefont {L.~J.}\ \bibnamefont
  {Heyderman}}\ and\ \bibinfo {author} {\bibfnamefont {R.~L.}\ \bibnamefont
  {Stamps}},\ }\href@noop {} {\bibfield  {journal} {\bibinfo  {journal}
  {Journal of Physics: Condensed Matter}\ }\textbf {\bibinfo {volume} {25}},\
  \bibinfo {pages} {363201} (\bibinfo {year} {2013})}\BibitemShut {NoStop}%
\bibitem [{\citenamefont {Wang}\ \emph {et~al.}(2006)\citenamefont {Wang},
  \citenamefont {Nisoli}, \citenamefont {Freitas}, \citenamefont {Li},
  \citenamefont {McConville}, \citenamefont {Cooley}, \citenamefont {Lund},
  \citenamefont {Samarth}, \citenamefont {Leighton}, \citenamefont {Crespi},\
  and\ \citenamefont {Schiffer}}]{Wang2006}%
  \BibitemOpen
  \bibfield  {author} {\bibinfo {author} {\bibfnamefont {R.~F.}\ \bibnamefont
  {Wang}}, \bibinfo {author} {\bibfnamefont {C.}~\bibnamefont {Nisoli}},
  \bibinfo {author} {\bibfnamefont {R.~S.}\ \bibnamefont {Freitas}}, \bibinfo
  {author} {\bibfnamefont {J.}~\bibnamefont {Li}}, \bibinfo {author}
  {\bibfnamefont {W.}~\bibnamefont {McConville}}, \bibinfo {author}
  {\bibfnamefont {B.~J.}\ \bibnamefont {Cooley}}, \bibinfo {author}
  {\bibfnamefont {M.~S.}\ \bibnamefont {Lund}}, \bibinfo {author}
  {\bibfnamefont {N.}~\bibnamefont {Samarth}}, \bibinfo {author} {\bibfnamefont
  {C.}~\bibnamefont {Leighton}}, \bibinfo {author} {\bibfnamefont {V.~H.}\
  \bibnamefont {Crespi}}, \ and\ \bibinfo {author} {\bibfnamefont
  {P.}~\bibnamefont {Schiffer}},\ }\href@noop {} {\bibfield  {journal}
  {\bibinfo  {journal} {Nature}\ }\textbf {\bibinfo {volume} {439}},\ \bibinfo
  {pages} {303} (\bibinfo {year} {2006})}\BibitemShut {NoStop}%
\bibitem [{\citenamefont {Castelnovo}\ \emph {et~al.}(2008)\citenamefont
  {Castelnovo}, \citenamefont {Moessner},\ and\ \citenamefont
  {Sondhi}}]{Castelnovo2008}%
  \BibitemOpen
  \bibfield  {author} {\bibinfo {author} {\bibfnamefont {C.}~\bibnamefont
  {Castelnovo}}, \bibinfo {author} {\bibfnamefont {R.}~\bibnamefont
  {Moessner}}, \ and\ \bibinfo {author} {\bibfnamefont {S.~L.}\ \bibnamefont
  {Sondhi}},\ }\href@noop {} {\bibfield  {journal} {\bibinfo  {journal}
  {Nature}\ }\textbf {\bibinfo {volume} {451}},\ \bibinfo {pages} {42 }
  (\bibinfo {year} {2008})}\BibitemShut {NoStop}%
\bibitem [{\citenamefont {Morgan}\ \emph {et~al.}(2011)\citenamefont {Morgan},
  \citenamefont {Stein}, \citenamefont {Langridge},\ and\ \citenamefont
  {Marrows}}]{morgan2011}%
  \BibitemOpen
  \bibfield  {author} {\bibinfo {author} {\bibfnamefont {J.~P.}\ \bibnamefont
  {Morgan}}, \bibinfo {author} {\bibfnamefont {A.}~\bibnamefont {Stein}},
  \bibinfo {author} {\bibfnamefont {S.}~\bibnamefont {Langridge}}, \ and\
  \bibinfo {author} {\bibfnamefont {C.~H.}\ \bibnamefont {Marrows}},\
  }\href@noop {} {\bibfield  {journal} {\bibinfo  {journal} {Nature Physics}\
  }\textbf {\bibinfo {volume} {7}},\ \bibinfo {pages} {75} (\bibinfo {year}
  {2011})}\BibitemShut {NoStop}%
\bibitem [{\citenamefont {Farhan}\ \emph {et~al.}(2013)\citenamefont {Farhan},
  \citenamefont {Derlet}, \citenamefont {Kleibert}, \citenamefont {Balan},
  \citenamefont {Chopdekar}, \citenamefont {Wyss}, \citenamefont {Perron},
  \citenamefont {Scholl}, \citenamefont {Nolting},\ and\ \citenamefont
  {Heyderman}}]{Farhan2013}%
  \BibitemOpen
  \bibfield  {author} {\bibinfo {author} {\bibfnamefont {A.}~\bibnamefont
  {Farhan}}, \bibinfo {author} {\bibfnamefont {P.~M.}\ \bibnamefont {Derlet}},
  \bibinfo {author} {\bibfnamefont {A.}~\bibnamefont {Kleibert}}, \bibinfo
  {author} {\bibfnamefont {A.}~\bibnamefont {Balan}}, \bibinfo {author}
  {\bibfnamefont {R.~V.}\ \bibnamefont {Chopdekar}}, \bibinfo {author}
  {\bibfnamefont {M.}~\bibnamefont {Wyss}}, \bibinfo {author} {\bibfnamefont
  {J.}~\bibnamefont {Perron}}, \bibinfo {author} {\bibfnamefont
  {A.}~\bibnamefont {Scholl}}, \bibinfo {author} {\bibfnamefont
  {F.}~\bibnamefont {Nolting}}, \ and\ \bibinfo {author} {\bibfnamefont
  {L.~J.}\ \bibnamefont {Heyderman}},\ }\href {\doibase
  10.1103/PhysRevLett.111.057204} {\bibfield  {journal} {\bibinfo  {journal}
  {Phys. Rev. Lett.}\ }\textbf {\bibinfo {volume} {111}},\ \bibinfo {pages}
  {057204} (\bibinfo {year} {2013})}\BibitemShut {NoStop}%
\bibitem [{\citenamefont {Sklenar}\ \emph {et~al.}(2019)\citenamefont
  {Sklenar}, \citenamefont {Lao}, \citenamefont {Albrecht}, \citenamefont
  {Watts}, \citenamefont {Nisoli}, \citenamefont {Chern},\ and\ \citenamefont
  {Schiffer}}]{Sklenar2018}%
  \BibitemOpen
  \bibfield  {author} {\bibinfo {author} {\bibfnamefont {J.}~\bibnamefont
  {Sklenar}}, \bibinfo {author} {\bibfnamefont {Y.}~\bibnamefont {Lao}},
  \bibinfo {author} {\bibfnamefont {A.}~\bibnamefont {Albrecht}}, \bibinfo
  {author} {\bibfnamefont {J.~D.}\ \bibnamefont {Watts}}, \bibinfo {author}
  {\bibfnamefont {C.}~\bibnamefont {Nisoli}}, \bibinfo {author} {\bibfnamefont
  {G.-W.}\ \bibnamefont {Chern}}, \ and\ \bibinfo {author} {\bibfnamefont
  {P.}~\bibnamefont {Schiffer}},\ }\href@noop {} {\bibfield  {journal}
  {\bibinfo  {journal} {Nature Physics}\ }\textbf {\bibinfo {volume} {15}},\
  \bibinfo {pages} {191–195} (\bibinfo {year} {2019})}\BibitemShut {NoStop}%
\bibitem [{\citenamefont {Gilbert}\ \emph {et~al.}(2014)\citenamefont
  {Gilbert}, \citenamefont {Chern}, \citenamefont {Zhange}, \citenamefont
  {O'Brien}, \citenamefont {Foe}, \citenamefont {Nisoli},\ and\ \citenamefont
  {Schiffer}}]{Gilbert2014}%
  \BibitemOpen
  \bibfield  {author} {\bibinfo {author} {\bibfnamefont {I.}~\bibnamefont
  {Gilbert}}, \bibinfo {author} {\bibfnamefont {G.-W.}\ \bibnamefont {Chern}},
  \bibinfo {author} {\bibfnamefont {S.}~\bibnamefont {Zhange}}, \bibinfo
  {author} {\bibfnamefont {L.}~\bibnamefont {O'Brien}}, \bibinfo {author}
  {\bibfnamefont {B.}~\bibnamefont {Foe}}, \bibinfo {author} {\bibfnamefont
  {C.}~\bibnamefont {Nisoli}}, \ and\ \bibinfo {author} {\bibfnamefont
  {P.}~\bibnamefont {Schiffer}},\ }\href@noop {} {\bibfield  {journal}
  {\bibinfo  {journal} {Nature Physics}\ }\textbf {\bibinfo {volume} {10}},\
  \bibinfo {pages} {670} (\bibinfo {year} {2014})}\BibitemShut {NoStop}%
\bibitem [{\citenamefont {Gilbert}\ \emph {et~al.}(2016)\citenamefont
  {Gilbert}, \citenamefont {Lao}, \citenamefont {Carrasquillo}, \citenamefont
  {O'Brian}, \citenamefont {Watts}, \citenamefont {Manno}, \citenamefont
  {Leighton}, \citenamefont {Scholl}, \citenamefont {Nisoli},\ and\
  \citenamefont {Schiffer}}]{Gilbert2016}%
  \BibitemOpen
  \bibfield  {author} {\bibinfo {author} {\bibfnamefont {I.}~\bibnamefont
  {Gilbert}}, \bibinfo {author} {\bibfnamefont {Y.}~\bibnamefont {Lao}},
  \bibinfo {author} {\bibfnamefont {I.}~\bibnamefont {Carrasquillo}}, \bibinfo
  {author} {\bibfnamefont {L.}~\bibnamefont {O'Brian}}, \bibinfo {author}
  {\bibfnamefont {J.~D.}\ \bibnamefont {Watts}}, \bibinfo {author}
  {\bibfnamefont {M.}~\bibnamefont {Manno}}, \bibinfo {author} {\bibfnamefont
  {C.}~\bibnamefont {Leighton}}, \bibinfo {author} {\bibfnamefont
  {A.}~\bibnamefont {Scholl}}, \bibinfo {author} {\bibfnamefont
  {C.}~\bibnamefont {Nisoli}}, \ and\ \bibinfo {author} {\bibfnamefont
  {P.}~\bibnamefont {Schiffer}},\ }\href@noop {} {\bibfield  {journal}
  {\bibinfo  {journal} {Nature Physics}\ }\textbf {\bibinfo {volume} {12}},\
  \bibinfo {pages} {162} (\bibinfo {year} {2016})}\BibitemShut {NoStop}%
\bibitem [{\citenamefont {Wang}\ \emph {et~al.}(2016)\citenamefont {Wang},
  \citenamefont {Xiao}, \citenamefont {Snezhko}, \citenamefont {Xu},
  \citenamefont {Ocoloa}, \citenamefont {Divan}, \citenamefont {Peason},
  \citenamefont {Crabtree},\ and\ \citenamefont {Kwok}}]{Wang2016}%
  \BibitemOpen
  \bibfield  {author} {\bibinfo {author} {\bibfnamefont {Y.-L.}\ \bibnamefont
  {Wang}}, \bibinfo {author} {\bibfnamefont {Z.-L.}\ \bibnamefont {Xiao}},
  \bibinfo {author} {\bibfnamefont {A.}~\bibnamefont {Snezhko}}, \bibinfo
  {author} {\bibfnamefont {J.}~\bibnamefont {Xu}}, \bibinfo {author}
  {\bibfnamefont {L.~E.}\ \bibnamefont {Ocoloa}}, \bibinfo {author}
  {\bibfnamefont {R.}~\bibnamefont {Divan}}, \bibinfo {author} {\bibfnamefont
  {J.~E.}\ \bibnamefont {Peason}}, \bibinfo {author} {\bibfnamefont {G.~W.}\
  \bibnamefont {Crabtree}}, \ and\ \bibinfo {author} {\bibfnamefont {W.-K.}\
  \bibnamefont {Kwok}},\ }\href@noop {} {\bibfield  {journal} {\bibinfo
  {journal} {Science}\ }\textbf {\bibinfo {volume} {352}},\ \bibinfo {pages}
  {962} (\bibinfo {year} {2016})}\BibitemShut {NoStop}%
\bibitem [{\citenamefont {Gliga}\ \emph {et~al.}(2017)\citenamefont {Gliga},
  \citenamefont {Hrkac}, \citenamefont {Donnelly}, \citenamefont {B{\"u}chi},
  \citenamefont {Kleibert}, \citenamefont {Cui}, \citenamefont {Farhan},
  \citenamefont {Kirk}, \citenamefont {Chopdekar}, \citenamefont {Masaki},
  \citenamefont {Bingham}, \citenamefont {Scholl}, \citenamefont {Stamps},\
  and\ \citenamefont {Heyderman}}]{Gliga2017}%
  \BibitemOpen
  \bibfield  {author} {\bibinfo {author} {\bibfnamefont {S.}~\bibnamefont
  {Gliga}}, \bibinfo {author} {\bibfnamefont {G.}~\bibnamefont {Hrkac}},
  \bibinfo {author} {\bibfnamefont {C.}~\bibnamefont {Donnelly}}, \bibinfo
  {author} {\bibfnamefont {J.}~\bibnamefont {B{\"u}chi}}, \bibinfo {author}
  {\bibfnamefont {A.}~\bibnamefont {Kleibert}}, \bibinfo {author}
  {\bibfnamefont {J.}~\bibnamefont {Cui}}, \bibinfo {author} {\bibfnamefont
  {A.}~\bibnamefont {Farhan}}, \bibinfo {author} {\bibfnamefont
  {E.}~\bibnamefont {Kirk}}, \bibinfo {author} {\bibfnamefont {R.~V.}\
  \bibnamefont {Chopdekar}}, \bibinfo {author} {\bibfnamefont {Y.}~\bibnamefont
  {Masaki}}, \bibinfo {author} {\bibfnamefont {N.~S.}\ \bibnamefont {Bingham}},
  \bibinfo {author} {\bibfnamefont {A.}~\bibnamefont {Scholl}}, \bibinfo
  {author} {\bibfnamefont {R.~L.}\ \bibnamefont {Stamps}}, \ and\ \bibinfo
  {author} {\bibfnamefont {L.~J.}\ \bibnamefont {Heyderman}},\ }\href@noop {}
  {\bibfield  {journal} {\bibinfo  {journal} {Nature Materials}\ }\textbf
  {\bibinfo {volume} {16}},\ \bibinfo {pages} {1106–1111} (\bibinfo {year}
  {2017})}\BibitemShut {NoStop}%
\bibitem [{\citenamefont {Luo}\ \emph {et~al.}(2019)\citenamefont {Luo},
  \citenamefont {Dao}, \citenamefont {Hrabec}, \citenamefont {Vijayakumar},
  \citenamefont {Kleibert}, \citenamefont {Baumgartner}, \citenamefont {Kirk},
  \citenamefont {Cui}, \citenamefont {Savchenko}, \citenamefont {Krishnaswamy},
  \citenamefont {Heyderman},\ and\ \citenamefont {Gambardella}}]{Luo2019}%
  \BibitemOpen
  \bibfield  {author} {\bibinfo {author} {\bibfnamefont {Z.}~\bibnamefont
  {Luo}}, \bibinfo {author} {\bibfnamefont {T.~P.}\ \bibnamefont {Dao}},
  \bibinfo {author} {\bibfnamefont {A.}~\bibnamefont {Hrabec}}, \bibinfo
  {author} {\bibfnamefont {J.}~\bibnamefont {Vijayakumar}}, \bibinfo {author}
  {\bibfnamefont {A.}~\bibnamefont {Kleibert}}, \bibinfo {author}
  {\bibfnamefont {M.}~\bibnamefont {Baumgartner}}, \bibinfo {author}
  {\bibfnamefont {E.}~\bibnamefont {Kirk}}, \bibinfo {author} {\bibfnamefont
  {J.}~\bibnamefont {Cui}}, \bibinfo {author} {\bibfnamefont {T.}~\bibnamefont
  {Savchenko}}, \bibinfo {author} {\bibfnamefont {G.}~\bibnamefont
  {Krishnaswamy}}, \bibinfo {author} {\bibfnamefont {L.~J.}\ \bibnamefont
  {Heyderman}}, \ and\ \bibinfo {author} {\bibfnamefont {P.}~\bibnamefont
  {Gambardella}},\ }\href@noop {} {\bibfield  {journal} {\bibinfo  {journal}
  {Science}\ }\textbf {\bibinfo {volume} {363}},\ \bibinfo {pages} {1435}
  (\bibinfo {year} {2019})}\BibitemShut {NoStop}%
\bibitem [{\citenamefont {Saccone}\ \emph {et~al.}(2019)\citenamefont
  {Saccone}, \citenamefont {Hofhuis}, \citenamefont {Huang}, \citenamefont
  {Dhuey}, \citenamefont {Chen}, \citenamefont {Scholl}, \citenamefont
  {Chopdekar}, \citenamefont {van Dijken},\ and\ \citenamefont
  {Farhan}}]{Saccone2019}%
  \BibitemOpen
  \bibfield  {author} {\bibinfo {author} {\bibfnamefont {M.}~\bibnamefont
  {Saccone}}, \bibinfo {author} {\bibfnamefont {K.}~\bibnamefont {Hofhuis}},
  \bibinfo {author} {\bibfnamefont {Y.-L.}\ \bibnamefont {Huang}}, \bibinfo
  {author} {\bibfnamefont {S.}~\bibnamefont {Dhuey}}, \bibinfo {author}
  {\bibfnamefont {Z.}~\bibnamefont {Chen}}, \bibinfo {author} {\bibfnamefont
  {A.}~\bibnamefont {Scholl}}, \bibinfo {author} {\bibfnamefont {R.~V.}\
  \bibnamefont {Chopdekar}}, \bibinfo {author} {\bibfnamefont {S.}~\bibnamefont
  {van Dijken}}, \ and\ \bibinfo {author} {\bibfnamefont {A.}~\bibnamefont
  {Farhan}},\ }\href@noop {} {\bibfield  {journal} {\bibinfo  {journal} {Phys.
  Rev. Materials}\ }\textbf {\bibinfo {volume} {3}},\ \bibinfo {pages} {104402}
  (\bibinfo {year} {2019})}\BibitemShut {NoStop}%
\bibitem [{\citenamefont {Nisoli}\ \emph {et~al.}(2017)\citenamefont {Nisoli},
  \citenamefont {Kapaklis},\ and\ \citenamefont {Schiffer}}]{Nisoli2017}%
  \BibitemOpen
  \bibfield  {author} {\bibinfo {author} {\bibfnamefont {C.}~\bibnamefont
  {Nisoli}}, \bibinfo {author} {\bibfnamefont {V.}~\bibnamefont {Kapaklis}}, \
  and\ \bibinfo {author} {\bibfnamefont {P.}~\bibnamefont {Schiffer}},\
  }\href@noop {} {\bibfield  {journal} {\bibinfo  {journal} {Nature Physics}\
  }\textbf {\bibinfo {volume} {13}},\ \bibinfo {pages} {200} (\bibinfo {year}
  {2017})}\BibitemShut {NoStop}%
\bibitem [{\citenamefont {Nikitov}\ \emph {et~al.}(2001)\citenamefont
  {Nikitov}, \citenamefont {Tailhades},\ and\ \citenamefont
  {Tsai}}]{Nikitov2001}%
  \BibitemOpen
  \bibfield  {author} {\bibinfo {author} {\bibfnamefont {S.}~\bibnamefont
  {Nikitov}}, \bibinfo {author} {\bibfnamefont {P.}~\bibnamefont {Tailhades}},
  \ and\ \bibinfo {author} {\bibfnamefont {C.}~\bibnamefont {Tsai}},\
  }\href@noop {} {\bibfield  {journal} {\bibinfo  {journal} {Journal of
  Magnetism and Magnetic Materials}\ }\textbf {\bibinfo {volume} {236}},\
  \bibinfo {pages} {320 } (\bibinfo {year} {2001})}\BibitemShut {NoStop}%
\bibitem [{\citenamefont {Kruglyak}\ \emph {et~al.}(2010)\citenamefont
  {Kruglyak}, \citenamefont {Demokritov},\ and\ \citenamefont
  {Grundler}}]{Kruglyak2010}%
  \BibitemOpen
  \bibfield  {author} {\bibinfo {author} {\bibfnamefont {V.~V.}\ \bibnamefont
  {Kruglyak}}, \bibinfo {author} {\bibfnamefont {S.~O.}\ \bibnamefont
  {Demokritov}}, \ and\ \bibinfo {author} {\bibfnamefont {D.}~\bibnamefont
  {Grundler}},\ }\href@noop {} {\bibfield  {journal} {\bibinfo  {journal}
  {Journal of Physics D: Applied Physics}\ }\textbf {\bibinfo {volume} {43}},\
  \bibinfo {pages} {264001} (\bibinfo {year} {2010})}\BibitemShut {NoStop}%
\bibitem [{\citenamefont {Lenk}\ \emph {et~al.}(2011)\citenamefont {Lenk},
  \citenamefont {Ulrichs}, \citenamefont {Garbs},\ and\ \citenamefont
  {M\"{u}nzenberg}}]{Lenk2011}%
  \BibitemOpen
  \bibfield  {author} {\bibinfo {author} {\bibfnamefont {B.}~\bibnamefont
  {Lenk}}, \bibinfo {author} {\bibfnamefont {H.}~\bibnamefont {Ulrichs}},
  \bibinfo {author} {\bibfnamefont {F.}~\bibnamefont {Garbs}}, \ and\ \bibinfo
  {author} {\bibfnamefont {M.}~\bibnamefont {M\"{u}nzenberg}},\ }\href@noop {}
  {\bibfield  {journal} {\bibinfo  {journal} {Physics Reports}\ }\textbf
  {\bibinfo {volume} {507}},\ \bibinfo {pages} {107 } (\bibinfo {year}
  {2011})}\BibitemShut {NoStop}%
\bibitem [{\citenamefont {Demokritov}\ and\ \citenamefont
  {Slavin}(2013)}]{Demokritov2013}%
  \BibitemOpen
  \bibfield  {author} {\bibinfo {author} {\bibfnamefont {S.~O.}\ \bibnamefont
  {Demokritov}}\ and\ \bibinfo {author} {\bibfnamefont {A.~N.}\ \bibnamefont
  {Slavin}},\ }\href@noop {} {\emph {\bibinfo {title} {Magnonics: From
  Fundamentals to Applications}}}\ (\bibinfo  {publisher} {Springer},\ \bibinfo
  {year} {2013})\BibitemShut {NoStop}%
\bibitem [{\citenamefont {Chumak}\ \emph {et~al.}(2015)\citenamefont {Chumak},
  \citenamefont {Vasyuchka}, \citenamefont {Serga},\ and\ \citenamefont
  {Hillebrands}}]{Chumak2015}%
  \BibitemOpen
  \bibfield  {author} {\bibinfo {author} {\bibfnamefont {A.~V.}\ \bibnamefont
  {Chumak}}, \bibinfo {author} {\bibfnamefont {V.~I.}\ \bibnamefont
  {Vasyuchka}}, \bibinfo {author} {\bibfnamefont {A.~A.}\ \bibnamefont
  {Serga}}, \ and\ \bibinfo {author} {\bibfnamefont {B.}~\bibnamefont
  {Hillebrands}},\ }\href@noop {} {\bibfield  {journal} {\bibinfo  {journal}
  {Nature physics}\ }\textbf {\bibinfo {volume} {11}},\ \bibinfo {pages} {453}
  (\bibinfo {year} {2015})}\BibitemShut {NoStop}%
\bibitem [{\citenamefont {Chumak}(2019)}]{chumak2019fundamentals}%
  \BibitemOpen
  \bibfield  {author} {\bibinfo {author} {\bibfnamefont {A.~V.}\ \bibnamefont
  {Chumak}},\ }\href@noop {} {\bibfield  {journal} {\bibinfo  {journal} {arXiv
  preprint arXiv:1901.08934}\ } (\bibinfo {year} {2019})}\BibitemShut {NoStop}%
\bibitem [{\citenamefont {Grundler}(2015)}]{Grundler2015}%
  \BibitemOpen
  \bibfield  {author} {\bibinfo {author} {\bibfnamefont {D.}~\bibnamefont
  {Grundler}},\ }\href@noop {} {\bibfield  {journal} {\bibinfo  {journal}
  {Nature Physics}\ }\textbf {\bibinfo {volume} {11}},\ \bibinfo {pages} {438 }
  (\bibinfo {year} {2015})}\BibitemShut {NoStop}%
\bibitem [{\citenamefont {Lendinez}\ and\ \citenamefont
  {Jungfleisch}(2019)}]{Lendinez2019}%
  \BibitemOpen
  \bibfield  {author} {\bibinfo {author} {\bibfnamefont {S.}~\bibnamefont
  {Lendinez}}\ and\ \bibinfo {author} {\bibfnamefont {M.~B.}\ \bibnamefont
  {Jungfleisch}},\ }\href@noop {} {\bibfield  {journal} {\bibinfo  {journal}
  {J. Phys.: Condens. Matter}\ }\textbf {\bibinfo {volume} {32}},\ \bibinfo
  {pages} {013001} (\bibinfo {year} {2019})}\BibitemShut {NoStop}%
\bibitem [{\citenamefont {Skj{\ae}rv{\o}}\ \emph {et~al.}(2019)\citenamefont
  {Skj{\ae}rv{\o}}, \citenamefont {Marrows}, \citenamefont {Stamps},\ and\
  \citenamefont {Heyderman}}]{Skaervo2019}%
  \BibitemOpen
  \bibfield  {author} {\bibinfo {author} {\bibfnamefont {S.~H.}\ \bibnamefont
  {Skj{\ae}rv{\o}}}, \bibinfo {author} {\bibfnamefont {C.~H.}\ \bibnamefont
  {Marrows}}, \bibinfo {author} {\bibfnamefont {R.~L.}\ \bibnamefont {Stamps}},
  \ and\ \bibinfo {author} {\bibfnamefont {L.~J.}\ \bibnamefont {Heyderman}},\
  }\href {\doibase 10.1038/s42254-019-0118-3} {\bibfield  {journal} {\bibinfo
  {journal} {Nature Reviews Physics}\ } (\bibinfo {year} {2019}),\
  10.1038/s42254-019-0118-3}\BibitemShut {NoStop}%
\bibitem [{\citenamefont {Nisoli}\ \emph {et~al.}(2007)\citenamefont {Nisoli},
  \citenamefont {Wang}, \citenamefont {Li}, \citenamefont {McConville},
  \citenamefont {Lammert}, \citenamefont {Schiffer},\ and\ \citenamefont
  {Crespi}}]{Nisoli2007GS}%
  \BibitemOpen
  \bibfield  {author} {\bibinfo {author} {\bibfnamefont {C.}~\bibnamefont
  {Nisoli}}, \bibinfo {author} {\bibfnamefont {R.}~\bibnamefont {Wang}},
  \bibinfo {author} {\bibfnamefont {J.}~\bibnamefont {Li}}, \bibinfo {author}
  {\bibfnamefont {W.~F.}\ \bibnamefont {McConville}}, \bibinfo {author}
  {\bibfnamefont {P.~E.}\ \bibnamefont {Lammert}}, \bibinfo {author}
  {\bibfnamefont {P.}~\bibnamefont {Schiffer}}, \ and\ \bibinfo {author}
  {\bibfnamefont {V.~H.}\ \bibnamefont {Crespi}},\ }\href {\doibase
  10.1103/PhysRevLett.98.217203} {\bibfield  {journal} {\bibinfo  {journal}
  {Phys. Rev. Lett.}\ }\textbf {\bibinfo {volume} {98}},\ \bibinfo {pages}
  {217203} (\bibinfo {year} {2007})}\BibitemShut {NoStop}%
\bibitem [{\citenamefont {Porro}\ \emph {et~al.}(2013)\citenamefont {Porro},
  \citenamefont {Bedoya-Pinto}, \citenamefont {Berger},\ and\ \citenamefont
  {Vavassori}}]{Porro2013}%
  \BibitemOpen
  \bibfield  {author} {\bibinfo {author} {\bibfnamefont {J.-M.}\ \bibnamefont
  {Porro}}, \bibinfo {author} {\bibfnamefont {A.}~\bibnamefont {Bedoya-Pinto}},
  \bibinfo {author} {\bibfnamefont {A.}~\bibnamefont {Berger}}, \ and\ \bibinfo
  {author} {\bibfnamefont {P.}~\bibnamefont {Vavassori}},\ }\href {\doibase
  10.1088/1367-2630/15/5/055012} {\bibfield  {journal} {\bibinfo  {journal}
  {New Journal of Physics}\ }\textbf {\bibinfo {volume} {15}},\ \bibinfo
  {pages} {055012} (\bibinfo {year} {2013})}\BibitemShut {NoStop}%
\bibitem [{\citenamefont {Zhang}\ \emph
  {et~al.}(2013{\natexlab{a}})\citenamefont {Zhang}, \citenamefont {Ren},
  \citenamefont {Wang},\ and\ \citenamefont {Li}}]{Zhang2013}%
  \BibitemOpen
  \bibfield  {author} {\bibinfo {author} {\bibfnamefont {L.}~\bibnamefont
  {Zhang}}, \bibinfo {author} {\bibfnamefont {J.}~\bibnamefont {Ren}}, \bibinfo
  {author} {\bibfnamefont {J.-S.}\ \bibnamefont {Wang}}, \ and\ \bibinfo
  {author} {\bibfnamefont {B.}~\bibnamefont {Li}},\ }\href@noop {} {\bibfield
  {journal} {\bibinfo  {journal} {Phys. Rev. B}\ }\textbf {\bibinfo {volume}
  {87}},\ \bibinfo {pages} {144101} (\bibinfo {year}
  {2013}{\natexlab{a}})}\BibitemShut {NoStop}%
\bibitem [{\citenamefont {Zhang}\ \emph
  {et~al.}(2019{\natexlab{a}})\citenamefont {Zhang}, \citenamefont {Lao},
  \citenamefont {Sklenar}, \citenamefont {Bingham}, \citenamefont {Batley},
  \citenamefont {Watts}, \citenamefont {Nisoli}, \citenamefont {Leighton},\
  and\ \citenamefont {Schiffer}}]{Zhang2019_2}%
  \BibitemOpen
  \bibfield  {author} {\bibinfo {author} {\bibfnamefont {X.}~\bibnamefont
  {Zhang}}, \bibinfo {author} {\bibfnamefont {Y.}~\bibnamefont {Lao}}, \bibinfo
  {author} {\bibfnamefont {J.}~\bibnamefont {Sklenar}}, \bibinfo {author}
  {\bibfnamefont {N.~S.}\ \bibnamefont {Bingham}}, \bibinfo {author}
  {\bibfnamefont {J.~T.}\ \bibnamefont {Batley}}, \bibinfo {author}
  {\bibfnamefont {J.~D.}\ \bibnamefont {Watts}}, \bibinfo {author}
  {\bibfnamefont {C.}~\bibnamefont {Nisoli}}, \bibinfo {author} {\bibfnamefont
  {C.}~\bibnamefont {Leighton}}, \ and\ \bibinfo {author} {\bibfnamefont
  {P.}~\bibnamefont {Schiffer}},\ }\href {\doibase 10.1063/1.5126713}
  {\bibfield  {journal} {\bibinfo  {journal} {APL Materials}\ }\textbf
  {\bibinfo {volume} {7}},\ \bibinfo {pages} {111112} (\bibinfo {year}
  {2019}{\natexlab{a}})},\ \Eprint
  {http://arxiv.org/abs/https://doi.org/10.1063/1.5126713}
  {https://doi.org/10.1063/1.5126713} \BibitemShut {NoStop}%
\bibitem [{\citenamefont {Morley}\ \emph {et~al.}(2018)\citenamefont {Morley},
  \citenamefont {Riley}, \citenamefont {Porro}, \citenamefont {Rosamond},
  \citenamefont {Linfield}, \citenamefont {Cunningham}, \citenamefont
  {Langridge},\ and\ \citenamefont {Marrows}}]{Morley2018}%
  \BibitemOpen
  \bibfield  {author} {\bibinfo {author} {\bibfnamefont {S.~A.}\ \bibnamefont
  {Morley}}, \bibinfo {author} {\bibfnamefont {S.~T.}\ \bibnamefont {Riley}},
  \bibinfo {author} {\bibfnamefont {J.-M.}\ \bibnamefont {Porro}}, \bibinfo
  {author} {\bibfnamefont {M.~C.}\ \bibnamefont {Rosamond}}, \bibinfo {author}
  {\bibfnamefont {E.~H.}\ \bibnamefont {Linfield}}, \bibinfo {author}
  {\bibfnamefont {J.~E.}\ \bibnamefont {Cunningham}}, \bibinfo {author}
  {\bibfnamefont {S.}~\bibnamefont {Langridge}}, \ and\ \bibinfo {author}
  {\bibfnamefont {C.~H.}\ \bibnamefont {Marrows}},\ }\href {\doibase
  10.1038/s41598-018-23208-6} {\bibfield  {journal} {\bibinfo  {journal} {Sci.
  Rep.}\ }\textbf {\bibinfo {volume} {8}},\ \bibinfo {pages} {4750} (\bibinfo
  {year} {2018})}\BibitemShut {NoStop}%
\bibitem [{\citenamefont {Gliga}\ \emph {et~al.}(2013)\citenamefont {Gliga},
  \citenamefont {K\'akay}, \citenamefont {Hertel},\ and\ \citenamefont
  {Heinonen}}]{Gliga2013}%
  \BibitemOpen
  \bibfield  {author} {\bibinfo {author} {\bibfnamefont {S.}~\bibnamefont
  {Gliga}}, \bibinfo {author} {\bibfnamefont {A.}~\bibnamefont {K\'akay}},
  \bibinfo {author} {\bibfnamefont {R.}~\bibnamefont {Hertel}}, \ and\ \bibinfo
  {author} {\bibfnamefont {O.~G.}\ \bibnamefont {Heinonen}},\ }\href {\doibase
  10.1103/PhysRevLett.110.117205} {\bibfield  {journal} {\bibinfo  {journal}
  {Phys. Rev. Lett.}\ }\textbf {\bibinfo {volume} {110}},\ \bibinfo {pages}
  {117205} (\bibinfo {year} {2013})}\BibitemShut {NoStop}%
\bibitem [{\citenamefont {Anghinolfi}\ \emph {et~al.}(2015)\citenamefont
  {Anghinolfi}, \citenamefont {Luetkens}, \citenamefont {Perron}, \citenamefont
  {Flokstra}, \citenamefont {Sendetskyi}, \citenamefont {Suter}, \citenamefont
  {Prokscha}, \citenamefont {Derlet}, \citenamefont {L},\ and\ \citenamefont
  {Heyderman}}]{Anghinolfi2015}%
  \BibitemOpen
  \bibfield  {author} {\bibinfo {author} {\bibfnamefont {L.}~\bibnamefont
  {Anghinolfi}}, \bibinfo {author} {\bibfnamefont {H.}~\bibnamefont
  {Luetkens}}, \bibinfo {author} {\bibfnamefont {J.}~\bibnamefont {Perron}},
  \bibinfo {author} {\bibfnamefont {M.~G.}\ \bibnamefont {Flokstra}}, \bibinfo
  {author} {\bibfnamefont {O.}~\bibnamefont {Sendetskyi}}, \bibinfo {author}
  {\bibfnamefont {A.}~\bibnamefont {Suter}}, \bibinfo {author} {\bibfnamefont
  {T.}~\bibnamefont {Prokscha}}, \bibinfo {author} {\bibfnamefont {P.~M.}\
  \bibnamefont {Derlet}}, \bibinfo {author} {\bibfnamefont {L.~S.}\
  \bibnamefont {L}}, \ and\ \bibinfo {author} {\bibfnamefont {L.~J.}\
  \bibnamefont {Heyderman}},\ }\href {\doibase 10.1038/ncomms9278} {\bibfield
  {journal} {\bibinfo  {journal} {Nat. Commun.}\ }\textbf {\bibinfo {volume}
  {6}},\ \bibinfo {pages} {8278} (\bibinfo {year} {2015})}\BibitemShut
  {NoStop}%
\bibitem [{\citenamefont {Gartside}\ \emph {et~al.}(2018)\citenamefont
  {Gartside}, \citenamefont {Arroo}, \citenamefont {Burn}, \citenamefont
  {Bemmer}, \citenamefont {Moskalenko}, \citenamefont {Cohen},\ and\
  \citenamefont {Branford}}]{Gartside2018}%
  \BibitemOpen
  \bibfield  {author} {\bibinfo {author} {\bibfnamefont {J.~C.}\ \bibnamefont
  {Gartside}}, \bibinfo {author} {\bibfnamefont {D.~M.}\ \bibnamefont {Arroo}},
  \bibinfo {author} {\bibfnamefont {D.~M.}\ \bibnamefont {Burn}}, \bibinfo
  {author} {\bibfnamefont {V.~L.}\ \bibnamefont {Bemmer}}, \bibinfo {author}
  {\bibfnamefont {A.}~\bibnamefont {Moskalenko}}, \bibinfo {author}
  {\bibfnamefont {L.~F.}\ \bibnamefont {Cohen}}, \ and\ \bibinfo {author}
  {\bibfnamefont {W.~R.}\ \bibnamefont {Branford}},\ }\href@noop {} {\bibfield
  {journal} {\bibinfo  {journal} {Nature Nanotechnology}\ }\textbf {\bibinfo
  {volume} {13}},\ \bibinfo {pages} {53–58} (\bibinfo {year}
  {2018})}\BibitemShut {NoStop}%
\bibitem [{\citenamefont {Morrison}\ \emph {et~al.}(2013)\citenamefont
  {Morrison}, \citenamefont {Nelson},\ and\ \citenamefont
  {Nisoli}}]{Morrison2013}%
  \BibitemOpen
  \bibfield  {author} {\bibinfo {author} {\bibfnamefont {M.~J.}\ \bibnamefont
  {Morrison}}, \bibinfo {author} {\bibfnamefont {T.~R.}\ \bibnamefont
  {Nelson}}, \ and\ \bibinfo {author} {\bibfnamefont {C.}~\bibnamefont
  {Nisoli}},\ }\href@noop {} {\bibfield  {journal} {\bibinfo  {journal} {New
  Journal of Physics}\ }\textbf {\bibinfo {volume} {15}},\ \bibinfo {pages}
  {045009} (\bibinfo {year} {2013})}\BibitemShut {NoStop}%
\bibitem [{\citenamefont {Lao}\ \emph {et~al.}(2018)\citenamefont {Lao},
  \citenamefont {Caravelli}, \citenamefont {Sheikh}, \citenamefont {Sklenar},
  \citenamefont {Gardeazabal}, \citenamefont {Watts}, \citenamefont {Albrecht},
  \citenamefont {Scholl}, \citenamefont {Dahmen}, \citenamefont {Nisoli} \emph
  {et~al.}}]{Lao2018}%
  \BibitemOpen
  \bibfield  {author} {\bibinfo {author} {\bibfnamefont {Y.}~\bibnamefont
  {Lao}}, \bibinfo {author} {\bibfnamefont {F.}~\bibnamefont {Caravelli}},
  \bibinfo {author} {\bibfnamefont {M.}~\bibnamefont {Sheikh}}, \bibinfo
  {author} {\bibfnamefont {J.}~\bibnamefont {Sklenar}}, \bibinfo {author}
  {\bibfnamefont {D.}~\bibnamefont {Gardeazabal}}, \bibinfo {author}
  {\bibfnamefont {J.~D.}\ \bibnamefont {Watts}}, \bibinfo {author}
  {\bibfnamefont {A.~M.}\ \bibnamefont {Albrecht}}, \bibinfo {author}
  {\bibfnamefont {A.}~\bibnamefont {Scholl}}, \bibinfo {author} {\bibfnamefont
  {K.}~\bibnamefont {Dahmen}}, \bibinfo {author} {\bibfnamefont
  {C.}~\bibnamefont {Nisoli}},  \emph {et~al.},\ }\href@noop {} {\bibfield
  {journal} {\bibinfo  {journal} {Nature Physics}\ }\textbf {\bibinfo {volume}
  {14}},\ \bibinfo {pages} {723} (\bibinfo {year} {2018})}\BibitemShut
  {NoStop}%
\bibitem [{\citenamefont {Zhang}\ \emph
  {et~al.}(2013{\natexlab{b}})\citenamefont {Zhang}, \citenamefont {Gilbert},
  \citenamefont {Nisoli}, \citenamefont {Chern}, \citenamefont {Erikson},
  \citenamefont {O'Brien}, \citenamefont {Leighton}, \citenamefont {Lammert},
  \citenamefont {Crespi},\ and\ \citenamefont
  {Schiffer}}]{Zhang2013-crystallites}%
  \BibitemOpen
  \bibfield  {author} {\bibinfo {author} {\bibfnamefont {S.}~\bibnamefont
  {Zhang}}, \bibinfo {author} {\bibfnamefont {I.}~\bibnamefont {Gilbert}},
  \bibinfo {author} {\bibfnamefont {C.}~\bibnamefont {Nisoli}}, \bibinfo
  {author} {\bibfnamefont {G.-W.}\ \bibnamefont {Chern}}, \bibinfo {author}
  {\bibfnamefont {M.~J.}\ \bibnamefont {Erikson}}, \bibinfo {author}
  {\bibfnamefont {L.}~\bibnamefont {O'Brien}}, \bibinfo {author} {\bibfnamefont
  {C.}~\bibnamefont {Leighton}}, \bibinfo {author} {\bibfnamefont {P.~E.}\
  \bibnamefont {Lammert}}, \bibinfo {author} {\bibfnamefont {V.~H.}\
  \bibnamefont {Crespi}}, \ and\ \bibinfo {author} {\bibfnamefont
  {P.}~\bibnamefont {Schiffer}},\ }\href {\doibase 10.1038/nature12399}
  {\bibfield  {journal} {\bibinfo  {journal} {Nature}\ }\textbf {\bibinfo
  {volume} {500}},\ \bibinfo {pages} {553} (\bibinfo {year}
  {2013}{\natexlab{b}})}\BibitemShut {NoStop}%
\bibitem [{\citenamefont {Phatak}\ \emph {et~al.}(2011)\citenamefont {Phatak},
  \citenamefont {Petford-Long}, \citenamefont {Heinonen}, \citenamefont
  {Tanase},\ and\ \citenamefont {De~Graef}}]{Phatak2011}%
  \BibitemOpen
  \bibfield  {author} {\bibinfo {author} {\bibfnamefont {C.}~\bibnamefont
  {Phatak}}, \bibinfo {author} {\bibfnamefont {A.~K.}\ \bibnamefont
  {Petford-Long}}, \bibinfo {author} {\bibfnamefont {O.}~\bibnamefont
  {Heinonen}}, \bibinfo {author} {\bibfnamefont {M.}~\bibnamefont {Tanase}}, \
  and\ \bibinfo {author} {\bibfnamefont {M.}~\bibnamefont {De~Graef}},\ }\href
  {\doibase 10.1103/PhysRevB.83.174431} {\bibfield  {journal} {\bibinfo
  {journal} {Phys. Rev. B}\ }\textbf {\bibinfo {volume} {83}},\ \bibinfo
  {pages} {174431} (\bibinfo {year} {2011})}\BibitemShut {NoStop}%
\bibitem [{\citenamefont {Li}\ \emph {et~al.}(2019{\natexlab{a}})\citenamefont
  {Li}, \citenamefont {Paterson}, \citenamefont {Macauley}, \citenamefont
  {Nascimento}, \citenamefont {Ferguson}, \citenamefont {Morley}, \citenamefont
  {Rosamond}, \citenamefont {Linfield}, \citenamefont {MacLaren}, \citenamefont
  {Macêdo}, \citenamefont {Marrows}, \citenamefont {McVitie},\ and\
  \citenamefont {Stamps}}]{Li2019_pin}%
  \BibitemOpen
  \bibfield  {author} {\bibinfo {author} {\bibfnamefont {Y.}~\bibnamefont
  {Li}}, \bibinfo {author} {\bibfnamefont {G.~W.}\ \bibnamefont {Paterson}},
  \bibinfo {author} {\bibfnamefont {G.~M.}\ \bibnamefont {Macauley}}, \bibinfo
  {author} {\bibfnamefont {F.~S.}\ \bibnamefont {Nascimento}}, \bibinfo
  {author} {\bibfnamefont {C.}~\bibnamefont {Ferguson}}, \bibinfo {author}
  {\bibfnamefont {S.~A.}\ \bibnamefont {Morley}}, \bibinfo {author}
  {\bibfnamefont {M.~C.}\ \bibnamefont {Rosamond}}, \bibinfo {author}
  {\bibfnamefont {E.~H.}\ \bibnamefont {Linfield}}, \bibinfo {author}
  {\bibfnamefont {D.~A.}\ \bibnamefont {MacLaren}}, \bibinfo {author}
  {\bibfnamefont {R.}~\bibnamefont {Macêdo}}, \bibinfo {author} {\bibfnamefont
  {C.~H.}\ \bibnamefont {Marrows}}, \bibinfo {author} {\bibfnamefont
  {S.}~\bibnamefont {McVitie}}, \ and\ \bibinfo {author} {\bibfnamefont
  {R.~L.}\ \bibnamefont {Stamps}},\ }\href {\doibase 10.1021/acsnano.8b08884}
  {\bibfield  {journal} {\bibinfo  {journal} {ACS Nano}\ }\textbf {\bibinfo
  {volume} {13}},\ \bibinfo {pages} {2213} (\bibinfo {year}
  {2019}{\natexlab{a}})}\BibitemShut {NoStop}%
\bibitem [{\citenamefont {Sebastian}\ \emph {et~al.}(2015)\citenamefont
  {Sebastian}, \citenamefont {Schultheiss}, \citenamefont {Obry}, \citenamefont
  {Hillebrands},\ and\ \citenamefont {Schultheiss}}]{SebastianFPhy}%
  \BibitemOpen
  \bibfield  {author} {\bibinfo {author} {\bibfnamefont {T.}~\bibnamefont
  {Sebastian}}, \bibinfo {author} {\bibfnamefont {K.}~\bibnamefont
  {Schultheiss}}, \bibinfo {author} {\bibfnamefont {B.}~\bibnamefont {Obry}},
  \bibinfo {author} {\bibfnamefont {B.}~\bibnamefont {Hillebrands}}, \ and\
  \bibinfo {author} {\bibfnamefont {H.}~\bibnamefont {Schultheiss}},\
  }\href@noop {} {\bibfield  {journal} {\bibinfo  {journal} {Frontiers in
  Physics}\ }\textbf {\bibinfo {volume} {3}},\ \bibinfo {pages} {35} (\bibinfo
  {year} {2015})}\BibitemShut {NoStop}%
\bibitem [{\citenamefont {Li}\ \emph {et~al.}(2016{\natexlab{a}})\citenamefont
  {Li}, \citenamefont {Gubbiotti}, \citenamefont {Casoli}, \citenamefont
  {Gon{\c{c}}alves}, \citenamefont {Morley}, \citenamefont {Rosamond},
  \citenamefont {Linfield}, \citenamefont {Marrows}, \citenamefont {McVitie},\
  and\ \citenamefont {Stamps}}]{Li2016modes}%
  \BibitemOpen
  \bibfield  {author} {\bibinfo {author} {\bibfnamefont {Y.}~\bibnamefont
  {Li}}, \bibinfo {author} {\bibfnamefont {G.}~\bibnamefont {Gubbiotti}},
  \bibinfo {author} {\bibfnamefont {F.}~\bibnamefont {Casoli}}, \bibinfo
  {author} {\bibfnamefont {F.~J.~T.}\ \bibnamefont {Gon{\c{c}}alves}}, \bibinfo
  {author} {\bibfnamefont {S.~A.}\ \bibnamefont {Morley}}, \bibinfo {author}
  {\bibfnamefont {M.~C.}\ \bibnamefont {Rosamond}}, \bibinfo {author}
  {\bibfnamefont {E.~H.}\ \bibnamefont {Linfield}}, \bibinfo {author}
  {\bibfnamefont {C.~H.}\ \bibnamefont {Marrows}}, \bibinfo {author}
  {\bibfnamefont {S.}~\bibnamefont {McVitie}}, \ and\ \bibinfo {author}
  {\bibfnamefont {R.~L.}\ \bibnamefont {Stamps}},\ }\href {\doibase
  10.1088/1361-6463/50/1/015003} {\bibfield  {journal} {\bibinfo  {journal}
  {Journal of Physics D: Applied Physics}\ }\textbf {\bibinfo {volume} {50}},\
  \bibinfo {pages} {015003} (\bibinfo {year} {2016}{\natexlab{a}})}\BibitemShut
  {NoStop}%
\bibitem [{\citenamefont {Sch\"utz}\ \emph {et~al.}(1987)\citenamefont
  {Sch\"utz}, \citenamefont {Wagner}, \citenamefont {Wilhelm}, \citenamefont
  {Kienle}, \citenamefont {Zeller}, \citenamefont {Frahm},\ and\ \citenamefont
  {Materlik}}]{Schuetz1987}%
  \BibitemOpen
  \bibfield  {author} {\bibinfo {author} {\bibfnamefont {G.}~\bibnamefont
  {Sch\"utz}}, \bibinfo {author} {\bibfnamefont {W.}~\bibnamefont {Wagner}},
  \bibinfo {author} {\bibfnamefont {W.}~\bibnamefont {Wilhelm}}, \bibinfo
  {author} {\bibfnamefont {P.}~\bibnamefont {Kienle}}, \bibinfo {author}
  {\bibfnamefont {R.}~\bibnamefont {Zeller}}, \bibinfo {author} {\bibfnamefont
  {R.}~\bibnamefont {Frahm}}, \ and\ \bibinfo {author} {\bibfnamefont
  {G.}~\bibnamefont {Materlik}},\ }\href {\doibase 10.1103/PhysRevLett.58.737}
  {\bibfield  {journal} {\bibinfo  {journal} {Phys. Rev. Lett.}\ }\textbf
  {\bibinfo {volume} {58}},\ \bibinfo {pages} {737} (\bibinfo {year}
  {1987})}\BibitemShut {NoStop}%
\bibitem [{\citenamefont {Mengotti}\ \emph {et~al.}(2011)\citenamefont
  {Mengotti}, \citenamefont {Heyderman}, \citenamefont {Fraile~Rodr/'{i}guez},
  \citenamefont {Nolting}, \citenamefont {H\"{u}gli},\ and\ \citenamefont
  {Braun}}]{Mengotti2011}%
  \BibitemOpen
  \bibfield  {author} {\bibinfo {author} {\bibfnamefont {E.}~\bibnamefont
  {Mengotti}}, \bibinfo {author} {\bibfnamefont {L.}~\bibnamefont {Heyderman}},
  \bibinfo {author} {\bibfnamefont {A.}~\bibnamefont {Fraile~Rodr/'{i}guez}},
  \bibinfo {author} {\bibfnamefont {F.}~\bibnamefont {Nolting}}, \bibinfo
  {author} {\bibfnamefont {R.}~\bibnamefont {H\"{u}gli}}, \ and\ \bibinfo
  {author} {\bibfnamefont {H.~B.}\ \bibnamefont {Braun}},\ }\href@noop {}
  {\bibfield  {journal} {\bibinfo  {journal} {Nature Physics}\ }\textbf
  {\bibinfo {volume} {7}},\ \bibinfo {pages} {68} (\bibinfo {year}
  {2011})}\BibitemShut {NoStop}%
\bibitem [{\citenamefont {Arnalds}\ \emph {et~al.}(2012)\citenamefont
  {Arnalds}, \citenamefont {Farhan}, \citenamefont {Chopdekar}, \citenamefont
  {Kapaklis}, \citenamefont {Balan}, \citenamefont {Papaioannou}, \citenamefont
  {Ahlberg}, \citenamefont {Nolting}, \citenamefont {Heyderman},\ and\
  \citenamefont {Hj\"{o}rvarsson}}]{Arnalds2013}%
  \BibitemOpen
  \bibfield  {author} {\bibinfo {author} {\bibfnamefont {U.~B.}\ \bibnamefont
  {Arnalds}}, \bibinfo {author} {\bibfnamefont {A.}~\bibnamefont {Farhan}},
  \bibinfo {author} {\bibfnamefont {R.~V.}\ \bibnamefont {Chopdekar}}, \bibinfo
  {author} {\bibfnamefont {V.}~\bibnamefont {Kapaklis}}, \bibinfo {author}
  {\bibfnamefont {A.}~\bibnamefont {Balan}}, \bibinfo {author} {\bibfnamefont
  {E.~T.}\ \bibnamefont {Papaioannou}}, \bibinfo {author} {\bibfnamefont
  {M.}~\bibnamefont {Ahlberg}}, \bibinfo {author} {\bibfnamefont
  {F.}~\bibnamefont {Nolting}}, \bibinfo {author} {\bibfnamefont {L.~J.}\
  \bibnamefont {Heyderman}}, \ and\ \bibinfo {author} {\bibfnamefont
  {B.}~\bibnamefont {Hj\"{o}rvarsson}},\ }\href {\doibase 10.1063/1.4751844}
  {\bibfield  {journal} {\bibinfo  {journal} {Appl. Phys. Lett.}\ }\textbf
  {\bibinfo {volume} {101}},\ \bibinfo {pages} {112404} (\bibinfo {year}
  {2012})}\BibitemShut {NoStop}%
\bibitem [{\citenamefont {Farhan}\ \emph {et~al.}(2014)\citenamefont {Farhan},
  \citenamefont {Kleibert}, \citenamefont {Derlet}, \citenamefont {Anghinolfi},
  \citenamefont {Balan}, \citenamefont {Chopdekar}, \citenamefont {Wyss},
  \citenamefont {Gliga}, \citenamefont {Nolting},\ and\ \citenamefont
  {Heyderman}}]{Farhan2014}%
  \BibitemOpen
  \bibfield  {author} {\bibinfo {author} {\bibfnamefont {A.}~\bibnamefont
  {Farhan}}, \bibinfo {author} {\bibfnamefont {A.}~\bibnamefont {Kleibert}},
  \bibinfo {author} {\bibfnamefont {P.~M.}\ \bibnamefont {Derlet}}, \bibinfo
  {author} {\bibfnamefont {L.}~\bibnamefont {Anghinolfi}}, \bibinfo {author}
  {\bibfnamefont {A.}~\bibnamefont {Balan}}, \bibinfo {author} {\bibfnamefont
  {R.~V.}\ \bibnamefont {Chopdekar}}, \bibinfo {author} {\bibfnamefont
  {M.}~\bibnamefont {Wyss}}, \bibinfo {author} {\bibfnamefont {S.}~\bibnamefont
  {Gliga}}, \bibinfo {author} {\bibfnamefont {F.}~\bibnamefont {Nolting}}, \
  and\ \bibinfo {author} {\bibfnamefont {L.~J.}\ \bibnamefont {Heyderman}},\
  }\href {\doibase 10.1103/PhysRevB.89.214405} {\bibfield  {journal} {\bibinfo
  {journal} {Phys. Rev. B}\ }\textbf {\bibinfo {volume} {89}},\ \bibinfo
  {pages} {214405} (\bibinfo {year} {2014})}\BibitemShut {NoStop}%
\bibitem [{\citenamefont {Finizio}\ \emph {et~al.}(2018)\citenamefont
  {Finizio}, \citenamefont {Wintz}, \citenamefont {Watts},\ and\ \citenamefont
  {Raabe}}]{Finizio2018}%
  \BibitemOpen
  \bibfield  {author} {\bibinfo {author} {\bibfnamefont {S.}~\bibnamefont
  {Finizio}}, \bibinfo {author} {\bibfnamefont {S.}~\bibnamefont {Wintz}},
  \bibinfo {author} {\bibfnamefont {B.}~\bibnamefont {Watts}}, \ and\ \bibinfo
  {author} {\bibfnamefont {J.}~\bibnamefont {Raabe}},\ }\href {\doibase
  10.1017/S1431927618014502} {\bibfield  {journal} {\bibinfo  {journal}
  {Microscopy and Microanalysis}\ }\textbf {\bibinfo {volume} {24}},\ \bibinfo
  {pages} {452} (\bibinfo {year} {2018})}\BibitemShut {NoStop}%
\bibitem [{\citenamefont {Wintz}\ \emph {et~al.}(2016)\citenamefont {Wintz},
  \citenamefont {Tiberkevich}, \citenamefont {Weigand}, \citenamefont {Raabe},
  \citenamefont {Lindner}, \citenamefont {Erbe}, \citenamefont {Slavin},\ and\
  \citenamefont {Fassbender}}]{Wintz2016}%
  \BibitemOpen
  \bibfield  {author} {\bibinfo {author} {\bibfnamefont {S.}~\bibnamefont
  {Wintz}}, \bibinfo {author} {\bibfnamefont {V.}~\bibnamefont {Tiberkevich}},
  \bibinfo {author} {\bibfnamefont {M.}~\bibnamefont {Weigand}}, \bibinfo
  {author} {\bibfnamefont {J.}~\bibnamefont {Raabe}}, \bibinfo {author}
  {\bibfnamefont {J.}~\bibnamefont {Lindner}}, \bibinfo {author} {\bibfnamefont
  {A.}~\bibnamefont {Erbe}}, \bibinfo {author} {\bibfnamefont {A.}~\bibnamefont
  {Slavin}}, \ and\ \bibinfo {author} {\bibfnamefont {J.}~\bibnamefont
  {Fassbender}},\ }\href {\doibase 10.1038/nnano.2016.117} {\bibfield
  {journal} {\bibinfo  {journal} {Nature Nanotechnology}\ }\textbf {\bibinfo
  {volume} {11}},\ \bibinfo {pages} {948} (\bibinfo {year} {2016})}\BibitemShut
  {NoStop}%
\bibitem [{\citenamefont {F\"orster}\ \emph {et~al.}(2019)\citenamefont
  {F\"orster}, \citenamefont {Gr\"afe}, \citenamefont {Bailey}, \citenamefont
  {Finizio}, \citenamefont {Tr\"ager}, \citenamefont {Gro\ss{}}, \citenamefont
  {Mayr}, \citenamefont {Stoll}, \citenamefont {Dubs}, \citenamefont
  {Surzhenko}, \citenamefont {Liebing}, \citenamefont {Woltersdorf},
  \citenamefont {Raabe}, \citenamefont {Weigand}, \citenamefont {Sch\"utz},\
  and\ \citenamefont {Wintz}}]{Foerster2019}%
  \BibitemOpen
  \bibfield  {author} {\bibinfo {author} {\bibfnamefont {J.}~\bibnamefont
  {F\"orster}}, \bibinfo {author} {\bibfnamefont {J.}~\bibnamefont {Gr\"afe}},
  \bibinfo {author} {\bibfnamefont {J.}~\bibnamefont {Bailey}}, \bibinfo
  {author} {\bibfnamefont {S.}~\bibnamefont {Finizio}}, \bibinfo {author}
  {\bibfnamefont {N.}~\bibnamefont {Tr\"ager}}, \bibinfo {author}
  {\bibfnamefont {F.}~\bibnamefont {Gro\ss{}}}, \bibinfo {author}
  {\bibfnamefont {S.}~\bibnamefont {Mayr}}, \bibinfo {author} {\bibfnamefont
  {H.}~\bibnamefont {Stoll}}, \bibinfo {author} {\bibfnamefont
  {C.}~\bibnamefont {Dubs}}, \bibinfo {author} {\bibfnamefont {O.}~\bibnamefont
  {Surzhenko}}, \bibinfo {author} {\bibfnamefont {N.}~\bibnamefont {Liebing}},
  \bibinfo {author} {\bibfnamefont {G.}~\bibnamefont {Woltersdorf}}, \bibinfo
  {author} {\bibfnamefont {J.}~\bibnamefont {Raabe}}, \bibinfo {author}
  {\bibfnamefont {M.}~\bibnamefont {Weigand}}, \bibinfo {author} {\bibfnamefont
  {G.}~\bibnamefont {Sch\"utz}}, \ and\ \bibinfo {author} {\bibfnamefont
  {S.}~\bibnamefont {Wintz}},\ }\href {\doibase 10.1103/PhysRevB.100.214416}
  {\bibfield  {journal} {\bibinfo  {journal} {Phys. Rev. B}\ }\textbf {\bibinfo
  {volume} {100}},\ \bibinfo {pages} {214416} (\bibinfo {year}
  {2019})}\BibitemShut {NoStop}%
\bibitem [{\citenamefont {Albisetti}\ \emph {et~al.}(2020)\citenamefont
  {Albisetti}, \citenamefont {Tacchi}, \citenamefont {Silvani}, \citenamefont
  {Scaramuzzi}, \citenamefont {Finizio}, \citenamefont {Wintz}, \citenamefont
  {Rinaldi}, \citenamefont {Cantoni}, \citenamefont {Raabe}, \citenamefont
  {Carlotti}, \citenamefont {Bertacco}, \citenamefont {Riedo},\ and\
  \citenamefont {Petti}}]{Albisetti20}%
  \BibitemOpen
  \bibfield  {author} {\bibinfo {author} {\bibfnamefont {E.}~\bibnamefont
  {Albisetti}}, \bibinfo {author} {\bibfnamefont {S.}~\bibnamefont {Tacchi}},
  \bibinfo {author} {\bibfnamefont {R.}~\bibnamefont {Silvani}}, \bibinfo
  {author} {\bibfnamefont {G.}~\bibnamefont {Scaramuzzi}}, \bibinfo {author}
  {\bibfnamefont {S.}~\bibnamefont {Finizio}}, \bibinfo {author} {\bibfnamefont
  {S.}~\bibnamefont {Wintz}}, \bibinfo {author} {\bibfnamefont
  {C.}~\bibnamefont {Rinaldi}}, \bibinfo {author} {\bibfnamefont
  {M.}~\bibnamefont {Cantoni}}, \bibinfo {author} {\bibfnamefont
  {J.}~\bibnamefont {Raabe}}, \bibinfo {author} {\bibfnamefont
  {G.}~\bibnamefont {Carlotti}}, \bibinfo {author} {\bibfnamefont
  {R.}~\bibnamefont {Bertacco}}, \bibinfo {author} {\bibfnamefont
  {E.}~\bibnamefont {Riedo}}, \ and\ \bibinfo {author} {\bibfnamefont
  {D.}~\bibnamefont {Petti}},\ }\href {\doibase 10.1002/adma.201906439}
  {\bibfield  {journal} {\bibinfo  {journal} {Advanced Materials}\ }\textbf
  {\bibinfo {volume} {n/a}},\ \bibinfo {pages} {1906439} (\bibinfo {year}
  {2020})},\ \Eprint
  {http://arxiv.org/abs/https://onlinelibrary.wiley.com/doi/pdf/10.1002/adma.201906439}
  {https://onlinelibrary.wiley.com/doi/pdf/10.1002/adma.201906439} \BibitemShut
  {NoStop}%
\bibitem [{\citenamefont {Boero}\ \emph {et~al.}(2009)\citenamefont {Boero},
  \citenamefont {Rusponi}, \citenamefont {Kavich}, \citenamefont
  {Lodi~Rizzini}, \citenamefont {Piamonteze}, \citenamefont {Nolting},
  \citenamefont {Tieg}, \citenamefont {Thiele},\ and\ \citenamefont
  {Gambardella}}]{Boero2009}%
  \BibitemOpen
  \bibfield  {author} {\bibinfo {author} {\bibfnamefont {G.}~\bibnamefont
  {Boero}}, \bibinfo {author} {\bibfnamefont {S.}~\bibnamefont {Rusponi}},
  \bibinfo {author} {\bibfnamefont {J.}~\bibnamefont {Kavich}}, \bibinfo
  {author} {\bibfnamefont {A.}~\bibnamefont {Lodi~Rizzini}}, \bibinfo {author}
  {\bibfnamefont {C.}~\bibnamefont {Piamonteze}}, \bibinfo {author}
  {\bibfnamefont {F.}~\bibnamefont {Nolting}}, \bibinfo {author} {\bibfnamefont
  {C.}~\bibnamefont {Tieg}}, \bibinfo {author} {\bibfnamefont {J.-U.}\
  \bibnamefont {Thiele}}, \ and\ \bibinfo {author} {\bibfnamefont
  {P.}~\bibnamefont {Gambardella}},\ }\href {\doibase 10.1063/1.3267192}
  {\bibfield  {journal} {\bibinfo  {journal} {Rev. Sci. Instrum.}\ }\textbf
  {\bibinfo {volume} {80}},\ \bibinfo {pages} {123902} (\bibinfo {year}
  {2009})}\BibitemShut {NoStop}%
\bibitem [{\citenamefont {Wyss}\ \emph {et~al.}(0)\citenamefont {Wyss},
  \citenamefont {Gliga}, \citenamefont {Vasyukov}, \citenamefont {Ceccarelli},
  \citenamefont {Romagnoli}, \citenamefont {Cui}, \citenamefont {Kleibert},
  \citenamefont {Stamps},\ and\ \citenamefont {Poggio}}]{Wyss2019}%
  \BibitemOpen
  \bibfield  {author} {\bibinfo {author} {\bibfnamefont {M.}~\bibnamefont
  {Wyss}}, \bibinfo {author} {\bibfnamefont {S.}~\bibnamefont {Gliga}},
  \bibinfo {author} {\bibfnamefont {D.}~\bibnamefont {Vasyukov}}, \bibinfo
  {author} {\bibfnamefont {L.}~\bibnamefont {Ceccarelli}}, \bibinfo {author}
  {\bibfnamefont {G.}~\bibnamefont {Romagnoli}}, \bibinfo {author}
  {\bibfnamefont {J.}~\bibnamefont {Cui}}, \bibinfo {author} {\bibfnamefont
  {A.}~\bibnamefont {Kleibert}}, \bibinfo {author} {\bibfnamefont {R.~L.}\
  \bibnamefont {Stamps}}, \ and\ \bibinfo {author} {\bibfnamefont
  {M.}~\bibnamefont {Poggio}},\ }\href {\doibase 10.1021/acsnano.9b05428}
  {\bibfield  {journal} {\bibinfo  {journal} {ACS Nano}\ }\textbf {\bibinfo
  {volume} {0}},\ \bibinfo {pages} {null} (\bibinfo {year} {0})},\ \bibinfo
  {note} {pMID: 31820931},\ \Eprint
  {http://arxiv.org/abs/https://doi.org/10.1021/acsnano.9b05428}
  {https://doi.org/10.1021/acsnano.9b05428} \BibitemShut {NoStop}%
\bibitem [{\citenamefont {Finkler}\ \emph {et~al.}(2010)\citenamefont
  {Finkler}, \citenamefont {Segev}, \citenamefont {Myasoedov}, \citenamefont
  {Rappaport}, \citenamefont {Ne`eman}, \citenamefont {Vasyukov}, \citenamefont
  {Zeldov}, \citenamefont {Huber}, \citenamefont {Martin},\ and\ \citenamefont
  {Yacoby}}]{Finkler2010}%
  \BibitemOpen
  \bibfield  {author} {\bibinfo {author} {\bibfnamefont {A.}~\bibnamefont
  {Finkler}}, \bibinfo {author} {\bibfnamefont {Y.}~\bibnamefont {Segev}},
  \bibinfo {author} {\bibfnamefont {Y.}~\bibnamefont {Myasoedov}}, \bibinfo
  {author} {\bibfnamefont {M.~L.}\ \bibnamefont {Rappaport}}, \bibinfo {author}
  {\bibfnamefont {L.}~\bibnamefont {Ne`eman}}, \bibinfo {author} {\bibfnamefont
  {D.}~\bibnamefont {Vasyukov}}, \bibinfo {author} {\bibfnamefont
  {E.}~\bibnamefont {Zeldov}}, \bibinfo {author} {\bibfnamefont {M.~E.}\
  \bibnamefont {Huber}}, \bibinfo {author} {\bibfnamefont {J.}~\bibnamefont
  {Martin}}, \ and\ \bibinfo {author} {\bibfnamefont {A.}~\bibnamefont
  {Yacoby}},\ }\href {\doibase 10.1021/nl100009r} {\bibfield  {journal}
  {\bibinfo  {journal} {Nano Letters}\ }\textbf {\bibinfo {volume} {10}},\
  \bibinfo {pages} {1046} (\bibinfo {year} {2010})},\ \bibinfo {note} {pMID:
  20131810},\ \Eprint {http://arxiv.org/abs/https://doi.org/10.1021/nl100009r}
  {https://doi.org/10.1021/nl100009r} \BibitemShut {NoStop}%
\bibitem [{\citenamefont {Vasyukov}\ \emph {et~al.}(2013)\citenamefont
  {Vasyukov}, \citenamefont {Anahory}, \citenamefont {Embon}, \citenamefont
  {Halbertal}, \citenamefont {Cuppens}, \citenamefont {Neeman}, \citenamefont
  {Finkler}, \citenamefont {Segev}, \citenamefont {Myasoedov}, \citenamefont
  {Rappaport}, \citenamefont {Huber},\ and\ \citenamefont
  {Zeldov}}]{Vasyukov2013}%
  \BibitemOpen
  \bibfield  {author} {\bibinfo {author} {\bibfnamefont {D.}~\bibnamefont
  {Vasyukov}}, \bibinfo {author} {\bibfnamefont {Y.}~\bibnamefont {Anahory}},
  \bibinfo {author} {\bibfnamefont {L.}~\bibnamefont {Embon}}, \bibinfo
  {author} {\bibfnamefont {D.}~\bibnamefont {Halbertal}}, \bibinfo {author}
  {\bibfnamefont {J.}~\bibnamefont {Cuppens}}, \bibinfo {author} {\bibfnamefont
  {L.}~\bibnamefont {Neeman}}, \bibinfo {author} {\bibfnamefont
  {A.}~\bibnamefont {Finkler}}, \bibinfo {author} {\bibfnamefont
  {Y.}~\bibnamefont {Segev}}, \bibinfo {author} {\bibfnamefont
  {Y.}~\bibnamefont {Myasoedov}}, \bibinfo {author} {\bibfnamefont {M.~L.}\
  \bibnamefont {Rappaport}}, \bibinfo {author} {\bibfnamefont {M.~E.}\
  \bibnamefont {Huber}}, \ and\ \bibinfo {author} {\bibfnamefont
  {E.}~\bibnamefont {Zeldov}},\ }\href {\doibase 10.1038/NNANO.2013.169}
  {\bibfield  {journal} {\bibinfo  {journal} {Nature Nanotechnology}\ }\textbf
  {\bibinfo {volume} {8}},\ \bibinfo {pages} {639} (\bibinfo {year}
  {2013})}\BibitemShut {NoStop}%
\bibitem [{\citenamefont {Vasyukov}\ \emph {et~al.}(2018)\citenamefont
  {Vasyukov}, \citenamefont {Ceccarelli}, \citenamefont {Wyss}, \citenamefont
  {Gross}, \citenamefont {Schwarb}, \citenamefont {Mehlin}, \citenamefont
  {Rossi}, \citenamefont {Tütüncüoglu}, \citenamefont {Heimbach},
  \citenamefont {Zamani}, \citenamefont {Kovács}, \citenamefont {Fontcuberta~i
  Morral}, \citenamefont {Grundler},\ and\ \citenamefont
  {Poggio}}]{Vasyukov2018}%
  \BibitemOpen
  \bibfield  {author} {\bibinfo {author} {\bibfnamefont {D.}~\bibnamefont
  {Vasyukov}}, \bibinfo {author} {\bibfnamefont {L.}~\bibnamefont
  {Ceccarelli}}, \bibinfo {author} {\bibfnamefont {M.}~\bibnamefont {Wyss}},
  \bibinfo {author} {\bibfnamefont {B.}~\bibnamefont {Gross}}, \bibinfo
  {author} {\bibfnamefont {A.}~\bibnamefont {Schwarb}}, \bibinfo {author}
  {\bibfnamefont {A.}~\bibnamefont {Mehlin}}, \bibinfo {author} {\bibfnamefont
  {N.}~\bibnamefont {Rossi}}, \bibinfo {author} {\bibfnamefont
  {G.}~\bibnamefont {Tütüncüoglu}}, \bibinfo {author} {\bibfnamefont
  {F.}~\bibnamefont {Heimbach}}, \bibinfo {author} {\bibfnamefont {R.~R.}\
  \bibnamefont {Zamani}}, \bibinfo {author} {\bibfnamefont {A.}~\bibnamefont
  {Kovács}}, \bibinfo {author} {\bibfnamefont {A.}~\bibnamefont {Fontcuberta~i
  Morral}}, \bibinfo {author} {\bibfnamefont {D.}~\bibnamefont {Grundler}}, \
  and\ \bibinfo {author} {\bibfnamefont {M.}~\bibnamefont {Poggio}},\ }\href
  {\doibase 10.1021/acs.nanolett.7b04386} {\bibfield  {journal} {\bibinfo
  {journal} {Nano Letters}\ }\textbf {\bibinfo {volume} {18}},\ \bibinfo
  {pages} {964} (\bibinfo {year} {2018})},\ \bibinfo {note} {pMID: 29293345},\
  \Eprint {http://arxiv.org/abs/https://doi.org/10.1021/acs.nanolett.7b04386}
  {https://doi.org/10.1021/acs.nanolett.7b04386} \BibitemShut {NoStop}%
\bibitem [{\citenamefont {Maletinsky}\ \emph {et~al.}(2012)\citenamefont
  {Maletinsky}, \citenamefont {Hong}, \citenamefont {Grinolds}, \citenamefont
  {Hausmann}, \citenamefont {Lukin}, \citenamefont {Walsworth}, \citenamefont
  {Loncar},\ and\ \citenamefont {Yacoby}}]{Maletinsky2012}%
  \BibitemOpen
  \bibfield  {author} {\bibinfo {author} {\bibfnamefont {P.}~\bibnamefont
  {Maletinsky}}, \bibinfo {author} {\bibfnamefont {S.}~\bibnamefont {Hong}},
  \bibinfo {author} {\bibfnamefont {M.~S.}\ \bibnamefont {Grinolds}}, \bibinfo
  {author} {\bibfnamefont {B.}~\bibnamefont {Hausmann}}, \bibinfo {author}
  {\bibfnamefont {M.~D.}\ \bibnamefont {Lukin}}, \bibinfo {author}
  {\bibfnamefont {R.~L.}\ \bibnamefont {Walsworth}}, \bibinfo {author}
  {\bibfnamefont {M.}~\bibnamefont {Loncar}}, \ and\ \bibinfo {author}
  {\bibfnamefont {A.}~\bibnamefont {Yacoby}},\ }\href {\doibase
  10.1038/NNANO.2012.50} {\bibfield  {journal} {\bibinfo  {journal} {Nature
  Nanotechnology}\ }\textbf {\bibinfo {volume} {7}},\ \bibinfo {pages} {320}
  (\bibinfo {year} {2012})}\BibitemShut {NoStop}%
\bibitem [{\citenamefont {Rondin}\ \emph {et~al.}(2012)\citenamefont {Rondin},
  \citenamefont {Tetienne}, \citenamefont {Spinicelli}, \citenamefont
  {Dal~Savio}, \citenamefont {Karrai}, \citenamefont {Dantelle}, \citenamefont
  {Thiaville}, \citenamefont {Rohart}, \citenamefont {Roch},\ and\
  \citenamefont {Jacques}}]{Rondin2012}%
  \BibitemOpen
  \bibfield  {author} {\bibinfo {author} {\bibfnamefont {L.}~\bibnamefont
  {Rondin}}, \bibinfo {author} {\bibfnamefont {J.-P.}\ \bibnamefont
  {Tetienne}}, \bibinfo {author} {\bibfnamefont {P.}~\bibnamefont
  {Spinicelli}}, \bibinfo {author} {\bibfnamefont {C.}~\bibnamefont
  {Dal~Savio}}, \bibinfo {author} {\bibfnamefont {K.}~\bibnamefont {Karrai}},
  \bibinfo {author} {\bibfnamefont {G.}~\bibnamefont {Dantelle}}, \bibinfo
  {author} {\bibfnamefont {A.}~\bibnamefont {Thiaville}}, \bibinfo {author}
  {\bibfnamefont {S.}~\bibnamefont {Rohart}}, \bibinfo {author} {\bibfnamefont
  {J.-F.}\ \bibnamefont {Roch}}, \ and\ \bibinfo {author} {\bibfnamefont
  {V.}~\bibnamefont {Jacques}},\ }\href {\doibase 10.1063/1.3703128} {\bibfield
   {journal} {\bibinfo  {journal} {Appl. Phys. Lett.}\ }\textbf {\bibinfo
  {volume} {100}},\ \bibinfo {pages} {153118} (\bibinfo {year}
  {2012})}\BibitemShut {NoStop}%
\bibitem [{\citenamefont {McVitie}\ \emph {et~al.}(2015)\citenamefont
  {McVitie}, \citenamefont {McGrouther}, \citenamefont {McFadzean},
  \citenamefont {MacLaren}, \citenamefont {O’Shea},\ and\ \citenamefont
  {Benitez}}]{McVitie2015}%
  \BibitemOpen
  \bibfield  {author} {\bibinfo {author} {\bibfnamefont {S.}~\bibnamefont
  {McVitie}}, \bibinfo {author} {\bibfnamefont {D.}~\bibnamefont {McGrouther}},
  \bibinfo {author} {\bibfnamefont {S.}~\bibnamefont {McFadzean}}, \bibinfo
  {author} {\bibfnamefont {D.}~\bibnamefont {MacLaren}}, \bibinfo {author}
  {\bibfnamefont {K.}~\bibnamefont {O’Shea}}, \ and\ \bibinfo {author}
  {\bibfnamefont {M.}~\bibnamefont {Benitez}},\ }\href {\doibase
  https://doi.org/10.1016/j.ultramic.2015.01.003} {\bibfield  {journal}
  {\bibinfo  {journal} {Ultramicroscopy}\ }\textbf {\bibinfo {volume} {152}},\
  \bibinfo {pages} {57 } (\bibinfo {year} {2015})}\BibitemShut {NoStop}%
\bibitem [{\citenamefont {Goncalves}\ \emph {et~al.}(2017)\citenamefont
  {Goncalves}, \citenamefont {Paterson}, \citenamefont {McGrouther},
  \citenamefont {Drysdale}, \citenamefont {Togawa}, \citenamefont {Schmool},\
  and\ \citenamefont {Stamps}}]{Goncalves2017}%
  \BibitemOpen
  \bibfield  {author} {\bibinfo {author} {\bibfnamefont {F.~J.~T.}\
  \bibnamefont {Goncalves}}, \bibinfo {author} {\bibfnamefont {G.~W.}\
  \bibnamefont {Paterson}}, \bibinfo {author} {\bibfnamefont {D.}~\bibnamefont
  {McGrouther}}, \bibinfo {author} {\bibfnamefont {T.}~\bibnamefont
  {Drysdale}}, \bibinfo {author} {\bibfnamefont {Y.}~\bibnamefont {Togawa}},
  \bibinfo {author} {\bibfnamefont {D.~S.}\ \bibnamefont {Schmool}}, \ and\
  \bibinfo {author} {\bibfnamefont {R.~L.}\ \bibnamefont {Stamps}},\ }\href
  {\doibase 10.1038/s41598-017-11009-2} {\bibfield  {journal} {\bibinfo
  {journal} {Sci. Rep.}\ }\textbf {\bibinfo {volume} {7}},\ \bibinfo {pages}
  {11064} (\bibinfo {year} {2017})}\BibitemShut {NoStop}%
\bibitem [{\citenamefont {Gilbert}(2004)}]{Gilbert2004}%
  \BibitemOpen
  \bibfield  {author} {\bibinfo {author} {\bibfnamefont {T.~L.}\ \bibnamefont
  {Gilbert}},\ }\href@noop {} {\bibfield  {journal} {\bibinfo  {journal}
  {Magnetics, IEEE Transactions on}\ }\textbf {\bibinfo {volume} {40}},\
  \bibinfo {pages} {3443 } (\bibinfo {year} {2004})}\BibitemShut {NoStop}%
\bibitem [{\citenamefont {Landau}\ and\ \citenamefont
  {Lifshitz}(1953)}]{Landau1953}%
  \BibitemOpen
  \bibfield  {author} {\bibinfo {author} {\bibfnamefont {L.~D.}\ \bibnamefont
  {Landau}}\ and\ \bibinfo {author} {\bibfnamefont {E.}~\bibnamefont
  {Lifshitz}},\ }\href@noop {} {\bibfield  {journal} {\bibinfo  {journal}
  {Phys. Z. Sowjet.}\ }\textbf {\bibinfo {volume} {8}},\ \bibinfo {pages} {153}
  (\bibinfo {year} {1953})}\BibitemShut {NoStop}%
\bibitem [{\citenamefont {St\"{o}hr}\ and\ \citenamefont
  {Siegmann}(2006)}]{Stohr2006}%
  \BibitemOpen
  \bibfield  {author} {\bibinfo {author} {\bibfnamefont {J.}~\bibnamefont
  {St\"{o}hr}}\ and\ \bibinfo {author} {\bibfnamefont {H.~C.}\ \bibnamefont
  {Siegmann}},\ }\href@noop {} {\emph {\bibinfo {title} {Magnetism: from
  fundamentals to nanoscale dynamics}}}\ (\bibinfo  {publisher} {Springer},\
  \bibinfo {year} {2006})\BibitemShut {NoStop}%
\bibitem [{\citenamefont {Garc\'{i}a-Cervera}(1999)}]{GarciaCervera1999}%
  \BibitemOpen
  \bibfield  {author} {\bibinfo {author} {\bibfnamefont {C.~J.}\ \bibnamefont
  {Garc\'{i}a-Cervera}},\ }\emph {\bibinfo {title} {Magnetic domains and
  magnetic domain walls}},\ \href@noop {} {Ph.D. thesis},\ \bibinfo  {school}
  {New York University}, \bibinfo {address} {NewYork} (\bibinfo {year}
  {1999})\BibitemShut {NoStop}%
\bibitem [{\citenamefont {Stancil}\ and\ \citenamefont
  {Prabhakar}(2009)}]{Stancil2009}%
  \BibitemOpen
  \bibfield  {author} {\bibinfo {author} {\bibfnamefont {D.}~\bibnamefont
  {Stancil}}\ and\ \bibinfo {author} {\bibfnamefont {A.}~\bibnamefont
  {Prabhakar}},\ }\href@noop {} {\emph {\bibinfo {title} {Spin waves: Theory
  and applications}}}\ (\bibinfo  {publisher} {Springer},\ \bibinfo {year}
  {2009})\BibitemShut {NoStop}%
\bibitem [{\citenamefont {Brown}(1963)}]{Brown1963b}%
  \BibitemOpen
  \bibfield  {author} {\bibinfo {author} {\bibfnamefont {W.~F.}\ \bibnamefont
  {Brown}},\ }\href@noop {} {\emph {\bibinfo {title} {Micromagnetics}}}\
  (\bibinfo  {publisher} {Interscience publishers},\ \bibinfo {year}
  {1963})\BibitemShut {NoStop}%
\bibitem [{\citenamefont {Abert}(2019)}]{Abert2019}%
  \BibitemOpen
  \bibfield  {author} {\bibinfo {author} {\bibfnamefont {C.}~\bibnamefont
  {Abert}},\ }\href@noop {} {\bibfield  {journal} {\bibinfo  {journal} {The
  European Physical Journal B}\ }\textbf {\bibinfo {volume} {92}},\ \bibinfo
  {pages} {120} (\bibinfo {year} {2019})}\BibitemShut {NoStop}%
\bibitem [{\citenamefont {Gliga}\ \emph {et~al.}(2015)\citenamefont {Gliga},
  \citenamefont {K\'akay}, \citenamefont {Heyderman}, \citenamefont {Hertel},\
  and\ \citenamefont {Heinonen}}]{Gliga2015}%
  \BibitemOpen
  \bibfield  {author} {\bibinfo {author} {\bibfnamefont {S.}~\bibnamefont
  {Gliga}}, \bibinfo {author} {\bibfnamefont {A.}~\bibnamefont {K\'akay}},
  \bibinfo {author} {\bibfnamefont {L.~J.}\ \bibnamefont {Heyderman}}, \bibinfo
  {author} {\bibfnamefont {R.}~\bibnamefont {Hertel}}, \ and\ \bibinfo {author}
  {\bibfnamefont {O.~G.}\ \bibnamefont {Heinonen}},\ }\href@noop {} {\bibfield
  {journal} {\bibinfo  {journal} {Phys. Rev. B}\ }\textbf {\bibinfo {volume}
  {92}},\ \bibinfo {pages} {060413} (\bibinfo {year} {2015})}\BibitemShut
  {NoStop}%
\bibitem [{\citenamefont {Jungfleisch}\ \emph {et~al.}(2016)\citenamefont
  {Jungfleisch}, \citenamefont {Zhang}, \citenamefont {Iacocca}, \citenamefont
  {Sklenar}, \citenamefont {Ding}, \citenamefont {Jiang}, \citenamefont
  {Zhang}, \citenamefont {Pearson}, \citenamefont {Novosad}, \citenamefont
  {Ketterson}, \citenamefont {Heinonen},\ and\ \citenamefont
  {Hoffmann}}]{Jungfleisch2016}%
  \BibitemOpen
  \bibfield  {author} {\bibinfo {author} {\bibfnamefont {M.~B.}\ \bibnamefont
  {Jungfleisch}}, \bibinfo {author} {\bibfnamefont {W.}~\bibnamefont {Zhang}},
  \bibinfo {author} {\bibfnamefont {E.}~\bibnamefont {Iacocca}}, \bibinfo
  {author} {\bibfnamefont {J.}~\bibnamefont {Sklenar}}, \bibinfo {author}
  {\bibfnamefont {J.}~\bibnamefont {Ding}}, \bibinfo {author} {\bibfnamefont
  {W.}~\bibnamefont {Jiang}}, \bibinfo {author} {\bibfnamefont
  {S.}~\bibnamefont {Zhang}}, \bibinfo {author} {\bibfnamefont {J.~E.}\
  \bibnamefont {Pearson}}, \bibinfo {author} {\bibfnamefont {V.}~\bibnamefont
  {Novosad}}, \bibinfo {author} {\bibfnamefont {J.~B.}\ \bibnamefont
  {Ketterson}}, \bibinfo {author} {\bibfnamefont {O.}~\bibnamefont {Heinonen}},
  \ and\ \bibinfo {author} {\bibfnamefont {A.}~\bibnamefont {Hoffmann}},\
  }\href {\doibase 10.1103/PhysRevB.93.100401} {\bibfield  {journal} {\bibinfo
  {journal} {Phys. Rev. B}\ }\textbf {\bibinfo {volume} {93}},\ \bibinfo
  {pages} {100401} (\bibinfo {year} {2016})}\BibitemShut {NoStop}%
\bibitem [{\citenamefont {Iacocca}\ and\ \citenamefont
  {Heinonen}(2017)}]{Iacocca2017c}%
  \BibitemOpen
  \bibfield  {author} {\bibinfo {author} {\bibfnamefont {E.}~\bibnamefont
  {Iacocca}}\ and\ \bibinfo {author} {\bibfnamefont {O.}~\bibnamefont
  {Heinonen}},\ }\href@noop {} {\bibfield  {journal} {\bibinfo  {journal}
  {Phys. Rev. Applied}\ }\textbf {\bibinfo {volume} {8}},\ \bibinfo {pages}
  {034015} (\bibinfo {year} {2017})}\BibitemShut {NoStop}%
\bibitem [{\citenamefont {Iacocca}\ \emph {et~al.}(2016)\citenamefont
  {Iacocca}, \citenamefont {Gliga}, \citenamefont {Stamps},\ and\ \citenamefont
  {Heinonen}}]{Iacocca2016}%
  \BibitemOpen
  \bibfield  {author} {\bibinfo {author} {\bibfnamefont {E.}~\bibnamefont
  {Iacocca}}, \bibinfo {author} {\bibfnamefont {S.}~\bibnamefont {Gliga}},
  \bibinfo {author} {\bibfnamefont {R.~L.}\ \bibnamefont {Stamps}}, \ and\
  \bibinfo {author} {\bibfnamefont {O.}~\bibnamefont {Heinonen}},\ }\href
  {\doibase 10.1103/PhysRevB.93.134420} {\bibfield  {journal} {\bibinfo
  {journal} {Phys. Rev. B}\ }\textbf {\bibinfo {volume} {93}},\ \bibinfo
  {pages} {134420} (\bibinfo {year} {2016})}\BibitemShut {NoStop}%
\bibitem [{\citenamefont {Iacocca}\ \emph {et~al.}(2019)\citenamefont
  {Iacocca}, \citenamefont {Gliga},\ and\ \citenamefont
  {Heinonen}}]{Iacocca2019}%
  \BibitemOpen
  \bibfield  {author} {\bibinfo {author} {\bibfnamefont {E.}~\bibnamefont
  {Iacocca}}, \bibinfo {author} {\bibfnamefont {S.}~\bibnamefont {Gliga}}, \
  and\ \bibinfo {author} {\bibfnamefont {O.}~\bibnamefont {Heinonen}},\
  }\href@noop {} {\ ,\ \bibinfo {pages} {arXiv:1911.05354} (\bibinfo {year}
  {2019})}\BibitemShut {NoStop}%
\bibitem [{\citenamefont {Dion}\ \emph {et~al.}(2019)\citenamefont {Dion},
  \citenamefont {Arroo}, \citenamefont {Yamanoi}, \citenamefont {Kimura},
  \citenamefont {Gartside}, \citenamefont {Cohen}, \citenamefont
  {Kurebayashi},\ and\ \citenamefont {Branford}}]{Dion2019}%
  \BibitemOpen
  \bibfield  {author} {\bibinfo {author} {\bibfnamefont {T.}~\bibnamefont
  {Dion}}, \bibinfo {author} {\bibfnamefont {D.~M.}\ \bibnamefont {Arroo}},
  \bibinfo {author} {\bibfnamefont {K.}~\bibnamefont {Yamanoi}}, \bibinfo
  {author} {\bibfnamefont {T.}~\bibnamefont {Kimura}}, \bibinfo {author}
  {\bibfnamefont {J.~C.}\ \bibnamefont {Gartside}}, \bibinfo {author}
  {\bibfnamefont {L.~F.}\ \bibnamefont {Cohen}}, \bibinfo {author}
  {\bibfnamefont {H.}~\bibnamefont {Kurebayashi}}, \ and\ \bibinfo {author}
  {\bibfnamefont {W.~R.}\ \bibnamefont {Branford}},\ }\href@noop {} {\bibfield
  {journal} {\bibinfo  {journal} {Phys. Rev. B}\ }\textbf {\bibinfo {volume}
  {100}},\ \bibinfo {pages} {054433} (\bibinfo {year} {2019})}\BibitemShut
  {NoStop}%
\bibitem [{\citenamefont {Mengotti}\ \emph {et~al.}(2009)\citenamefont
  {Mengotti}, \citenamefont {Heyderman}, \citenamefont {Bisig}, \citenamefont
  {Fraile~RodrÃ­guez}, \citenamefont {Le~Guyader}, \citenamefont {Nolting},\
  and\ \citenamefont {Braun}}]{Mengotti2009}%
  \BibitemOpen
  \bibfield  {author} {\bibinfo {author} {\bibfnamefont {E.}~\bibnamefont
  {Mengotti}}, \bibinfo {author} {\bibfnamefont {L.~J.}\ \bibnamefont
  {Heyderman}}, \bibinfo {author} {\bibfnamefont {A.}~\bibnamefont {Bisig}},
  \bibinfo {author} {\bibfnamefont {A.}~\bibnamefont {Fraile~RodrÃ­guez}},
  \bibinfo {author} {\bibfnamefont {L.}~\bibnamefont {Le~Guyader}}, \bibinfo
  {author} {\bibfnamefont {F.}~\bibnamefont {Nolting}}, \ and\ \bibinfo
  {author} {\bibfnamefont {H.~B.}\ \bibnamefont {Braun}},\ }\href@noop {}
  {\bibfield  {journal} {\bibinfo  {journal} {Journal of Applied Physics}\
  }\textbf {\bibinfo {volume} {105}},\ \bibinfo {pages} {113113} (\bibinfo
  {year} {2009})}\BibitemShut {NoStop}%
\bibitem [{\citenamefont {Grimsditch}\ \emph {et~al.}(2004)\citenamefont
  {Grimsditch}, \citenamefont {Giovannini}, \citenamefont {Montoncello},
  \citenamefont {Nizzoli}, \citenamefont {Leaf},\ and\ \citenamefont
  {Kaper}}]{Grimsditch2004}%
  \BibitemOpen
  \bibfield  {author} {\bibinfo {author} {\bibfnamefont {M.}~\bibnamefont
  {Grimsditch}}, \bibinfo {author} {\bibfnamefont {L.}~\bibnamefont
  {Giovannini}}, \bibinfo {author} {\bibfnamefont {F.}~\bibnamefont
  {Montoncello}}, \bibinfo {author} {\bibfnamefont {F.}~\bibnamefont
  {Nizzoli}}, \bibinfo {author} {\bibfnamefont {G.~K.}\ \bibnamefont {Leaf}}, \
  and\ \bibinfo {author} {\bibfnamefont {H.~G.}\ \bibnamefont {Kaper}},\
  }\href@noop {} {\bibfield  {journal} {\bibinfo  {journal} {Phys. Rev. B}\
  }\textbf {\bibinfo {volume} {70}},\ \bibinfo {pages} {054409} (\bibinfo
  {year} {2004})}\BibitemShut {NoStop}%
\bibitem [{\citenamefont {Dvornik}\ \emph {et~al.}(2011)\citenamefont
  {Dvornik}, \citenamefont {Bondarenko}, \citenamefont {Ivanov},\ and\
  \citenamefont {Kruglyak}}]{Dvornik2011}%
  \BibitemOpen
  \bibfield  {author} {\bibinfo {author} {\bibfnamefont {M.}~\bibnamefont
  {Dvornik}}, \bibinfo {author} {\bibfnamefont {P.~V.}\ \bibnamefont
  {Bondarenko}}, \bibinfo {author} {\bibfnamefont {B.~A.}\ \bibnamefont
  {Ivanov}}, \ and\ \bibinfo {author} {\bibfnamefont {V.~V.}\ \bibnamefont
  {Kruglyak}},\ }\href@noop {} {\bibfield  {journal} {\bibinfo  {journal}
  {Journal of Applied Physics}\ }\textbf {\bibinfo {volume} {109}},\ \bibinfo
  {pages} {07B912} (\bibinfo {year} {2011})}\BibitemShut {NoStop}%
\bibitem [{\citenamefont {Bang}\ \emph {et~al.}(2019)\citenamefont {Bang},
  \citenamefont {Montoncello}, \citenamefont {Jungfleisch}, \citenamefont
  {Hoffmann}, \citenamefont {Giovannini},\ and\ \citenamefont
  {Ketterson}}]{Bang2019}%
  \BibitemOpen
  \bibfield  {author} {\bibinfo {author} {\bibfnamefont {W.}~\bibnamefont
  {Bang}}, \bibinfo {author} {\bibfnamefont {F.}~\bibnamefont {Montoncello}},
  \bibinfo {author} {\bibfnamefont {M.~B.}\ \bibnamefont {Jungfleisch}},
  \bibinfo {author} {\bibfnamefont {A.}~\bibnamefont {Hoffmann}}, \bibinfo
  {author} {\bibfnamefont {L.}~\bibnamefont {Giovannini}}, \ and\ \bibinfo
  {author} {\bibfnamefont {J.~B.}\ \bibnamefont {Ketterson}},\ }\href@noop {}
  {\bibfield  {journal} {\bibinfo  {journal} {Phys. Rev. B}\ }\textbf {\bibinfo
  {volume} {99}},\ \bibinfo {pages} {014415} (\bibinfo {year}
  {2019})}\BibitemShut {NoStop}%
\bibitem [{\citenamefont {Paterson}\ \emph {et~al.}(2019)\citenamefont
  {Paterson}, \citenamefont {Macauley}, \citenamefont {Li}, \citenamefont
  {Mac\^edo}, \citenamefont {Ferguson}, \citenamefont {Morley}, \citenamefont
  {Rosamond}, \citenamefont {Linfield}, \citenamefont {Marrows}, \citenamefont
  {Stamps},\ and\ \citenamefont {McVitie}}]{Paterson2019}%
  \BibitemOpen
  \bibfield  {author} {\bibinfo {author} {\bibfnamefont {G.~W.}\ \bibnamefont
  {Paterson}}, \bibinfo {author} {\bibfnamefont {G.~M.}\ \bibnamefont
  {Macauley}}, \bibinfo {author} {\bibfnamefont {Y.}~\bibnamefont {Li}},
  \bibinfo {author} {\bibfnamefont {R.}~\bibnamefont {Mac\^edo}}, \bibinfo
  {author} {\bibfnamefont {C.}~\bibnamefont {Ferguson}}, \bibinfo {author}
  {\bibfnamefont {S.~A.}\ \bibnamefont {Morley}}, \bibinfo {author}
  {\bibfnamefont {M.~C.}\ \bibnamefont {Rosamond}}, \bibinfo {author}
  {\bibfnamefont {E.~H.}\ \bibnamefont {Linfield}}, \bibinfo {author}
  {\bibfnamefont {C.~H.}\ \bibnamefont {Marrows}}, \bibinfo {author}
  {\bibfnamefont {R.~L.}\ \bibnamefont {Stamps}}, \ and\ \bibinfo {author}
  {\bibfnamefont {S.}~\bibnamefont {McVitie}},\ }\href@noop {} {\bibfield
  {journal} {\bibinfo  {journal} {Phys. Rev. B}\ }\textbf {\bibinfo {volume}
  {100}},\ \bibinfo {pages} {174410} (\bibinfo {year} {2019})}\BibitemShut
  {NoStop}%
\bibitem [{\citenamefont {Balkanski}\ and\ \citenamefont
  {Wallis}(2000)}]{Balkanski2000}%
  \BibitemOpen
  \bibfield  {author} {\bibinfo {author} {\bibfnamefont {M.}~\bibnamefont
  {Balkanski}}\ and\ \bibinfo {author} {\bibfnamefont {R.}~\bibnamefont
  {Wallis}},\ }\href@noop {} {\emph {\bibinfo {title} {Semiconductor physics
  and applications}}}\ (\bibinfo  {publisher} {Oxford University Press},\
  \bibinfo {year} {2000})\BibitemShut {NoStop}%
\bibitem [{\citenamefont {Bondarenko}\ \emph {et~al.}(2010)\citenamefont
  {Bondarenko}, \citenamefont {Galkin}, \citenamefont {Ivanov},\ and\
  \citenamefont {Zaspel}}]{Bondarenko2010}%
  \BibitemOpen
  \bibfield  {author} {\bibinfo {author} {\bibfnamefont {P.~V.}\ \bibnamefont
  {Bondarenko}}, \bibinfo {author} {\bibfnamefont {A.~Y.}\ \bibnamefont
  {Galkin}}, \bibinfo {author} {\bibfnamefont {B.~A.}\ \bibnamefont {Ivanov}},
  \ and\ \bibinfo {author} {\bibfnamefont {C.~E.}\ \bibnamefont {Zaspel}},\
  }\href@noop {} {\bibfield  {journal} {\bibinfo  {journal} {Phys. Rev. B}\
  }\textbf {\bibinfo {volume} {81}},\ \bibinfo {pages} {224415} (\bibinfo
  {year} {2010})}\BibitemShut {NoStop}%
\bibitem [{\citenamefont {Verba}\ \emph {et~al.}(2012)\citenamefont {Verba},
  \citenamefont {Melkov}, \citenamefont {Tiberkevich},\ and\ \citenamefont
  {Slavin}}]{Verba2012}%
  \BibitemOpen
  \bibfield  {author} {\bibinfo {author} {\bibfnamefont {R.}~\bibnamefont
  {Verba}}, \bibinfo {author} {\bibfnamefont {G.}~\bibnamefont {Melkov}},
  \bibinfo {author} {\bibfnamefont {V.}~\bibnamefont {Tiberkevich}}, \ and\
  \bibinfo {author} {\bibfnamefont {A.}~\bibnamefont {Slavin}},\ }\href@noop {}
  {\bibfield  {journal} {\bibinfo  {journal} {Phys. Rev. B}\ }\textbf {\bibinfo
  {volume} {85}},\ \bibinfo {pages} {014427} (\bibinfo {year}
  {2012})}\BibitemShut {NoStop}%
\bibitem [{\citenamefont {Lisenkov}\ \emph {et~al.}(2016)\citenamefont
  {Lisenkov}, \citenamefont {Tyberkevych}, \citenamefont {Nikitov},\ and\
  \citenamefont {Slavin}}]{Lisenkov2016}%
  \BibitemOpen
  \bibfield  {author} {\bibinfo {author} {\bibfnamefont {I.}~\bibnamefont
  {Lisenkov}}, \bibinfo {author} {\bibfnamefont {V.}~\bibnamefont
  {Tyberkevych}}, \bibinfo {author} {\bibfnamefont {S.}~\bibnamefont
  {Nikitov}}, \ and\ \bibinfo {author} {\bibfnamefont {A.}~\bibnamefont
  {Slavin}},\ }\href@noop {} {\bibfield  {journal} {\bibinfo  {journal} {Phys.
  Rev. B}\ }\textbf {\bibinfo {volume} {93}},\ \bibinfo {pages} {214441}
  (\bibinfo {year} {2016})}\BibitemShut {NoStop}%
\bibitem [{\citenamefont {Shindou}\ \emph
  {et~al.}(2013{\natexlab{a}})\citenamefont {Shindou}, \citenamefont
  {Matsumoto}, \citenamefont {Murakami},\ and\ \citenamefont
  {Ohe}}]{Shindou2013}%
  \BibitemOpen
  \bibfield  {author} {\bibinfo {author} {\bibfnamefont {R.}~\bibnamefont
  {Shindou}}, \bibinfo {author} {\bibfnamefont {R.}~\bibnamefont {Matsumoto}},
  \bibinfo {author} {\bibfnamefont {S.}~\bibnamefont {Murakami}}, \ and\
  \bibinfo {author} {\bibfnamefont {J.-i.}\ \bibnamefont {Ohe}},\ }\href
  {\doibase 10.1103/PhysRevB.87.174427} {\bibfield  {journal} {\bibinfo
  {journal} {Phys. Rev. B}\ }\textbf {\bibinfo {volume} {87}},\ \bibinfo
  {pages} {174427} (\bibinfo {year} {2013}{\natexlab{a}})}\BibitemShut
  {NoStop}%
\bibitem [{\citenamefont {Shindou}\ \emph
  {et~al.}(2013{\natexlab{b}})\citenamefont {Shindou}, \citenamefont {Ohe},
  \citenamefont {Matsumoto}, \citenamefont {Murakami},\ and\ \citenamefont
  {Saitoh}}]{Shindou2013b}%
  \BibitemOpen
  \bibfield  {author} {\bibinfo {author} {\bibfnamefont {R.}~\bibnamefont
  {Shindou}}, \bibinfo {author} {\bibfnamefont {J.-i.}\ \bibnamefont {Ohe}},
  \bibinfo {author} {\bibfnamefont {R.}~\bibnamefont {Matsumoto}}, \bibinfo
  {author} {\bibfnamefont {S.}~\bibnamefont {Murakami}}, \ and\ \bibinfo
  {author} {\bibfnamefont {E.}~\bibnamefont {Saitoh}},\ }\href {\doibase
  10.1103/PhysRevB.87.174402} {\bibfield  {journal} {\bibinfo  {journal} {Phys.
  Rev. B}\ }\textbf {\bibinfo {volume} {87}},\ \bibinfo {pages} {174402}
  (\bibinfo {year} {2013}{\natexlab{b}})}\BibitemShut {NoStop}%
\bibitem [{\citenamefont {Budrikis}\ \emph {et~al.}(2012)\citenamefont
  {Budrikis}, \citenamefont {Livesey}, \citenamefont {Morgan}, \citenamefont
  {Akerman}, \citenamefont {Stein}, \citenamefont {Langridge}, \citenamefont
  {Marrows},\ and\ \citenamefont {Stamps}}]{Budrikis2012}%
  \BibitemOpen
  \bibfield  {author} {\bibinfo {author} {\bibfnamefont {Z.}~\bibnamefont
  {Budrikis}}, \bibinfo {author} {\bibfnamefont {K.~L.}\ \bibnamefont
  {Livesey}}, \bibinfo {author} {\bibfnamefont {J.~P.}\ \bibnamefont {Morgan}},
  \bibinfo {author} {\bibfnamefont {J.}~\bibnamefont {Akerman}}, \bibinfo
  {author} {\bibfnamefont {A.}~\bibnamefont {Stein}}, \bibinfo {author}
  {\bibfnamefont {S.}~\bibnamefont {Langridge}}, \bibinfo {author}
  {\bibfnamefont {C.~H.}\ \bibnamefont {Marrows}}, \ and\ \bibinfo {author}
  {\bibfnamefont {R.~L.}\ \bibnamefont {Stamps}},\ }\href {\doibase
  10.1088/1367-2630/14/3/035014} {\bibfield  {journal} {\bibinfo  {journal}
  {New Journal of Physics}\ }\textbf {\bibinfo {volume} {14}},\ \bibinfo
  {pages} {035014} (\bibinfo {year} {2012})}\BibitemShut {NoStop}%
\bibitem [{\citenamefont {Westphalen}\ \emph {et~al.}(2008)\citenamefont
  {Westphalen}, \citenamefont {Schumann}, \citenamefont {Remhof}, \citenamefont
  {Zabel}, \citenamefont {Karolak}, \citenamefont {Baxevanis}, \citenamefont
  {Vedmedenko}, \citenamefont {Last}, \citenamefont {Kunze},\ and\
  \citenamefont {Eim\"uller}}]{Westphalen2008}%
  \BibitemOpen
  \bibfield  {author} {\bibinfo {author} {\bibfnamefont {A.}~\bibnamefont
  {Westphalen}}, \bibinfo {author} {\bibfnamefont {A.}~\bibnamefont
  {Schumann}}, \bibinfo {author} {\bibfnamefont {A.}~\bibnamefont {Remhof}},
  \bibinfo {author} {\bibfnamefont {H.}~\bibnamefont {Zabel}}, \bibinfo
  {author} {\bibfnamefont {M.}~\bibnamefont {Karolak}}, \bibinfo {author}
  {\bibfnamefont {B.}~\bibnamefont {Baxevanis}}, \bibinfo {author}
  {\bibfnamefont {E.~Y.}\ \bibnamefont {Vedmedenko}}, \bibinfo {author}
  {\bibfnamefont {T.}~\bibnamefont {Last}}, \bibinfo {author} {\bibfnamefont
  {U.}~\bibnamefont {Kunze}}, \ and\ \bibinfo {author} {\bibfnamefont
  {T.}~\bibnamefont {Eim\"uller}},\ }\href {\doibase
  10.1103/PhysRevB.77.174407} {\bibfield  {journal} {\bibinfo  {journal} {Phys.
  Rev. B}\ }\textbf {\bibinfo {volume} {77}},\ \bibinfo {pages} {174407}
  (\bibinfo {year} {2008})}\BibitemShut {NoStop}%
\bibitem [{\citenamefont {Ghosh}\ \emph {et~al.}(2019)\citenamefont {Ghosh},
  \citenamefont {Ma}, \citenamefont {Lourembam}, \citenamefont {Xiangjun},
  \citenamefont {Maddu}, \citenamefont {Yap},\ and\ \citenamefont
  {Lim}}]{Ghosh2019}%
  \BibitemOpen
  \bibfield  {author} {\bibinfo {author} {\bibfnamefont {A.}~\bibnamefont
  {Ghosh}}, \bibinfo {author} {\bibfnamefont {F.}~\bibnamefont {Ma}}, \bibinfo
  {author} {\bibfnamefont {J.}~\bibnamefont {Lourembam}}, \bibinfo {author}
  {\bibfnamefont {J.}~\bibnamefont {Xiangjun}}, \bibinfo {author}
  {\bibfnamefont {R.}~\bibnamefont {Maddu}}, \bibinfo {author} {\bibfnamefont
  {Q.~J.}\ \bibnamefont {Yap}}, \ and\ \bibinfo {author} {\bibfnamefont
  {S.~T.}\ \bibnamefont {Lim}},\ }\href@noop {} {\bibfield  {journal} {\bibinfo
   {journal} {Nano letters}\ } (\bibinfo {year} {2019})}\BibitemShut {NoStop}%
\bibitem [{\citenamefont {Li}\ \emph {et~al.}(2016{\natexlab{b}})\citenamefont
  {Li}, \citenamefont {Gubbiotti}, \citenamefont {Casoli}, \citenamefont
  {Gon{\c{c}}alves}, \citenamefont {Morley}, \citenamefont {Rosamond},
  \citenamefont {Linfield}, \citenamefont {Marrows}, \citenamefont {McVitie},\
  and\ \citenamefont {Stamps}}]{Li2016}%
  \BibitemOpen
  \bibfield  {author} {\bibinfo {author} {\bibfnamefont {Y.}~\bibnamefont
  {Li}}, \bibinfo {author} {\bibfnamefont {G.}~\bibnamefont {Gubbiotti}},
  \bibinfo {author} {\bibfnamefont {F.}~\bibnamefont {Casoli}}, \bibinfo
  {author} {\bibfnamefont {F.~J.~T.}\ \bibnamefont {Gon{\c{c}}alves}}, \bibinfo
  {author} {\bibfnamefont {S.~A.}\ \bibnamefont {Morley}}, \bibinfo {author}
  {\bibfnamefont {M.~C.}\ \bibnamefont {Rosamond}}, \bibinfo {author}
  {\bibfnamefont {E.~H.}\ \bibnamefont {Linfield}}, \bibinfo {author}
  {\bibfnamefont {C.~H.}\ \bibnamefont {Marrows}}, \bibinfo {author}
  {\bibfnamefont {S.}~\bibnamefont {McVitie}}, \ and\ \bibinfo {author}
  {\bibfnamefont {R.~L.}\ \bibnamefont {Stamps}},\ }\href {\doibase
  10.1088/1361-6463/50/1/015003} {\bibfield  {journal} {\bibinfo  {journal}
  {Journal of Physics D: Applied Physics}\ }\textbf {\bibinfo {volume} {50}},\
  \bibinfo {pages} {015003} (\bibinfo {year} {2016}{\natexlab{b}})}\BibitemShut
  {NoStop}%
\bibitem [{\citenamefont {Mamica}\ \emph {et~al.}(2018)\citenamefont {Mamica},
  \citenamefont {Zhou}, \citenamefont {Adeyeye}, \citenamefont {Krawczyk},\
  and\ \citenamefont {Gubbiotti}}]{Mamica2018}%
  \BibitemOpen
  \bibfield  {author} {\bibinfo {author} {\bibfnamefont {S.}~\bibnamefont
  {Mamica}}, \bibinfo {author} {\bibfnamefont {X.}~\bibnamefont {Zhou}},
  \bibinfo {author} {\bibfnamefont {A.}~\bibnamefont {Adeyeye}}, \bibinfo
  {author} {\bibfnamefont {M.}~\bibnamefont {Krawczyk}}, \ and\ \bibinfo
  {author} {\bibfnamefont {G.}~\bibnamefont {Gubbiotti}},\ }\href {\doibase
  10.1103/PhysRevB.98.054405} {\bibfield  {journal} {\bibinfo  {journal} {Phys.
  Rev. B}\ }\textbf {\bibinfo {volume} {98}},\ \bibinfo {pages} {054405}
  (\bibinfo {year} {2018})}\BibitemShut {NoStop}%
\bibitem [{\citenamefont {Rougemaille}\ \emph {et~al.}(2013)\citenamefont
  {Rougemaille}, \citenamefont {Montaigne}, \citenamefont {Canals},
  \citenamefont {Hehn}, \citenamefont {Riahi}, \citenamefont {Lacour},\ and\
  \citenamefont {Toussaint}}]{Rougemaille2013}%
  \BibitemOpen
  \bibfield  {author} {\bibinfo {author} {\bibfnamefont {N.}~\bibnamefont
  {Rougemaille}}, \bibinfo {author} {\bibfnamefont {F.}~\bibnamefont
  {Montaigne}}, \bibinfo {author} {\bibfnamefont {B.}~\bibnamefont {Canals}},
  \bibinfo {author} {\bibfnamefont {M.}~\bibnamefont {Hehn}}, \bibinfo {author}
  {\bibfnamefont {H.}~\bibnamefont {Riahi}}, \bibinfo {author} {\bibfnamefont
  {D.}~\bibnamefont {Lacour}}, \ and\ \bibinfo {author} {\bibfnamefont {J.-C.}\
  \bibnamefont {Toussaint}},\ }\href@noop {} {\bibfield  {journal} {\bibinfo
  {journal} {New Journal of Physics}\ }\textbf {\bibinfo {volume} {15}},\
  \bibinfo {pages} {035026} (\bibinfo {year} {2013})}\BibitemShut {NoStop}%
\bibitem [{\citenamefont {Arroo}\ \emph {et~al.}(2018)\citenamefont {Arroo},
  \citenamefont {Gartside},\ and\ \citenamefont
  {Branford}}]{Arroo2018sculpting}%
  \BibitemOpen
  \bibfield  {author} {\bibinfo {author} {\bibfnamefont {D.~M.}\ \bibnamefont
  {Arroo}}, \bibinfo {author} {\bibfnamefont {J.~C.}\ \bibnamefont {Gartside}},
  \ and\ \bibinfo {author} {\bibfnamefont {W.~R.}\ \bibnamefont {Branford}},\
  }\href@noop {} {\enquote {\bibinfo {title} {Sculpting the spin-wave response
  of artificial spin ice via microstate selection},}\ } (\bibinfo {year}
  {2018}),\ \Eprint {http://arxiv.org/abs/1805.01397} {arXiv:1805.01397
  [cond-mat.mes-hall]} \BibitemShut {NoStop}%
\bibitem [{\citenamefont {Bhat}\ \emph {et~al.}(2019)\citenamefont {Bhat},
  \citenamefont {Watanabe}, \citenamefont {Baumgaertl},\ and\ \citenamefont
  {Grundler}}]{Bhat2019}%
  \BibitemOpen
  \bibfield  {author} {\bibinfo {author} {\bibfnamefont {V.}~\bibnamefont
  {Bhat}}, \bibinfo {author} {\bibfnamefont {S.}~\bibnamefont {Watanabe}},
  \bibinfo {author} {\bibfnamefont {K.}~\bibnamefont {Baumgaertl}}, \ and\
  \bibinfo {author} {\bibfnamefont {D.}~\bibnamefont {Grundler}},\ }\href@noop
  {} {\bibfield  {journal} {\bibinfo  {journal} {arXiv preprint
  arXiv:1910.00874}\ } (\bibinfo {year} {2019})}\BibitemShut {NoStop}%
\bibitem [{\citenamefont {Sklenar}\ \emph {et~al.}(2013)\citenamefont
  {Sklenar}, \citenamefont {Bhat}, \citenamefont {DeLong}, \citenamefont
  {Heinonen},\ and\ \citenamefont {Ketterson}}]{Sklenar2013}%
  \BibitemOpen
  \bibfield  {author} {\bibinfo {author} {\bibfnamefont {J.}~\bibnamefont
  {Sklenar}}, \bibinfo {author} {\bibfnamefont {V.~S.}\ \bibnamefont {Bhat}},
  \bibinfo {author} {\bibfnamefont {L.~E.}\ \bibnamefont {DeLong}}, \bibinfo
  {author} {\bibfnamefont {O.}~\bibnamefont {Heinonen}}, \ and\ \bibinfo
  {author} {\bibfnamefont {J.~B.}\ \bibnamefont {Ketterson}},\ }\href@noop {}
  {\bibfield  {journal} {\bibinfo  {journal} {Applied Physics Letters}\
  }\textbf {\bibinfo {volume} {102}} (\bibinfo {year} {2013})}\BibitemShut
  {NoStop}%
\bibitem [{\citenamefont {Fert}\ and\ \citenamefont
  {Levy}(1980)}]{fert1980role}%
  \BibitemOpen
  \bibfield  {author} {\bibinfo {author} {\bibfnamefont {A.}~\bibnamefont
  {Fert}}\ and\ \bibinfo {author} {\bibfnamefont {P.~M.}\ \bibnamefont
  {Levy}},\ }\href@noop {} {\bibfield  {journal} {\bibinfo  {journal} {Physical
  Review Letters}\ }\textbf {\bibinfo {volume} {44}},\ \bibinfo {pages} {1538}
  (\bibinfo {year} {1980})}\BibitemShut {NoStop}%
\bibitem [{\citenamefont {Fert}(1990)}]{fert1990magnetic}%
  \BibitemOpen
  \bibfield  {author} {\bibinfo {author} {\bibfnamefont {A.}~\bibnamefont
  {Fert}},\ }in\ \href@noop {} {\emph {\bibinfo {booktitle} {Materials Science
  Forum}}},\ Vol.~\bibinfo {volume} {59}\ (\bibinfo {organization} {Trans Tech
  Publ},\ \bibinfo {year} {1990})\ pp.\ \bibinfo {pages} {439--480}\BibitemShut
  {NoStop}%
\bibitem [{\citenamefont {Thiaville}\ \emph {et~al.}(2012)\citenamefont
  {Thiaville}, \citenamefont {Rohart}, \citenamefont {Ju{\'e}}, \citenamefont
  {Cros},\ and\ \citenamefont {Fert}}]{thiaville2012dynamics}%
  \BibitemOpen
  \bibfield  {author} {\bibinfo {author} {\bibfnamefont {A.}~\bibnamefont
  {Thiaville}}, \bibinfo {author} {\bibfnamefont {S.}~\bibnamefont {Rohart}},
  \bibinfo {author} {\bibfnamefont {{\'E}.}~\bibnamefont {Ju{\'e}}}, \bibinfo
  {author} {\bibfnamefont {V.}~\bibnamefont {Cros}}, \ and\ \bibinfo {author}
  {\bibfnamefont {A.}~\bibnamefont {Fert}},\ }\href@noop {} {\bibfield
  {journal} {\bibinfo  {journal} {EPL (Europhysics Letters)}\ }\textbf
  {\bibinfo {volume} {100}},\ \bibinfo {pages} {57002} (\bibinfo {year}
  {2012})}\BibitemShut {NoStop}%
\bibitem [{\citenamefont {Yanes}\ \emph {et~al.}(2013)\citenamefont {Yanes},
  \citenamefont {Jackson}, \citenamefont {Udvardi}, \citenamefont {Szunyogh},\
  and\ \citenamefont {Nowak}}]{yanes2013exchange}%
  \BibitemOpen
  \bibfield  {author} {\bibinfo {author} {\bibfnamefont {R.}~\bibnamefont
  {Yanes}}, \bibinfo {author} {\bibfnamefont {J.}~\bibnamefont {Jackson}},
  \bibinfo {author} {\bibfnamefont {L.}~\bibnamefont {Udvardi}}, \bibinfo
  {author} {\bibfnamefont {L.}~\bibnamefont {Szunyogh}}, \ and\ \bibinfo
  {author} {\bibfnamefont {U.}~\bibnamefont {Nowak}},\ }\href@noop {}
  {\bibfield  {journal} {\bibinfo  {journal} {Physical review letters}\
  }\textbf {\bibinfo {volume} {111}},\ \bibinfo {pages} {217202} (\bibinfo
  {year} {2013})}\BibitemShut {NoStop}%
\bibitem [{\citenamefont {Heide}\ \emph {et~al.}(2008)\citenamefont {Heide},
  \citenamefont {Bihlmayer},\ and\ \citenamefont
  {Bl{\"u}gel}}]{heide2008dzyaloshinskii}%
  \BibitemOpen
  \bibfield  {author} {\bibinfo {author} {\bibfnamefont {M.}~\bibnamefont
  {Heide}}, \bibinfo {author} {\bibfnamefont {G.}~\bibnamefont {Bihlmayer}}, \
  and\ \bibinfo {author} {\bibfnamefont {S.}~\bibnamefont {Bl{\"u}gel}},\
  }\href@noop {} {\bibfield  {journal} {\bibinfo  {journal} {Physical Review
  B}\ }\textbf {\bibinfo {volume} {78}},\ \bibinfo {pages} {140403} (\bibinfo
  {year} {2008})}\BibitemShut {NoStop}%
\bibitem [{\citenamefont {Bode}\ \emph {et~al.}(2007)\citenamefont {Bode},
  \citenamefont {Heide}, \citenamefont {von Bergmann}, \citenamefont
  {Ferriani}, \citenamefont {Heinze}, \citenamefont {Bihlmayer}, \citenamefont
  {Kubetzka}, \citenamefont {Pietzsch}, \citenamefont {Bl{\"u}gel},\ and\
  \citenamefont {Wiesendanger}}]{Bode2007DMI}%
  \BibitemOpen
  \bibfield  {author} {\bibinfo {author} {\bibfnamefont {M.}~\bibnamefont
  {Bode}}, \bibinfo {author} {\bibfnamefont {M.}~\bibnamefont {Heide}},
  \bibinfo {author} {\bibfnamefont {K.}~\bibnamefont {von Bergmann}}, \bibinfo
  {author} {\bibfnamefont {P.}~\bibnamefont {Ferriani}}, \bibinfo {author}
  {\bibfnamefont {S.}~\bibnamefont {Heinze}}, \bibinfo {author} {\bibfnamefont
  {G.}~\bibnamefont {Bihlmayer}}, \bibinfo {author} {\bibfnamefont
  {A.}~\bibnamefont {Kubetzka}}, \bibinfo {author} {\bibfnamefont
  {O.}~\bibnamefont {Pietzsch}}, \bibinfo {author} {\bibfnamefont
  {S.}~\bibnamefont {Bl{\"u}gel}}, \ and\ \bibinfo {author} {\bibfnamefont
  {R.}~\bibnamefont {Wiesendanger}},\ }\href@noop {} {\bibfield  {journal}
  {\bibinfo  {journal} {Nature}\ }\textbf {\bibinfo {volume} {447}},\ \bibinfo
  {pages} {190–193} (\bibinfo {year} {2007})}\BibitemShut {NoStop}%
\bibitem [{\citenamefont {Gartside}\ \emph {et~al.}(2016)\citenamefont
  {Gartside}, \citenamefont {Burn}, \citenamefont {Cohen},\ and\ \citenamefont
  {Branford}}]{Gartside2016}%
  \BibitemOpen
  \bibfield  {author} {\bibinfo {author} {\bibfnamefont {J.~C.}\ \bibnamefont
  {Gartside}}, \bibinfo {author} {\bibfnamefont {D.~M.}\ \bibnamefont {Burn}},
  \bibinfo {author} {\bibfnamefont {L.~F.}\ \bibnamefont {Cohen}}, \ and\
  \bibinfo {author} {\bibfnamefont {W.~R.}\ \bibnamefont {Branford}},\ }\href
  {\doibase 10.1038/s41565-017-0002-1} {\bibfield  {journal} {\bibinfo
  {journal} {Sci. Rep.}\ }\textbf {\bibinfo {volume} {6}},\ \bibinfo {pages}
  {32864} (\bibinfo {year} {2016})}\BibitemShut {NoStop}%
\bibitem [{\citenamefont {Lehmann}\ \emph {et~al.}(2019)\citenamefont
  {Lehmann}, \citenamefont {Donnelly}, \citenamefont {Derlet}, \citenamefont
  {Heyderman},\ and\ \citenamefont {Fiebig}}]{Lehmann2019}%
  \BibitemOpen
  \bibfield  {author} {\bibinfo {author} {\bibfnamefont {J.}~\bibnamefont
  {Lehmann}}, \bibinfo {author} {\bibfnamefont {C.}~\bibnamefont {Donnelly}},
  \bibinfo {author} {\bibfnamefont {P.~M.}\ \bibnamefont {Derlet}}, \bibinfo
  {author} {\bibfnamefont {L.~J.}\ \bibnamefont {Heyderman}}, \ and\ \bibinfo
  {author} {\bibfnamefont {M.}~\bibnamefont {Fiebig}},\ }\href {\doibase
  10.1038/s41565-018-0321-x} {\bibfield  {journal} {\bibinfo  {journal} {Nat.
  Nanotechnol.}\ }\textbf {\bibinfo {volume} {14}},\ \bibinfo {pages} {141}
  (\bibinfo {year} {2019})}\BibitemShut {NoStop}%
\bibitem [{\citenamefont {Apalkov}\ \emph {et~al.}(2013)\citenamefont
  {Apalkov}, \citenamefont {Khvalkovskiy}, \citenamefont {Watts}, \citenamefont
  {Nikitin}, \citenamefont {Tang}, \citenamefont {Lottis}, \citenamefont
  {Moon}, \citenamefont {Luo}, \citenamefont {Chen}, \citenamefont {Ong} \emph
  {et~al.}}]{Apalkov2013}%
  \BibitemOpen
  \bibfield  {author} {\bibinfo {author} {\bibfnamefont {D.}~\bibnamefont
  {Apalkov}}, \bibinfo {author} {\bibfnamefont {A.}~\bibnamefont
  {Khvalkovskiy}}, \bibinfo {author} {\bibfnamefont {S.}~\bibnamefont {Watts}},
  \bibinfo {author} {\bibfnamefont {V.}~\bibnamefont {Nikitin}}, \bibinfo
  {author} {\bibfnamefont {X.}~\bibnamefont {Tang}}, \bibinfo {author}
  {\bibfnamefont {D.}~\bibnamefont {Lottis}}, \bibinfo {author} {\bibfnamefont
  {K.}~\bibnamefont {Moon}}, \bibinfo {author} {\bibfnamefont {X.}~\bibnamefont
  {Luo}}, \bibinfo {author} {\bibfnamefont {E.}~\bibnamefont {Chen}}, \bibinfo
  {author} {\bibfnamefont {A.}~\bibnamefont {Ong}},  \emph {et~al.},\
  }\href@noop {} {\bibfield  {journal} {\bibinfo  {journal} {ACM Journal on
  Emerging Technologies in Computing Systems (JETC)}\ }\textbf {\bibinfo
  {volume} {9}},\ \bibinfo {pages} {13} (\bibinfo {year} {2013})}\BibitemShut
  {NoStop}%
\bibitem [{\citenamefont {Mac\^edo}\ \emph {et~al.}(2018)\citenamefont
  {Mac\^edo}, \citenamefont {Macauley}, \citenamefont {Nascimento},\ and\
  \citenamefont {Stamps}}]{Macedo2018}%
  \BibitemOpen
  \bibfield  {author} {\bibinfo {author} {\bibfnamefont {R.}~\bibnamefont
  {Mac\^edo}}, \bibinfo {author} {\bibfnamefont {G.~M.}\ \bibnamefont
  {Macauley}}, \bibinfo {author} {\bibfnamefont {F.~S.}\ \bibnamefont
  {Nascimento}}, \ and\ \bibinfo {author} {\bibfnamefont {R.~L.}\ \bibnamefont
  {Stamps}},\ }\href@noop {} {\bibfield  {journal} {\bibinfo  {journal} {Phys.
  Rev. B}\ }\textbf {\bibinfo {volume} {98}},\ \bibinfo {pages} {014437}
  (\bibinfo {year} {2018})}\BibitemShut {NoStop}%
\bibitem [{\citenamefont {Drisko}\ \emph {et~al.}(2016)\citenamefont {Drisko},
  \citenamefont {Marsh},\ and\ \citenamefont {Cumings}}]{Drisko2016}%
  \BibitemOpen
  \bibfield  {author} {\bibinfo {author} {\bibfnamefont {J.}~\bibnamefont
  {Drisko}}, \bibinfo {author} {\bibfnamefont {T.}~\bibnamefont {Marsh}}, \
  and\ \bibinfo {author} {\bibfnamefont {J.}~\bibnamefont {Cumings}},\
  }\href@noop {} {\bibfield  {journal} {\bibinfo  {journal} {Nature
  Communications}\ }\textbf {\bibinfo {volume} {8}},\ \bibinfo {pages} {14009}
  (\bibinfo {year} {2016})}\BibitemShut {NoStop}%
\bibitem [{\citenamefont {Chopdekar}\ \emph {et~al.}(2013)\citenamefont
  {Chopdekar}, \citenamefont {Duff}, \citenamefont {Hügli}, \citenamefont
  {Mengotti}, \citenamefont {Zanin}, \citenamefont {Heyderman},\ and\
  \citenamefont {Braun}}]{Chopdekar2013}%
  \BibitemOpen
  \bibfield  {author} {\bibinfo {author} {\bibfnamefont {R.~V.}\ \bibnamefont
  {Chopdekar}}, \bibinfo {author} {\bibfnamefont {G.}~\bibnamefont {Duff}},
  \bibinfo {author} {\bibfnamefont {R.~V.}\ \bibnamefont {Hügli}}, \bibinfo
  {author} {\bibfnamefont {E.}~\bibnamefont {Mengotti}}, \bibinfo {author}
  {\bibfnamefont {D.~A.}\ \bibnamefont {Zanin}}, \bibinfo {author}
  {\bibfnamefont {L.~J.}\ \bibnamefont {Heyderman}}, \ and\ \bibinfo {author}
  {\bibfnamefont {H.~B.}\ \bibnamefont {Braun}},\ }\href {\doibase
  10.1088/1367-2630/15/12/125033} {\bibfield  {journal} {\bibinfo  {journal}
  {New Journal of Physics}\ }\textbf {\bibinfo {volume} {15}},\ \bibinfo
  {pages} {125033} (\bibinfo {year} {2013})}\BibitemShut {NoStop}%
\bibitem [{\citenamefont {Bhat}\ \emph {et~al.}(2016)\citenamefont {Bhat},
  \citenamefont {Heimbach}, \citenamefont {Stasinopoulos},\ and\ \citenamefont
  {Grundler}}]{Bhat2016}%
  \BibitemOpen
  \bibfield  {author} {\bibinfo {author} {\bibfnamefont {V.~S.}\ \bibnamefont
  {Bhat}}, \bibinfo {author} {\bibfnamefont {F.}~\bibnamefont {Heimbach}},
  \bibinfo {author} {\bibfnamefont {I.}~\bibnamefont {Stasinopoulos}}, \ and\
  \bibinfo {author} {\bibfnamefont {D.}~\bibnamefont {Grundler}},\ }\href@noop
  {} {\bibfield  {journal} {\bibinfo  {journal} {Phys. Rev. B}\ }\textbf
  {\bibinfo {volume} {93}},\ \bibinfo {pages} {140401} (\bibinfo {year}
  {2016})}\BibitemShut {NoStop}%
\bibitem [{\citenamefont {Arava}\ \emph {et~al.}(2018)\citenamefont {Arava},
  \citenamefont {Derlet}, \citenamefont {Vijayakumar}, \citenamefont {Cui},
  \citenamefont {Bingham}, \citenamefont {Kleibert},\ and\ \citenamefont
  {Heyderman}}]{Arava2018}%
  \BibitemOpen
  \bibfield  {author} {\bibinfo {author} {\bibfnamefont {H.}~\bibnamefont
  {Arava}}, \bibinfo {author} {\bibfnamefont {P.~M.}\ \bibnamefont {Derlet}},
  \bibinfo {author} {\bibfnamefont {J.}~\bibnamefont {Vijayakumar}}, \bibinfo
  {author} {\bibfnamefont {J.}~\bibnamefont {Cui}}, \bibinfo {author}
  {\bibfnamefont {N.~S.}\ \bibnamefont {Bingham}}, \bibinfo {author}
  {\bibfnamefont {A.}~\bibnamefont {Kleibert}}, \ and\ \bibinfo {author}
  {\bibfnamefont {L.~J.}\ \bibnamefont {Heyderman}},\ }\href {\doibase
  10.1088/1361-6528/aabbc3} {\bibfield  {journal} {\bibinfo  {journal}
  {Nanotechnology}\ }\textbf {\bibinfo {volume} {29}},\ \bibinfo {pages}
  {265205} (\bibinfo {year} {2018})}\BibitemShut {NoStop}%
\bibitem [{\citenamefont {Jensen}\ \emph {et~al.}(2018)\citenamefont {Jensen},
  \citenamefont {Folven},\ and\ \citenamefont {Tufte}}]{Jensen2018}%
  \BibitemOpen
  \bibfield  {author} {\bibinfo {author} {\bibfnamefont {J.~H.}\ \bibnamefont
  {Jensen}}, \bibinfo {author} {\bibfnamefont {E.}~\bibnamefont {Folven}}, \
  and\ \bibinfo {author} {\bibfnamefont {G.}~\bibnamefont {Tufte}},\ }\href
  {\doibase 10.1162/isal\_a\_00011} {\bibfield  {journal} {\bibinfo  {journal}
  {The 2019 Conference on Artificial Life}\ ,\ \bibinfo {pages} {15}} (\bibinfo
  {year} {2018})}\BibitemShut {NoStop}%
\bibitem [{\citenamefont {Skadsem}\ \emph {et~al.}(2007)\citenamefont
  {Skadsem}, \citenamefont {Tserkovnyak}, \citenamefont {Brataas},\ and\
  \citenamefont {Bauer}}]{Skadsem2007}%
  \BibitemOpen
  \bibfield  {author} {\bibinfo {author} {\bibfnamefont {H.~J.}\ \bibnamefont
  {Skadsem}}, \bibinfo {author} {\bibfnamefont {Y.}~\bibnamefont
  {Tserkovnyak}}, \bibinfo {author} {\bibfnamefont {A.}~\bibnamefont
  {Brataas}}, \ and\ \bibinfo {author} {\bibfnamefont {G.~E.~W.}\ \bibnamefont
  {Bauer}},\ }\href {\doibase 10.1103/PhysRevB.75.094416} {\bibfield  {journal}
  {\bibinfo  {journal} {Phys. Rev. B}\ }\textbf {\bibinfo {volume} {75}},\
  \bibinfo {pages} {094416} (\bibinfo {year} {2007})}\BibitemShut {NoStop}%
\bibitem [{\citenamefont {Liu}\ \emph {et~al.}(2007)\citenamefont {Liu},
  \citenamefont {Giesen}, \citenamefont {Zhu}, \citenamefont {Sydora},\ and\
  \citenamefont {Freeman}}]{Liu2007PRL}%
  \BibitemOpen
  \bibfield  {author} {\bibinfo {author} {\bibfnamefont {Z.}~\bibnamefont
  {Liu}}, \bibinfo {author} {\bibfnamefont {F.}~\bibnamefont {Giesen}},
  \bibinfo {author} {\bibfnamefont {X.}~\bibnamefont {Zhu}}, \bibinfo {author}
  {\bibfnamefont {R.~D.}\ \bibnamefont {Sydora}}, \ and\ \bibinfo {author}
  {\bibfnamefont {M.~R.}\ \bibnamefont {Freeman}},\ }\href@noop {} {\bibfield
  {journal} {\bibinfo  {journal} {Phys. Rev. lett.}\ }\textbf {\bibinfo
  {volume} {98}},\ \bibinfo {pages} {087201} (\bibinfo {year}
  {2007})}\BibitemShut {NoStop}%
\bibitem [{\citenamefont {Madami}\ \emph {et~al.}(2011)\citenamefont {Madami},
  \citenamefont {Bonetti}, \citenamefont {Consolo}, \citenamefont {Tacchi},
  \citenamefont {Carlotti}, \citenamefont {Gubbiotti}, \citenamefont {Mancoff},
  \citenamefont {Yar},\ and\ \citenamefont {{\AA}kerman}}]{Madami2011}%
  \BibitemOpen
  \bibfield  {author} {\bibinfo {author} {\bibfnamefont {M.}~\bibnamefont
  {Madami}}, \bibinfo {author} {\bibfnamefont {S.}~\bibnamefont {Bonetti}},
  \bibinfo {author} {\bibfnamefont {G.}~\bibnamefont {Consolo}}, \bibinfo
  {author} {\bibfnamefont {S.}~\bibnamefont {Tacchi}}, \bibinfo {author}
  {\bibfnamefont {G.}~\bibnamefont {Carlotti}}, \bibinfo {author}
  {\bibfnamefont {G.}~\bibnamefont {Gubbiotti}}, \bibinfo {author}
  {\bibfnamefont {F.}~\bibnamefont {Mancoff}}, \bibinfo {author} {\bibfnamefont
  {M.~A.}\ \bibnamefont {Yar}}, \ and\ \bibinfo {author} {\bibfnamefont
  {J.}~\bibnamefont {{\AA}kerman}},\ }\href@noop {} {\bibfield  {journal}
  {\bibinfo  {journal} {Nature nanotechnology}\ }\textbf {\bibinfo {volume}
  {6}},\ \bibinfo {pages} {635} (\bibinfo {year} {2011})}\BibitemShut {NoStop}%
\bibitem [{\citenamefont {Schoen}\ \emph {et~al.}(2016)\citenamefont {Schoen},
  \citenamefont {Thonig}, \citenamefont {Schneider}, \citenamefont {Silva},
  \citenamefont {Nembach}, \citenamefont {Eriksson}, \citenamefont {Karis},\
  and\ \citenamefont {Shaw}}]{Schoen2016}%
  \BibitemOpen
  \bibfield  {author} {\bibinfo {author} {\bibfnamefont {M.~A.}\ \bibnamefont
  {Schoen}}, \bibinfo {author} {\bibfnamefont {D.}~\bibnamefont {Thonig}},
  \bibinfo {author} {\bibfnamefont {M.~L.}\ \bibnamefont {Schneider}}, \bibinfo
  {author} {\bibfnamefont {T.}~\bibnamefont {Silva}}, \bibinfo {author}
  {\bibfnamefont {H.~T.}\ \bibnamefont {Nembach}}, \bibinfo {author}
  {\bibfnamefont {O.}~\bibnamefont {Eriksson}}, \bibinfo {author}
  {\bibfnamefont {O.}~\bibnamefont {Karis}}, \ and\ \bibinfo {author}
  {\bibfnamefont {J.~M.}\ \bibnamefont {Shaw}},\ }\href@noop {} {\bibfield
  {journal} {\bibinfo  {journal} {Nature Physics}\ }\textbf {\bibinfo {volume}
  {12}},\ \bibinfo {pages} {839} (\bibinfo {year} {2016})}\BibitemShut
  {NoStop}%
\bibitem [{\citenamefont {Mendil}\ \emph {et~al.}(2019)\citenamefont {Mendil},
  \citenamefont {Trassin}, \citenamefont {Bu}, \citenamefont {Schaab},
  \citenamefont {Baumgartner}, \citenamefont {Murer}, \citenamefont {Dao},
  \citenamefont {Vijayakumar}, \citenamefont {Bracher}, \citenamefont
  {Bouillet}, \citenamefont {Vaz}, \citenamefont {Fiebig},\ and\ \citenamefont
  {Gambardella}}]{Mendil2019}%
  \BibitemOpen
  \bibfield  {author} {\bibinfo {author} {\bibfnamefont {J.}~\bibnamefont
  {Mendil}}, \bibinfo {author} {\bibfnamefont {M.}~\bibnamefont {Trassin}},
  \bibinfo {author} {\bibfnamefont {Q.}~\bibnamefont {Bu}}, \bibinfo {author}
  {\bibfnamefont {J.}~\bibnamefont {Schaab}}, \bibinfo {author} {\bibfnamefont
  {M.}~\bibnamefont {Baumgartner}}, \bibinfo {author} {\bibfnamefont
  {C.}~\bibnamefont {Murer}}, \bibinfo {author} {\bibfnamefont {P.~T.}\
  \bibnamefont {Dao}}, \bibinfo {author} {\bibfnamefont {J.}~\bibnamefont
  {Vijayakumar}}, \bibinfo {author} {\bibfnamefont {D.}~\bibnamefont
  {Bracher}}, \bibinfo {author} {\bibfnamefont {C.}~\bibnamefont {Bouillet}},
  \bibinfo {author} {\bibfnamefont {C.~A.~F.}\ \bibnamefont {Vaz}}, \bibinfo
  {author} {\bibfnamefont {M.}~\bibnamefont {Fiebig}}, \ and\ \bibinfo {author}
  {\bibfnamefont {P.}~\bibnamefont {Gambardella}},\ }\href {\doibase
  10.1103/PhysRevMaterials.3.034403} {\bibfield  {journal} {\bibinfo  {journal}
  {Phys. Rev. Materials}\ }\textbf {\bibinfo {volume} {3}},\ \bibinfo {pages}
  {034403} (\bibinfo {year} {2019})}\BibitemShut {NoStop}%
\bibitem [{\citenamefont {Felser}\ and\ \citenamefont
  {Hirohata}(2015)}]{Felser2015}%
  \BibitemOpen
  \bibfield  {author} {\bibinfo {author} {\bibfnamefont {C.}~\bibnamefont
  {Felser}}\ and\ \bibinfo {author} {\bibfnamefont {A.}~\bibnamefont
  {Hirohata}},\ }\href@noop {} {\emph {\bibinfo {title} {Heusler alloys}}}\
  (\bibinfo  {publisher} {Springer},\ \bibinfo {year} {2015})\BibitemShut
  {NoStop}%
\bibitem [{\citenamefont {Jourdan}\ \emph {et~al.}(2014)\citenamefont
  {Jourdan}, \citenamefont {Min{\'a}r}, \citenamefont {Braun}, \citenamefont
  {Kronenberg}, \citenamefont {Chadov}, \citenamefont {Balke}, \citenamefont
  {Gloskovskii}, \citenamefont {Kolbe}, \citenamefont {Elmers}, \citenamefont
  {Sch{\"o}nhense} \emph {et~al.}}]{Jourdan2014}%
  \BibitemOpen
  \bibfield  {author} {\bibinfo {author} {\bibfnamefont {M.}~\bibnamefont
  {Jourdan}}, \bibinfo {author} {\bibfnamefont {J.}~\bibnamefont {Min{\'a}r}},
  \bibinfo {author} {\bibfnamefont {J.}~\bibnamefont {Braun}}, \bibinfo
  {author} {\bibfnamefont {A.}~\bibnamefont {Kronenberg}}, \bibinfo {author}
  {\bibfnamefont {S.}~\bibnamefont {Chadov}}, \bibinfo {author} {\bibfnamefont
  {B.}~\bibnamefont {Balke}}, \bibinfo {author} {\bibfnamefont
  {A.}~\bibnamefont {Gloskovskii}}, \bibinfo {author} {\bibfnamefont
  {M.}~\bibnamefont {Kolbe}}, \bibinfo {author} {\bibfnamefont
  {H.}~\bibnamefont {Elmers}}, \bibinfo {author} {\bibfnamefont
  {G.}~\bibnamefont {Sch{\"o}nhense}},  \emph {et~al.},\ }\href@noop {}
  {\bibfield  {journal} {\bibinfo  {journal} {Nature communications}\ }\textbf
  {\bibinfo {volume} {5}},\ \bibinfo {pages} {3974} (\bibinfo {year}
  {2014})}\BibitemShut {NoStop}%
\bibitem [{\citenamefont {Emori}\ \emph {et~al.}(2017)\citenamefont {Emori},
  \citenamefont {Gray}, \citenamefont {Jeon}, \citenamefont {Peoples},
  \citenamefont {Schmitt}, \citenamefont {Mahalingam}, \citenamefont {Hill},
  \citenamefont {McConney}, \citenamefont {Gray}, \citenamefont {Alaan},
  \citenamefont {Bornstein}, \citenamefont {Shafer}, \citenamefont {N'Diaye},
  \citenamefont {Arenholz}, \citenamefont {Haugstad}, \citenamefont {Meng},
  \citenamefont {Yang}, \citenamefont {Li}, \citenamefont {Mahat},
  \citenamefont {Cahill}, \citenamefont {Dhagat}, \citenamefont {Jander},
  \citenamefont {Sun}, \citenamefont {Suzuki},\ and\ \citenamefont
  {Howe}}]{Emori2017}%
  \BibitemOpen
  \bibfield  {author} {\bibinfo {author} {\bibfnamefont {S.}~\bibnamefont
  {Emori}}, \bibinfo {author} {\bibfnamefont {B.~A.}\ \bibnamefont {Gray}},
  \bibinfo {author} {\bibfnamefont {H.-M.}\ \bibnamefont {Jeon}}, \bibinfo
  {author} {\bibfnamefont {J.}~\bibnamefont {Peoples}}, \bibinfo {author}
  {\bibfnamefont {M.}~\bibnamefont {Schmitt}}, \bibinfo {author} {\bibfnamefont
  {K.}~\bibnamefont {Mahalingam}}, \bibinfo {author} {\bibfnamefont
  {M.}~\bibnamefont {Hill}}, \bibinfo {author} {\bibfnamefont {M.~E.}\
  \bibnamefont {McConney}}, \bibinfo {author} {\bibfnamefont {M.~T.}\
  \bibnamefont {Gray}}, \bibinfo {author} {\bibfnamefont {U.~S.}\ \bibnamefont
  {Alaan}}, \bibinfo {author} {\bibfnamefont {A.~C.}\ \bibnamefont
  {Bornstein}}, \bibinfo {author} {\bibfnamefont {P.}~\bibnamefont {Shafer}},
  \bibinfo {author} {\bibfnamefont {A.~T.}\ \bibnamefont {N'Diaye}}, \bibinfo
  {author} {\bibfnamefont {E.}~\bibnamefont {Arenholz}}, \bibinfo {author}
  {\bibfnamefont {G.}~\bibnamefont {Haugstad}}, \bibinfo {author}
  {\bibfnamefont {K.-Y.}\ \bibnamefont {Meng}}, \bibinfo {author}
  {\bibfnamefont {F.}~\bibnamefont {Yang}}, \bibinfo {author} {\bibfnamefont
  {D.}~\bibnamefont {Li}}, \bibinfo {author} {\bibfnamefont {S.}~\bibnamefont
  {Mahat}}, \bibinfo {author} {\bibfnamefont {D.~G.}\ \bibnamefont {Cahill}},
  \bibinfo {author} {\bibfnamefont {P.}~\bibnamefont {Dhagat}}, \bibinfo
  {author} {\bibfnamefont {A.}~\bibnamefont {Jander}}, \bibinfo {author}
  {\bibfnamefont {N.~X.}\ \bibnamefont {Sun}}, \bibinfo {author} {\bibfnamefont
  {Y.}~\bibnamefont {Suzuki}}, \ and\ \bibinfo {author} {\bibfnamefont {B.~M.}\
  \bibnamefont {Howe}},\ }\href {\doibase 10.1002/adma.201701130} {\bibfield
  {journal} {\bibinfo  {journal} {Advanced Materials}\ }\textbf {\bibinfo
  {volume} {29}},\ \bibinfo {pages} {1701130} (\bibinfo {year} {2017})},\
  \Eprint
  {http://arxiv.org/abs/https://onlinelibrary.wiley.com/doi/pdf/10.1002/adma.201701130}
  {https://onlinelibrary.wiley.com/doi/pdf/10.1002/adma.201701130} \BibitemShut
  {NoStop}%
\bibitem [{\citenamefont {Wang}\ \emph
  {et~al.}(2018{\natexlab{a}})\citenamefont {Wang}, \citenamefont {Ma},
  \citenamefont {Xiao}, \citenamefont {Snezhko}, \citenamefont {Divan},
  \citenamefont {Ocola}, \citenamefont {Pearson}, \citenamefont {Janko},\ and\
  \citenamefont {Kwok}}]{Wang2018}%
  \BibitemOpen
  \bibfield  {author} {\bibinfo {author} {\bibfnamefont {Y.-L.}\ \bibnamefont
  {Wang}}, \bibinfo {author} {\bibfnamefont {X.}~\bibnamefont {Ma}}, \bibinfo
  {author} {\bibfnamefont {Z.-L.}\ \bibnamefont {Xiao}}, \bibinfo {author}
  {\bibfnamefont {A.}~\bibnamefont {Snezhko}}, \bibinfo {author} {\bibfnamefont
  {R.}~\bibnamefont {Divan}}, \bibinfo {author} {\bibfnamefont {L.~E.}\
  \bibnamefont {Ocola}}, \bibinfo {author} {\bibfnamefont {J.~E.}\ \bibnamefont
  {Pearson}}, \bibinfo {author} {\bibfnamefont {B.}~\bibnamefont {Janko}}, \
  and\ \bibinfo {author} {\bibfnamefont {W.-K.}\ \bibnamefont {Kwok}},\
  }\href@noop {} {\bibfield  {journal} {\bibinfo  {journal} {Nature
  Nanotechnology}\ }\textbf {\bibinfo {volume} {13}},\ \bibinfo {pages}
  {560–565} (\bibinfo {year} {2018}{\natexlab{a}})}\BibitemShut {NoStop}%
\bibitem [{\citenamefont {Golovchanskiy}\ \emph {et~al.}(2019)\citenamefont
  {Golovchanskiy}, \citenamefont {Abramov}, \citenamefont {Stolyarov},
  \citenamefont {Dzhumaev}, \citenamefont {Emelyanova}, \citenamefont
  {Golubov}, \citenamefont {Ryazanov},\ and\ \citenamefont
  {Ustinov}}]{golovchanskiy2019}%
  \BibitemOpen
  \bibfield  {author} {\bibinfo {author} {\bibfnamefont {I.~A.}\ \bibnamefont
  {Golovchanskiy}}, \bibinfo {author} {\bibfnamefont {N.~N.}\ \bibnamefont
  {Abramov}}, \bibinfo {author} {\bibfnamefont {V.~S.}\ \bibnamefont
  {Stolyarov}}, \bibinfo {author} {\bibfnamefont {P.~S.}\ \bibnamefont
  {Dzhumaev}}, \bibinfo {author} {\bibfnamefont {O.~V.}\ \bibnamefont
  {Emelyanova}}, \bibinfo {author} {\bibfnamefont {A.~A.}\ \bibnamefont
  {Golubov}}, \bibinfo {author} {\bibfnamefont {V.~V.}\ \bibnamefont
  {Ryazanov}}, \ and\ \bibinfo {author} {\bibfnamefont {A.~V.}\ \bibnamefont
  {Ustinov}},\ }\href@noop {} {\bibfield  {journal} {\bibinfo  {journal}
  {Advanced science}\ }\textbf {\bibinfo {volume} {6}},\ \bibinfo {pages}
  {1900435} (\bibinfo {year} {2019})}\BibitemShut {NoStop}%
\bibitem [{\citenamefont {Hasan}\ and\ \citenamefont {Kane}(2010)}]{Hasan2010}%
  \BibitemOpen
  \bibfield  {author} {\bibinfo {author} {\bibfnamefont {M.~Z.}\ \bibnamefont
  {Hasan}}\ and\ \bibinfo {author} {\bibfnamefont {C.~L.}\ \bibnamefont
  {Kane}},\ }\href@noop {} {\bibfield  {journal} {\bibinfo  {journal} {Reviews
  of Modern Physics}\ }\textbf {\bibinfo {volume} {82}},\ \bibinfo {pages}
  {3045} (\bibinfo {year} {2010})}\BibitemShut {NoStop}%
\bibitem [{\citenamefont {Bansil}\ \emph {et~al.}(2016)\citenamefont {Bansil},
  \citenamefont {Lin},\ and\ \citenamefont {Das}}]{Bansil2016}%
  \BibitemOpen
  \bibfield  {author} {\bibinfo {author} {\bibfnamefont {A.}~\bibnamefont
  {Bansil}}, \bibinfo {author} {\bibfnamefont {H.}~\bibnamefont {Lin}}, \ and\
  \bibinfo {author} {\bibfnamefont {T.}~\bibnamefont {Das}},\ }\href@noop {}
  {\bibfield  {journal} {\bibinfo  {journal} {Reviews of Modern Physics}\
  }\textbf {\bibinfo {volume} {88}},\ \bibinfo {pages} {021004} (\bibinfo
  {year} {2016})}\BibitemShut {NoStop}%
\bibitem [{\citenamefont {Kane}\ and\ \citenamefont
  {Mele}(2005{\natexlab{a}})}]{kane2005z}%
  \BibitemOpen
  \bibfield  {author} {\bibinfo {author} {\bibfnamefont {C.~L.}\ \bibnamefont
  {Kane}}\ and\ \bibinfo {author} {\bibfnamefont {E.~J.}\ \bibnamefont
  {Mele}},\ }\href@noop {} {\bibfield  {journal} {\bibinfo  {journal} {Physical
  review letters}\ }\textbf {\bibinfo {volume} {95}},\ \bibinfo {pages}
  {146802} (\bibinfo {year} {2005}{\natexlab{a}})}\BibitemShut {NoStop}%
\bibitem [{\citenamefont {Kane}\ and\ \citenamefont
  {Mele}(2005{\natexlab{b}})}]{kane2005}%
  \BibitemOpen
  \bibfield  {author} {\bibinfo {author} {\bibfnamefont {C.~L.}\ \bibnamefont
  {Kane}}\ and\ \bibinfo {author} {\bibfnamefont {E.~J.}\ \bibnamefont
  {Mele}},\ }\href@noop {} {\bibfield  {journal} {\bibinfo  {journal} {Physical
  review letters}\ }\textbf {\bibinfo {volume} {95}},\ \bibinfo {pages}
  {226801} (\bibinfo {year} {2005}{\natexlab{b}})}\BibitemShut {NoStop}%
\bibitem [{\citenamefont {Bernevig}\ \emph {et~al.}(2006)\citenamefont
  {Bernevig}, \citenamefont {Hughes},\ and\ \citenamefont
  {Zhang}}]{bernevig2006quantum}%
  \BibitemOpen
  \bibfield  {author} {\bibinfo {author} {\bibfnamefont {B.~A.}\ \bibnamefont
  {Bernevig}}, \bibinfo {author} {\bibfnamefont {T.~L.}\ \bibnamefont
  {Hughes}}, \ and\ \bibinfo {author} {\bibfnamefont {S.-C.}\ \bibnamefont
  {Zhang}},\ }\href@noop {} {\bibfield  {journal} {\bibinfo  {journal}
  {science}\ }\textbf {\bibinfo {volume} {314}},\ \bibinfo {pages} {1757}
  (\bibinfo {year} {2006})}\BibitemShut {NoStop}%
\bibitem [{\citenamefont {Fu}\ \emph {et~al.}(2007)\citenamefont {Fu},
  \citenamefont {Kane},\ and\ \citenamefont {Mele}}]{fu2007topological}%
  \BibitemOpen
  \bibfield  {author} {\bibinfo {author} {\bibfnamefont {L.}~\bibnamefont
  {Fu}}, \bibinfo {author} {\bibfnamefont {C.~L.}\ \bibnamefont {Kane}}, \ and\
  \bibinfo {author} {\bibfnamefont {E.~J.}\ \bibnamefont {Mele}},\ }\href@noop
  {} {\bibfield  {journal} {\bibinfo  {journal} {Physical review letters}\
  }\textbf {\bibinfo {volume} {98}},\ \bibinfo {pages} {106803} (\bibinfo
  {year} {2007})}\BibitemShut {NoStop}%
\bibitem [{\citenamefont {Moore}\ and\ \citenamefont
  {Balents}(2007)}]{moore2007topological}%
  \BibitemOpen
  \bibfield  {author} {\bibinfo {author} {\bibfnamefont {J.~E.}\ \bibnamefont
  {Moore}}\ and\ \bibinfo {author} {\bibfnamefont {L.}~\bibnamefont
  {Balents}},\ }\href@noop {} {\bibfield  {journal} {\bibinfo  {journal}
  {Physical Review B}\ }\textbf {\bibinfo {volume} {75}},\ \bibinfo {pages}
  {121306} (\bibinfo {year} {2007})}\BibitemShut {NoStop}%
\bibitem [{\citenamefont {Xia}\ \emph {et~al.}(2009)\citenamefont {Xia},
  \citenamefont {Qian}, \citenamefont {Hsieh}, \citenamefont {Wray},
  \citenamefont {Pal}, \citenamefont {Lin}, \citenamefont {Bansil},
  \citenamefont {Grauer}, \citenamefont {Hor}, \citenamefont {Cava} \emph
  {et~al.}}]{xia2009observation}%
  \BibitemOpen
  \bibfield  {author} {\bibinfo {author} {\bibfnamefont {Y.}~\bibnamefont
  {Xia}}, \bibinfo {author} {\bibfnamefont {D.}~\bibnamefont {Qian}}, \bibinfo
  {author} {\bibfnamefont {D.}~\bibnamefont {Hsieh}}, \bibinfo {author}
  {\bibfnamefont {L.}~\bibnamefont {Wray}}, \bibinfo {author} {\bibfnamefont
  {A.}~\bibnamefont {Pal}}, \bibinfo {author} {\bibfnamefont {H.}~\bibnamefont
  {Lin}}, \bibinfo {author} {\bibfnamefont {A.}~\bibnamefont {Bansil}},
  \bibinfo {author} {\bibfnamefont {D.}~\bibnamefont {Grauer}}, \bibinfo
  {author} {\bibfnamefont {Y.~S.}\ \bibnamefont {Hor}}, \bibinfo {author}
  {\bibfnamefont {R.~J.}\ \bibnamefont {Cava}},  \emph {et~al.},\ }\href@noop
  {} {\bibfield  {journal} {\bibinfo  {journal} {Nature physics}\ }\textbf
  {\bibinfo {volume} {5}},\ \bibinfo {pages} {398} (\bibinfo {year}
  {2009})}\BibitemShut {NoStop}%
\bibitem [{\citenamefont {Zhang}\ \emph {et~al.}(2009)\citenamefont {Zhang},
  \citenamefont {Liu}, \citenamefont {Qi}, \citenamefont {Dai}, \citenamefont
  {Fang},\ and\ \citenamefont {Zhang}}]{zhang2009topological}%
  \BibitemOpen
  \bibfield  {author} {\bibinfo {author} {\bibfnamefont {H.}~\bibnamefont
  {Zhang}}, \bibinfo {author} {\bibfnamefont {C.-X.}\ \bibnamefont {Liu}},
  \bibinfo {author} {\bibfnamefont {X.-L.}\ \bibnamefont {Qi}}, \bibinfo
  {author} {\bibfnamefont {X.}~\bibnamefont {Dai}}, \bibinfo {author}
  {\bibfnamefont {Z.}~\bibnamefont {Fang}}, \ and\ \bibinfo {author}
  {\bibfnamefont {S.-C.}\ \bibnamefont {Zhang}},\ }\href@noop {} {\bibfield
  {journal} {\bibinfo  {journal} {Nature physics}\ }\textbf {\bibinfo {volume}
  {5}},\ \bibinfo {pages} {438} (\bibinfo {year} {2009})}\BibitemShut {NoStop}%
\bibitem [{\citenamefont {Li}\ \emph {et~al.}(2019{\natexlab{b}})\citenamefont
  {Li}, \citenamefont {Kally}, \citenamefont {Zhang}, \citenamefont
  {Pillsbury}, \citenamefont {Ding}, \citenamefont {Csaba}, \citenamefont
  {Ding}, \citenamefont {Jiang}, \citenamefont {Liu}, \citenamefont {Sinclair}
  \emph {et~al.}}]{Li2019}%
  \BibitemOpen
  \bibfield  {author} {\bibinfo {author} {\bibfnamefont {P.}~\bibnamefont
  {Li}}, \bibinfo {author} {\bibfnamefont {J.}~\bibnamefont {Kally}}, \bibinfo
  {author} {\bibfnamefont {S.~S.-L.}\ \bibnamefont {Zhang}}, \bibinfo {author}
  {\bibfnamefont {T.}~\bibnamefont {Pillsbury}}, \bibinfo {author}
  {\bibfnamefont {J.}~\bibnamefont {Ding}}, \bibinfo {author} {\bibfnamefont
  {G.}~\bibnamefont {Csaba}}, \bibinfo {author} {\bibfnamefont
  {J.}~\bibnamefont {Ding}}, \bibinfo {author} {\bibfnamefont {J.}~\bibnamefont
  {Jiang}}, \bibinfo {author} {\bibfnamefont {Y.}~\bibnamefont {Liu}}, \bibinfo
  {author} {\bibfnamefont {R.}~\bibnamefont {Sinclair}},  \emph {et~al.},\
  }\href@noop {} {\bibfield  {journal} {\bibinfo  {journal} {Science advances}\
  }\textbf {\bibinfo {volume} {5}},\ \bibinfo {pages} {eaaw3415} (\bibinfo
  {year} {2019}{\natexlab{b}})}\BibitemShut {NoStop}%
\bibitem [{\citenamefont {Armitage}\ \emph {et~al.}(2018)\citenamefont
  {Armitage}, \citenamefont {Mele},\ and\ \citenamefont
  {Vishwanath}}]{Armitage2018}%
  \BibitemOpen
  \bibfield  {author} {\bibinfo {author} {\bibfnamefont {N.}~\bibnamefont
  {Armitage}}, \bibinfo {author} {\bibfnamefont {E.}~\bibnamefont {Mele}}, \
  and\ \bibinfo {author} {\bibfnamefont {A.}~\bibnamefont {Vishwanath}},\
  }\href@noop {} {\bibfield  {journal} {\bibinfo  {journal} {Reviews of Modern
  Physics}\ }\textbf {\bibinfo {volume} {90}},\ \bibinfo {pages} {015001}
  (\bibinfo {year} {2018})}\BibitemShut {NoStop}%
\bibitem [{\citenamefont {Wang}\ \emph {et~al.}(2012)\citenamefont {Wang},
  \citenamefont {Sun}, \citenamefont {Chen}, \citenamefont {Franchini},
  \citenamefont {Xu}, \citenamefont {Weng}, \citenamefont {Dai},\ and\
  \citenamefont {Fang}}]{wang2012dirac}%
  \BibitemOpen
  \bibfield  {author} {\bibinfo {author} {\bibfnamefont {Z.}~\bibnamefont
  {Wang}}, \bibinfo {author} {\bibfnamefont {Y.}~\bibnamefont {Sun}}, \bibinfo
  {author} {\bibfnamefont {X.-Q.}\ \bibnamefont {Chen}}, \bibinfo {author}
  {\bibfnamefont {C.}~\bibnamefont {Franchini}}, \bibinfo {author}
  {\bibfnamefont {G.}~\bibnamefont {Xu}}, \bibinfo {author} {\bibfnamefont
  {H.}~\bibnamefont {Weng}}, \bibinfo {author} {\bibfnamefont {X.}~\bibnamefont
  {Dai}}, \ and\ \bibinfo {author} {\bibfnamefont {Z.}~\bibnamefont {Fang}},\
  }\href@noop {} {\bibfield  {journal} {\bibinfo  {journal} {Physical Review
  B}\ }\textbf {\bibinfo {volume} {85}},\ \bibinfo {pages} {195320} (\bibinfo
  {year} {2012})}\BibitemShut {NoStop}%
\bibitem [{\citenamefont {Liu}\ \emph {et~al.}(2014)\citenamefont {Liu},
  \citenamefont {Jiang}, \citenamefont {Zhou}, \citenamefont {Wang},
  \citenamefont {Zhang}, \citenamefont {Weng}, \citenamefont {Prabhakaran},
  \citenamefont {Mo}, \citenamefont {Peng}, \citenamefont {Dudin} \emph
  {et~al.}}]{liu2014stable}%
  \BibitemOpen
  \bibfield  {author} {\bibinfo {author} {\bibfnamefont {Z.}~\bibnamefont
  {Liu}}, \bibinfo {author} {\bibfnamefont {J.}~\bibnamefont {Jiang}}, \bibinfo
  {author} {\bibfnamefont {B.}~\bibnamefont {Zhou}}, \bibinfo {author}
  {\bibfnamefont {Z.}~\bibnamefont {Wang}}, \bibinfo {author} {\bibfnamefont
  {Y.}~\bibnamefont {Zhang}}, \bibinfo {author} {\bibfnamefont
  {H.}~\bibnamefont {Weng}}, \bibinfo {author} {\bibfnamefont {D.}~\bibnamefont
  {Prabhakaran}}, \bibinfo {author} {\bibfnamefont {S.}~\bibnamefont {Mo}},
  \bibinfo {author} {\bibfnamefont {H.}~\bibnamefont {Peng}}, \bibinfo {author}
  {\bibfnamefont {P.}~\bibnamefont {Dudin}},  \emph {et~al.},\ }\href@noop {}
  {\bibfield  {journal} {\bibinfo  {journal} {Nature materials}\ }\textbf
  {\bibinfo {volume} {13}},\ \bibinfo {pages} {677} (\bibinfo {year}
  {2014})}\BibitemShut {NoStop}%
\bibitem [{\citenamefont {Young}\ \emph {et~al.}(2012)\citenamefont {Young},
  \citenamefont {Zaheer}, \citenamefont {Teo}, \citenamefont {Kane},
  \citenamefont {Mele},\ and\ \citenamefont {Rappe}}]{young2012dirac}%
  \BibitemOpen
  \bibfield  {author} {\bibinfo {author} {\bibfnamefont {S.~M.}\ \bibnamefont
  {Young}}, \bibinfo {author} {\bibfnamefont {S.}~\bibnamefont {Zaheer}},
  \bibinfo {author} {\bibfnamefont {J.~C.}\ \bibnamefont {Teo}}, \bibinfo
  {author} {\bibfnamefont {C.~L.}\ \bibnamefont {Kane}}, \bibinfo {author}
  {\bibfnamefont {E.~J.}\ \bibnamefont {Mele}}, \ and\ \bibinfo {author}
  {\bibfnamefont {A.~M.}\ \bibnamefont {Rappe}},\ }\href@noop {} {\bibfield
  {journal} {\bibinfo  {journal} {Physical review letters}\ }\textbf {\bibinfo
  {volume} {108}},\ \bibinfo {pages} {140405} (\bibinfo {year}
  {2012})}\BibitemShut {NoStop}%
\bibitem [{\citenamefont {Wan}\ \emph {et~al.}(2011)\citenamefont {Wan},
  \citenamefont {Turner}, \citenamefont {Vishwanath},\ and\ \citenamefont
  {Savrasov}}]{wan2011topological}%
  \BibitemOpen
  \bibfield  {author} {\bibinfo {author} {\bibfnamefont {X.}~\bibnamefont
  {Wan}}, \bibinfo {author} {\bibfnamefont {A.~M.}\ \bibnamefont {Turner}},
  \bibinfo {author} {\bibfnamefont {A.}~\bibnamefont {Vishwanath}}, \ and\
  \bibinfo {author} {\bibfnamefont {S.~Y.}\ \bibnamefont {Savrasov}},\
  }\href@noop {} {\bibfield  {journal} {\bibinfo  {journal} {Physical Review
  B}\ }\textbf {\bibinfo {volume} {83}},\ \bibinfo {pages} {205101} (\bibinfo
  {year} {2011})}\BibitemShut {NoStop}%
\bibitem [{\citenamefont {Xu}\ \emph {et~al.}(2011)\citenamefont {Xu},
  \citenamefont {Weng}, \citenamefont {Wang}, \citenamefont {Dai},\ and\
  \citenamefont {Fang}}]{xu2011chern}%
  \BibitemOpen
  \bibfield  {author} {\bibinfo {author} {\bibfnamefont {G.}~\bibnamefont
  {Xu}}, \bibinfo {author} {\bibfnamefont {H.}~\bibnamefont {Weng}}, \bibinfo
  {author} {\bibfnamefont {Z.}~\bibnamefont {Wang}}, \bibinfo {author}
  {\bibfnamefont {X.}~\bibnamefont {Dai}}, \ and\ \bibinfo {author}
  {\bibfnamefont {Z.}~\bibnamefont {Fang}},\ }\href@noop {} {\bibfield
  {journal} {\bibinfo  {journal} {Physical review letters}\ }\textbf {\bibinfo
  {volume} {107}},\ \bibinfo {pages} {186806} (\bibinfo {year}
  {2011})}\BibitemShut {NoStop}%
\bibitem [{\citenamefont {Lv}\ \emph {et~al.}(2015)\citenamefont {Lv},
  \citenamefont {Weng}, \citenamefont {Fu}, \citenamefont {Wang}, \citenamefont
  {Miao}, \citenamefont {Ma}, \citenamefont {Richard}, \citenamefont {Huang},
  \citenamefont {Zhao}, \citenamefont {Chen} \emph
  {et~al.}}]{lv2015experimental}%
  \BibitemOpen
  \bibfield  {author} {\bibinfo {author} {\bibfnamefont {B.}~\bibnamefont
  {Lv}}, \bibinfo {author} {\bibfnamefont {H.}~\bibnamefont {Weng}}, \bibinfo
  {author} {\bibfnamefont {B.}~\bibnamefont {Fu}}, \bibinfo {author}
  {\bibfnamefont {X.}~\bibnamefont {Wang}}, \bibinfo {author} {\bibfnamefont
  {H.}~\bibnamefont {Miao}}, \bibinfo {author} {\bibfnamefont {J.}~\bibnamefont
  {Ma}}, \bibinfo {author} {\bibfnamefont {P.}~\bibnamefont {Richard}},
  \bibinfo {author} {\bibfnamefont {X.}~\bibnamefont {Huang}}, \bibinfo
  {author} {\bibfnamefont {L.}~\bibnamefont {Zhao}}, \bibinfo {author}
  {\bibfnamefont {G.}~\bibnamefont {Chen}},  \emph {et~al.},\ }\href@noop {}
  {\bibfield  {journal} {\bibinfo  {journal} {Physical Review X}\ }\textbf
  {\bibinfo {volume} {5}},\ \bibinfo {pages} {031013} (\bibinfo {year}
  {2015})}\BibitemShut {NoStop}%
\bibitem [{\citenamefont {Zhang}\ \emph
  {et~al.}(2019{\natexlab{b}})\citenamefont {Zhang}, \citenamefont {Burkov},
  \citenamefont {Martin},\ and\ \citenamefont {Heinonen}}]{Zhang2019}%
  \BibitemOpen
  \bibfield  {author} {\bibinfo {author} {\bibfnamefont {S.~S.-L.}\
  \bibnamefont {Zhang}}, \bibinfo {author} {\bibfnamefont {A.~A.}\ \bibnamefont
  {Burkov}}, \bibinfo {author} {\bibfnamefont {I.}~\bibnamefont {Martin}}, \
  and\ \bibinfo {author} {\bibfnamefont {O.~G.}\ \bibnamefont {Heinonen}},\
  }\href@noop {} {\bibfield  {journal} {\bibinfo  {journal} {arXiv preprint
  arXiv:1904.07181}\ } (\bibinfo {year} {2019}{\natexlab{b}})}\BibitemShut
  {NoStop}%
\bibitem [{\citenamefont {Perrin}\ \emph {et~al.}(2016)\citenamefont {Perrin},
  \citenamefont {Canals},\ and\ \citenamefont {Rougemaille}}]{Perrin2016}%
  \BibitemOpen
  \bibfield  {author} {\bibinfo {author} {\bibfnamefont {Y.}~\bibnamefont
  {Perrin}}, \bibinfo {author} {\bibfnamefont {B.}~\bibnamefont {Canals}}, \
  and\ \bibinfo {author} {\bibfnamefont {N.}~\bibnamefont {Rougemaille}},\
  }\href@noop {} {\bibfield  {journal} {\bibinfo  {journal} {Nature}\ }\textbf
  {\bibinfo {volume} {540}},\ \bibinfo {pages} {410–413} (\bibinfo {year}
  {2016})}\BibitemShut {NoStop}%
\bibitem [{\citenamefont {Kirchner}\ and\ \citenamefont
  {Schift}(2015)}]{Kirchner2015}%
  \BibitemOpen
  \bibfield  {author} {\bibinfo {author} {\bibfnamefont {R.}~\bibnamefont
  {Kirchner}}\ and\ \bibinfo {author} {\bibfnamefont {H.}~\bibnamefont
  {Schift}},\ }\href {\doibase 10.1016/j.mee.2015.04.082} {\bibfield  {journal}
  {\bibinfo  {journal} {Microelectron. Eng.}\ }\textbf {\bibinfo {volume}
  {141}},\ \bibinfo {pages} {243} (\bibinfo {year} {2015})}\BibitemShut
  {NoStop}%
\bibitem [{\citenamefont {Maruo}\ \emph {et~al.}(1997)\citenamefont {Maruo},
  \citenamefont {Nakamura},\ and\ \citenamefont {Kawata}}]{Maruo97}%
  \BibitemOpen
  \bibfield  {author} {\bibinfo {author} {\bibfnamefont {S.}~\bibnamefont
  {Maruo}}, \bibinfo {author} {\bibfnamefont {O.}~\bibnamefont {Nakamura}}, \
  and\ \bibinfo {author} {\bibfnamefont {S.}~\bibnamefont {Kawata}},\ }\href
  {\doibase 10.1364/OL.22.000132} {\bibfield  {journal} {\bibinfo  {journal}
  {Opt. Lett.}\ }\textbf {\bibinfo {volume} {22}},\ \bibinfo {pages} {132}
  (\bibinfo {year} {1997})}\BibitemShut {NoStop}%
\bibitem [{\citenamefont {Keller}\ \emph {et~al.}(2018)\citenamefont {Keller},
  \citenamefont {Al~Mamoori}, \citenamefont {Pieper}, \citenamefont {Gspan},
  \citenamefont {Stockem}, \citenamefont {Schr\"{o}der}, \citenamefont {Sven},
  \citenamefont {Winkler}, \citenamefont {Plank}, \citenamefont {Pohlit},
  \citenamefont {M\"{u}ller},\ and\ \citenamefont {Huth}}]{Keller2018}%
  \BibitemOpen
  \bibfield  {author} {\bibinfo {author} {\bibfnamefont {L.}~\bibnamefont
  {Keller}}, \bibinfo {author} {\bibfnamefont {M.~K.}\ \bibnamefont
  {Al~Mamoori}}, \bibinfo {author} {\bibfnamefont {J.}~\bibnamefont {Pieper}},
  \bibinfo {author} {\bibfnamefont {C.}~\bibnamefont {Gspan}}, \bibinfo
  {author} {\bibfnamefont {I.}~\bibnamefont {Stockem}}, \bibinfo {author}
  {\bibfnamefont {C.}~\bibnamefont {Schr\"{o}der}}, \bibinfo {author}
  {\bibfnamefont {B.}~\bibnamefont {Sven}}, \bibinfo {author} {\bibfnamefont
  {R.}~\bibnamefont {Winkler}}, \bibinfo {author} {\bibfnamefont
  {H.}~\bibnamefont {Plank}}, \bibinfo {author} {\bibfnamefont
  {M.}~\bibnamefont {Pohlit}}, \bibinfo {author} {\bibfnamefont
  {J.}~\bibnamefont {M\"{u}ller}}, \ and\ \bibinfo {author} {\bibfnamefont
  {M.}~\bibnamefont {Huth}},\ }\href {\doibase 10.1038/s41598-018-24431-x}
  {\bibfield  {journal} {\bibinfo  {journal} {Sci. Rep.}\ }\textbf {\bibinfo
  {volume} {8}},\ \bibinfo {pages} {6160} (\bibinfo {year} {2018})}\BibitemShut
  {NoStop}%
\bibitem [{\citenamefont {Fowlkes}\ \emph {et~al.}(2018)\citenamefont
  {Fowlkes}, \citenamefont {Winkler}, \citenamefont {Lewis}, \citenamefont
  {Fernández-Pacheco}, \citenamefont {Skoric}, \citenamefont
  {Sanz-Hernández}, \citenamefont {Stanford}, \citenamefont {Mutunga},
  \citenamefont {Rack},\ and\ \citenamefont {Plank}}]{Fowlkes2018}%
  \BibitemOpen
  \bibfield  {author} {\bibinfo {author} {\bibfnamefont {J.~D.}\ \bibnamefont
  {Fowlkes}}, \bibinfo {author} {\bibfnamefont {R.}~\bibnamefont {Winkler}},
  \bibinfo {author} {\bibfnamefont {B.~B.}\ \bibnamefont {Lewis}}, \bibinfo
  {author} {\bibfnamefont {A.}~\bibnamefont {Fernández-Pacheco}}, \bibinfo
  {author} {\bibfnamefont {L.}~\bibnamefont {Skoric}}, \bibinfo {author}
  {\bibfnamefont {D.}~\bibnamefont {Sanz-Hernández}}, \bibinfo {author}
  {\bibfnamefont {M.~G.}\ \bibnamefont {Stanford}}, \bibinfo {author}
  {\bibfnamefont {E.}~\bibnamefont {Mutunga}}, \bibinfo {author} {\bibfnamefont
  {P.~D.}\ \bibnamefont {Rack}}, \ and\ \bibinfo {author} {\bibfnamefont
  {H.}~\bibnamefont {Plank}},\ }\href {\doibase 10.1021/acsanm.7b00342}
  {\bibfield  {journal} {\bibinfo  {journal} {ACS Applied Nano Materials}\
  }\textbf {\bibinfo {volume} {1}},\ \bibinfo {pages} {1028} (\bibinfo {year}
  {2018})},\ \Eprint
  {http://arxiv.org/abs/https://doi.org/10.1021/acsanm.7b00342}
  {https://doi.org/10.1021/acsanm.7b00342} \BibitemShut {NoStop}%
\bibitem [{\citenamefont {Farhan}\ \emph {et~al.}(2019)\citenamefont {Farhan},
  \citenamefont {Saccone}, \citenamefont {Petersen}, \citenamefont {Dhuey},
  \citenamefont {Chopdekar}, \citenamefont {Huang}, \citenamefont {Kent},
  \citenamefont {Zuhuang}, \citenamefont {Alava}, \citenamefont {Lippert},
  \citenamefont {Scholl},\ and\ \citenamefont {van Dijken}}]{Farhan2019}%
  \BibitemOpen
  \bibfield  {author} {\bibinfo {author} {\bibfnamefont {A.}~\bibnamefont
  {Farhan}}, \bibinfo {author} {\bibfnamefont {M.}~\bibnamefont {Saccone}},
  \bibinfo {author} {\bibfnamefont {C.~F.}\ \bibnamefont {Petersen}}, \bibinfo
  {author} {\bibfnamefont {S.}~\bibnamefont {Dhuey}}, \bibinfo {author}
  {\bibfnamefont {R.~V.}\ \bibnamefont {Chopdekar}}, \bibinfo {author}
  {\bibfnamefont {Y.-L.}\ \bibnamefont {Huang}}, \bibinfo {author}
  {\bibfnamefont {N.}~\bibnamefont {Kent}}, \bibinfo {author} {\bibfnamefont
  {C.}~\bibnamefont {Zuhuang}}, \bibinfo {author} {\bibfnamefont {M.~J.}\
  \bibnamefont {Alava}}, \bibinfo {author} {\bibfnamefont {T.}~\bibnamefont
  {Lippert}}, \bibinfo {author} {\bibfnamefont {A.}~\bibnamefont {Scholl}}, \
  and\ \bibinfo {author} {\bibfnamefont {S.}~\bibnamefont {van Dijken}},\
  }\href {\doibase 10.1126/sciadv.aav6380} {\bibfield  {journal} {\bibinfo
  {journal} {Science Advances}\ }\textbf {\bibinfo {volume} {5}},\ \bibinfo
  {pages} {eaav6380} (\bibinfo {year} {2019})}\BibitemShut {NoStop}%
\bibitem [{\citenamefont {Donnelly}\ \emph {et~al.}(2015)\citenamefont
  {Donnelly}, \citenamefont {Guizar-Sicairos}, \citenamefont {Scagnoli},
  \citenamefont {Holler}, \citenamefont {Huthwelker}, \citenamefont {Menzel},
  \citenamefont {Vartiainen}, \citenamefont {M\"uller}, \citenamefont {Kirk},
  \citenamefont {Gliga}, \citenamefont {Raabe},\ and\ \citenamefont
  {Heyderman}}]{Donnelly2015}%
  \BibitemOpen
  \bibfield  {author} {\bibinfo {author} {\bibfnamefont {C.}~\bibnamefont
  {Donnelly}}, \bibinfo {author} {\bibfnamefont {M.}~\bibnamefont
  {Guizar-Sicairos}}, \bibinfo {author} {\bibfnamefont {V.}~\bibnamefont
  {Scagnoli}}, \bibinfo {author} {\bibfnamefont {M.}~\bibnamefont {Holler}},
  \bibinfo {author} {\bibfnamefont {T.}~\bibnamefont {Huthwelker}}, \bibinfo
  {author} {\bibfnamefont {A.}~\bibnamefont {Menzel}}, \bibinfo {author}
  {\bibfnamefont {I.}~\bibnamefont {Vartiainen}}, \bibinfo {author}
  {\bibfnamefont {E.}~\bibnamefont {M\"uller}}, \bibinfo {author}
  {\bibfnamefont {E.}~\bibnamefont {Kirk}}, \bibinfo {author} {\bibfnamefont
  {S.}~\bibnamefont {Gliga}}, \bibinfo {author} {\bibfnamefont
  {J.}~\bibnamefont {Raabe}}, \ and\ \bibinfo {author} {\bibfnamefont {L.~J.}\
  \bibnamefont {Heyderman}},\ }\href {\doibase 10.1103/PhysRevLett.114.115501}
  {\bibfield  {journal} {\bibinfo  {journal} {Phys. Rev. Lett.}\ }\textbf
  {\bibinfo {volume} {114}},\ \bibinfo {pages} {115501} (\bibinfo {year}
  {2015})}\BibitemShut {NoStop}%
\bibitem [{\citenamefont {Gliga}\ \emph {et~al.}(2019)\citenamefont {Gliga},
  \citenamefont {Seniutinas}, \citenamefont {Weber},\ and\ \citenamefont
  {David}}]{Gliga2019MT}%
  \BibitemOpen
  \bibfield  {author} {\bibinfo {author} {\bibfnamefont {S.}~\bibnamefont
  {Gliga}}, \bibinfo {author} {\bibfnamefont {G.}~\bibnamefont {Seniutinas}},
  \bibinfo {author} {\bibfnamefont {A.}~\bibnamefont {Weber}}, \ and\ \bibinfo
  {author} {\bibfnamefont {C.}~\bibnamefont {David}},\ }\href {\doibase
  https://doi.org/10.1016/j.mattod.2019.05.001} {\bibfield  {journal} {\bibinfo
   {journal} {Materials Today}\ }\textbf {\bibinfo {volume} {26}},\ \bibinfo
  {pages} {100 } (\bibinfo {year} {2019})}\BibitemShut {NoStop}%
\bibitem [{\citenamefont {Streubel}\ \emph {et~al.}(2015)\citenamefont
  {Streubel}, \citenamefont {Kronast}, \citenamefont {Fischer}, \citenamefont
  {Parkinson}, \citenamefont {Schmidt},\ and\ \citenamefont
  {Makarov}}]{Streubel2015}%
  \BibitemOpen
  \bibfield  {author} {\bibinfo {author} {\bibfnamefont {R.}~\bibnamefont
  {Streubel}}, \bibinfo {author} {\bibfnamefont {F.}~\bibnamefont {Kronast}},
  \bibinfo {author} {\bibfnamefont {P.}~\bibnamefont {Fischer}}, \bibinfo
  {author} {\bibfnamefont {D.}~\bibnamefont {Parkinson}}, \bibinfo {author}
  {\bibfnamefont {O.~G.}\ \bibnamefont {Schmidt}}, \ and\ \bibinfo {author}
  {\bibfnamefont {D.}~\bibnamefont {Makarov}},\ }\href {\doibase
  10.1038/ncomms8612} {\bibfield  {journal} {\bibinfo  {journal} {Nat.
  Commun.}\ }\textbf {\bibinfo {volume} {6}},\ \bibinfo {pages} {7612}
  (\bibinfo {year} {2015})}\BibitemShut {NoStop}%
\bibitem [{\citenamefont {Donnelly}\ \emph {et~al.}(2017)\citenamefont
  {Donnelly}, \citenamefont {Guizar-Sicairos}, \citenamefont {Scagnoli},
  \citenamefont {Gliga}, \citenamefont {Holler}, , \citenamefont {Raabe},\ and\
  \citenamefont {Heyderman}}]{Donnelly2017}%
  \BibitemOpen
  \bibfield  {author} {\bibinfo {author} {\bibfnamefont {C.}~\bibnamefont
  {Donnelly}}, \bibinfo {author} {\bibfnamefont {M.}~\bibnamefont
  {Guizar-Sicairos}}, \bibinfo {author} {\bibfnamefont {V.}~\bibnamefont
  {Scagnoli}}, \bibinfo {author} {\bibfnamefont {S.}~\bibnamefont {Gliga}},
  \bibinfo {author} {\bibfnamefont {M.}~\bibnamefont {Holler}}, , \bibinfo
  {author} {\bibfnamefont {J.}~\bibnamefont {Raabe}}, \ and\ \bibinfo {author}
  {\bibfnamefont {L.~J.}\ \bibnamefont {Heyderman}},\ }\href {\doibase
  10.1038/nature23006} {\bibfield  {journal} {\bibinfo  {journal} {Nature}\
  }\textbf {\bibinfo {volume} {547}},\ \bibinfo {pages} {328} (\bibinfo {year}
  {2017})}\BibitemShut {NoStop}%
\bibitem [{\citenamefont {Witte}\ \emph {et~al.}(2020)\citenamefont {Witte},
  \citenamefont {Sp\"{a}th}, \citenamefont {Finizio}, \citenamefont {Donnelly},
  \citenamefont {Watts}, \citenamefont {Sarafimov}, \citenamefont {Odstrcil},
  \citenamefont {Guizar-Sicairos}, \citenamefont {Holler}, \citenamefont
  {Fink},\ and\ \citenamefont {Raabe}}]{Witte2020}%
  \BibitemOpen
  \bibfield  {author} {\bibinfo {author} {\bibfnamefont {K.}~\bibnamefont
  {Witte}}, \bibinfo {author} {\bibfnamefont {A.}~\bibnamefont {Sp\"{a}th}},
  \bibinfo {author} {\bibfnamefont {S.}~\bibnamefont {Finizio}}, \bibinfo
  {author} {\bibfnamefont {C.}~\bibnamefont {Donnelly}}, \bibinfo {author}
  {\bibfnamefont {B.}~\bibnamefont {Watts}}, \bibinfo {author} {\bibfnamefont
  {B.}~\bibnamefont {Sarafimov}}, \bibinfo {author} {\bibfnamefont
  {M.}~\bibnamefont {Odstrcil}}, \bibinfo {author} {\bibfnamefont
  {M.}~\bibnamefont {Guizar-Sicairos}}, \bibinfo {author} {\bibfnamefont
  {M.}~\bibnamefont {Holler}}, \bibinfo {author} {\bibfnamefont {R.~H.}\
  \bibnamefont {Fink}}, \ and\ \bibinfo {author} {\bibfnamefont
  {J.}~\bibnamefont {Raabe}},\ }\href {\doibase 10.1021/acs.nanolett.9b04782}
  {\bibfield  {journal} {\bibinfo  {journal} {Nano Lett.}\ }\textbf {\bibinfo
  {volume} {20}},\ \bibinfo {pages} {1305} (\bibinfo {year} {2020})},\ \Eprint
  {http://arxiv.org/abs/https://doi.org/10.1021/acs.nanolett.9b04782}
  {https://doi.org/10.1021/acs.nanolett.9b04782} \BibitemShut {NoStop}%
\bibitem [{\citenamefont {Manke}\ \emph {et~al.}(2010)\citenamefont {Manke},
  \citenamefont {Kardjilov}, \citenamefont {Schäfer}, \citenamefont {Hilger},
  \citenamefont {Strobl}, \citenamefont {Dawson}, \citenamefont
  {Gr\"{u}nzweig}, \citenamefont {Behr}, \citenamefont {Hentschel},
  \citenamefont {David}, \citenamefont {Kupsch}, \citenamefont {Lange},\ and\
  \citenamefont {Banhart}}]{Manke2010}%
  \BibitemOpen
  \bibfield  {author} {\bibinfo {author} {\bibfnamefont {I.}~\bibnamefont
  {Manke}}, \bibinfo {author} {\bibfnamefont {N.}~\bibnamefont {Kardjilov}},
  \bibinfo {author} {\bibfnamefont {R.}~\bibnamefont {Schäfer}}, \bibinfo
  {author} {\bibfnamefont {A.}~\bibnamefont {Hilger}}, \bibinfo {author}
  {\bibfnamefont {M.}~\bibnamefont {Strobl}}, \bibinfo {author} {\bibfnamefont
  {M.}~\bibnamefont {Dawson}}, \bibinfo {author} {\bibfnamefont
  {C.}~\bibnamefont {Gr\"{u}nzweig}}, \bibinfo {author} {\bibfnamefont
  {G.}~\bibnamefont {Behr}}, \bibinfo {author} {\bibfnamefont {M.}~\bibnamefont
  {Hentschel}}, \bibinfo {author} {\bibfnamefont {C.}~\bibnamefont {David}},
  \bibinfo {author} {\bibfnamefont {A.}~\bibnamefont {Kupsch}}, \bibinfo
  {author} {\bibfnamefont {A.}~\bibnamefont {Lange}}, \ and\ \bibinfo {author}
  {\bibfnamefont {J.}~\bibnamefont {Banhart}},\ }\href {\doibase
  10.1038/ncomms1125} {\bibfield  {journal} {\bibinfo  {journal} {Nat.
  Commun.}\ }\textbf {\bibinfo {volume} {1}},\ \bibinfo {pages} {125} (\bibinfo
  {year} {2010})}\BibitemShut {NoStop}%
\bibitem [{\citenamefont {Phatak}\ \emph {et~al.}(2014)\citenamefont {Phatak},
  \citenamefont {Liu}, \citenamefont {Gulsoy}, \citenamefont {Schmidt},
  \citenamefont {Franke-Schubert},\ and\ \citenamefont
  {Petford-Long}}]{Phatak2014}%
  \BibitemOpen
  \bibfield  {author} {\bibinfo {author} {\bibfnamefont {C.}~\bibnamefont
  {Phatak}}, \bibinfo {author} {\bibfnamefont {Y.}~\bibnamefont {Liu}},
  \bibinfo {author} {\bibfnamefont {E.~B.}\ \bibnamefont {Gulsoy}}, \bibinfo
  {author} {\bibfnamefont {D.}~\bibnamefont {Schmidt}}, \bibinfo {author}
  {\bibfnamefont {E.}~\bibnamefont {Franke-Schubert}}, \ and\ \bibinfo {author}
  {\bibfnamefont {A.}~\bibnamefont {Petford-Long}},\ }\href {\doibase
  10.1021/nl404071u} {\bibfield  {journal} {\bibinfo  {journal} {Nano Letters}\
  }\textbf {\bibinfo {volume} {14}},\ \bibinfo {pages} {759} (\bibinfo {year}
  {2014})},\ \Eprint {http://arxiv.org/abs/https://doi.org/10.1021/nl404071u}
  {https://doi.org/10.1021/nl404071u} \BibitemShut {NoStop}%
\bibitem [{\citenamefont {Wolf}\ \emph {et~al.}(2019)\citenamefont {Wolf},
  \citenamefont {Biziere}, \citenamefont {Sturm}, \citenamefont {Reyes},
  \citenamefont {Wade}, \citenamefont {Niermann}, \citenamefont {Krehl},
  \citenamefont {Warot-Fonrose}, \citenamefont {B\"{u}chner}, \citenamefont
  {Snoeck}, \citenamefont {Gatel},\ and\ \citenamefont {Lubk}}]{Biziere2019}%
  \BibitemOpen
  \bibfield  {author} {\bibinfo {author} {\bibfnamefont {D.}~\bibnamefont
  {Wolf}}, \bibinfo {author} {\bibfnamefont {N.}~\bibnamefont {Biziere}},
  \bibinfo {author} {\bibfnamefont {S.}~\bibnamefont {Sturm}}, \bibinfo
  {author} {\bibfnamefont {D.}~\bibnamefont {Reyes}}, \bibinfo {author}
  {\bibfnamefont {T.}~\bibnamefont {Wade}}, \bibinfo {author} {\bibfnamefont
  {T.}~\bibnamefont {Niermann}}, \bibinfo {author} {\bibfnamefont
  {J.}~\bibnamefont {Krehl}}, \bibinfo {author} {\bibfnamefont
  {B.}~\bibnamefont {Warot-Fonrose}}, \bibinfo {author} {\bibfnamefont
  {B.}~\bibnamefont {B\"{u}chner}}, \bibinfo {author} {\bibfnamefont
  {E.}~\bibnamefont {Snoeck}}, \bibinfo {author} {\bibfnamefont
  {C.}~\bibnamefont {Gatel}}, \ and\ \bibinfo {author} {\bibfnamefont
  {A.}~\bibnamefont {Lubk}},\ }\href {\doibase 10.1038/s42005-019-0187-8}
  {\bibfield  {journal} {\bibinfo  {journal} {Commun. Phys.}\ }\textbf
  {\bibinfo {volume} {2}},\ \bibinfo {pages} {87} (\bibinfo {year}
  {2019})}\BibitemShut {NoStop}%
\bibitem [{\citenamefont {Donnelly}\ \emph {et~al.}(2020)\citenamefont
  {Donnelly}, \citenamefont {Finizio}, \citenamefont {Gliga}, \citenamefont
  {Holler}, \citenamefont {Hrabec}, \citenamefont {Odstr\v{c}il}, \citenamefont
  {Mayr}, \citenamefont {Scagnoli}, \citenamefont {Heyderman}, \citenamefont
  {Guizar-Sicairos},\ and\ \citenamefont {Raabe}}]{Donnelly2020trl}%
  \BibitemOpen
  \bibfield  {author} {\bibinfo {author} {\bibfnamefont {C.}~\bibnamefont
  {Donnelly}}, \bibinfo {author} {\bibfnamefont {S.}~\bibnamefont {Finizio}},
  \bibinfo {author} {\bibfnamefont {S.}~\bibnamefont {Gliga}}, \bibinfo
  {author} {\bibfnamefont {M.}~\bibnamefont {Holler}}, \bibinfo {author}
  {\bibfnamefont {A.}~\bibnamefont {Hrabec}}, \bibinfo {author} {\bibfnamefont
  {M.}~\bibnamefont {Odstr\v{c}il}}, \bibinfo {author} {\bibfnamefont
  {S.}~\bibnamefont {Mayr}}, \bibinfo {author} {\bibfnamefont {V.}~\bibnamefont
  {Scagnoli}}, \bibinfo {author} {\bibfnamefont {L.~J.}\ \bibnamefont
  {Heyderman}}, \bibinfo {author} {\bibfnamefont {M.}~\bibnamefont
  {Guizar-Sicairos}}, \ and\ \bibinfo {author} {\bibfnamefont {J.}~\bibnamefont
  {Raabe}},\ }\href {\doibase 10.1038/s41565-020-0649-x} {\bibfield  {journal}
  {\bibinfo  {journal} {Nat. Nanotechnol.}\ } (\bibinfo {year} {2020}),\
  10.1038/s41565-020-0649-x}\BibitemShut {NoStop}%
\bibitem [{\citenamefont {Lenz}\ \emph {et~al.}(2019)\citenamefont {Lenz},
  \citenamefont {Narkowicz}, \citenamefont {Wagner}, \citenamefont {Reiche},
  \citenamefont {Körner}, \citenamefont {Schneider}, \citenamefont {Kákay},
  \citenamefont {Schultheiss}, \citenamefont {Weissker}, \citenamefont {Wolf},
  \citenamefont {Suter}, \citenamefont {Büchner}, \citenamefont {Fassbender},
  \citenamefont {Mühl},\ and\ \citenamefont {Lindner}}]{Lenz2019}%
  \BibitemOpen
  \bibfield  {author} {\bibinfo {author} {\bibfnamefont {K.}~\bibnamefont
  {Lenz}}, \bibinfo {author} {\bibfnamefont {R.}~\bibnamefont {Narkowicz}},
  \bibinfo {author} {\bibfnamefont {K.}~\bibnamefont {Wagner}}, \bibinfo
  {author} {\bibfnamefont {C.~F.}\ \bibnamefont {Reiche}}, \bibinfo {author}
  {\bibfnamefont {J.}~\bibnamefont {Körner}}, \bibinfo {author} {\bibfnamefont
  {T.}~\bibnamefont {Schneider}}, \bibinfo {author} {\bibfnamefont
  {A.}~\bibnamefont {Kákay}}, \bibinfo {author} {\bibfnamefont
  {H.}~\bibnamefont {Schultheiss}}, \bibinfo {author} {\bibfnamefont
  {U.}~\bibnamefont {Weissker}}, \bibinfo {author} {\bibfnamefont
  {D.}~\bibnamefont {Wolf}}, \bibinfo {author} {\bibfnamefont {D.}~\bibnamefont
  {Suter}}, \bibinfo {author} {\bibfnamefont {B.}~\bibnamefont {Büchner}},
  \bibinfo {author} {\bibfnamefont {J.}~\bibnamefont {Fassbender}}, \bibinfo
  {author} {\bibfnamefont {T.}~\bibnamefont {Mühl}}, \ and\ \bibinfo {author}
  {\bibfnamefont {J.}~\bibnamefont {Lindner}},\ }\href {\doibase
  10.1002/smll.201904315} {\bibfield  {journal} {\bibinfo  {journal} {Small}\
  }\textbf {\bibinfo {volume} {15}},\ \bibinfo {pages} {1904315} (\bibinfo
  {year} {2019})}\BibitemShut {NoStop}%
\bibitem [{\citenamefont {Vansteenkiste}\ \emph {et~al.}(2014)\citenamefont
  {Vansteenkiste}, \citenamefont {Leliaert}, \citenamefont {Dvornik},
  \citenamefont {Helsen}, \citenamefont {Garcia-Sanchez},\ and\ \citenamefont
  {Van~Waeyenberge}}]{Vansteenkiste2014}%
  \BibitemOpen
  \bibfield  {author} {\bibinfo {author} {\bibfnamefont {A.}~\bibnamefont
  {Vansteenkiste}}, \bibinfo {author} {\bibfnamefont {J.}~\bibnamefont
  {Leliaert}}, \bibinfo {author} {\bibfnamefont {M.}~\bibnamefont {Dvornik}},
  \bibinfo {author} {\bibfnamefont {M.}~\bibnamefont {Helsen}}, \bibinfo
  {author} {\bibfnamefont {F.}~\bibnamefont {Garcia-Sanchez}}, \ and\ \bibinfo
  {author} {\bibfnamefont {B.}~\bibnamefont {Van~Waeyenberge}},\ }\href@noop {}
  {\bibfield  {journal} {\bibinfo  {journal} {AIP Advances}\ }\textbf {\bibinfo
  {volume} {4}},\ \bibinfo {pages} {107133} (\bibinfo {year}
  {2014})}\BibitemShut {NoStop}%
\bibitem [{\citenamefont {Schrefl}\ \emph {et~al.}(2003)\citenamefont
  {Schrefl}, \citenamefont {Suess}, \citenamefont {Scholz}, \citenamefont
  {Forster}, \citenamefont {Tsiantos},\ and\ \citenamefont {J.}}]{Schrefl2003}%
  \BibitemOpen
  \bibfield  {author} {\bibinfo {author} {\bibfnamefont {T.}~\bibnamefont
  {Schrefl}}, \bibinfo {author} {\bibfnamefont {D.}~\bibnamefont {Suess}},
  \bibinfo {author} {\bibfnamefont {W.}~\bibnamefont {Scholz}}, \bibinfo
  {author} {\bibfnamefont {H.}~\bibnamefont {Forster}}, \bibinfo {author}
  {\bibfnamefont {V.}~\bibnamefont {Tsiantos}}, \ and\ \bibinfo {author}
  {\bibfnamefont {F.}~\bibnamefont {J.}},\ }in\ \href@noop {} {\emph {\bibinfo
  {booktitle} {Computational Electromagnetics. Lecture Notes in Computational
  Science and Engineering}}},\ \bibinfo {editor} {edited by\ \bibinfo {editor}
  {\bibfnamefont {P.}~\bibnamefont {Monk}}, \bibinfo {editor} {\bibfnamefont
  {C.}~\bibnamefont {Carstensen}}, \bibinfo {editor} {\bibfnamefont
  {S.}~\bibnamefont {Funken}}, \bibinfo {editor} {\bibfnamefont
  {W.}~\bibnamefont {Hackbusch}}, \ and\ \bibinfo {editor} {\bibfnamefont
  {H.}~\bibnamefont {R.H.W.}}}\ (\bibinfo  {publisher} {Springer},\ \bibinfo
  {address} {Berlin, Heidelberg},\ \bibinfo {year} {2003})\BibitemShut
  {NoStop}%
\bibitem [{\citenamefont {Miao}(2019)}]{miao2019}%
  \BibitemOpen
  \bibfield  {author} {\bibinfo {author} {\bibfnamefont {J.-Y.}\ \bibnamefont
  {Miao}},\ }\href@noop {} {\bibfield  {journal} {\bibinfo  {journal} {Chinese
  Physics Letters}\ }\textbf {\bibinfo {volume} {36}},\ \bibinfo {pages}
  {097501} (\bibinfo {year} {2019})}\BibitemShut {NoStop}%
\bibitem [{\citenamefont {Exl}\ \emph {et~al.}(2020)\citenamefont {Exl},
  \citenamefont {Mauser}, \citenamefont {Schrefl},\ and\ \citenamefont
  {Suess}}]{exl2019}%
  \BibitemOpen
  \bibfield  {author} {\bibinfo {author} {\bibfnamefont {L.}~\bibnamefont
  {Exl}}, \bibinfo {author} {\bibfnamefont {N.~J.}\ \bibnamefont {Mauser}},
  \bibinfo {author} {\bibfnamefont {T.}~\bibnamefont {Schrefl}}, \ and\
  \bibinfo {author} {\bibfnamefont {D.}~\bibnamefont {Suess}},\ }\href
  {\doibase https://doi.org/10.1016/j.cnsns.2020.105205} {\bibfield  {journal}
  {\bibinfo  {journal} {Communications in Nonlinear Science and Numerical
  Simulation}\ }\textbf {\bibinfo {volume} {84}},\ \bibinfo {pages} {105205}
  (\bibinfo {year} {2020})}\BibitemShut {NoStop}%
\bibitem [{\citenamefont {Rych\l{}y}\ \emph
  {et~al.}(2015{\natexlab{a}})\citenamefont {Rych\l{}y}, \citenamefont
  {Gruszecki}, \citenamefont {Mruczkiewicz}, \citenamefont {Kłos},
  \citenamefont {Mamica},\ and\ \citenamefont {Krawczyk}}]{Rychly2015MC}%
  \BibitemOpen
  \bibfield  {author} {\bibinfo {author} {\bibfnamefont {J.}~\bibnamefont
  {Rych\l{}y}}, \bibinfo {author} {\bibfnamefont {P.}~\bibnamefont
  {Gruszecki}}, \bibinfo {author} {\bibfnamefont {M.}~\bibnamefont
  {Mruczkiewicz}}, \bibinfo {author} {\bibfnamefont {J.~W.}\ \bibnamefont
  {Kłos}}, \bibinfo {author} {\bibfnamefont {S.}~\bibnamefont {Mamica}}, \
  and\ \bibinfo {author} {\bibfnamefont {M.}~\bibnamefont {Krawczyk}},\ }\href
  {\doibase 10.1063/1.4932348} {\bibfield  {journal} {\bibinfo  {journal} {Low
  Temperature Physics}\ }\textbf {\bibinfo {volume} {41}},\ \bibinfo {pages}
  {745} (\bibinfo {year} {2015}{\natexlab{a}})},\ \Eprint
  {http://arxiv.org/abs/https://doi.org/10.1063/1.4932348}
  {https://doi.org/10.1063/1.4932348} \BibitemShut {NoStop}%
\bibitem [{\citenamefont {Slavin}\ and\ \citenamefont
  {Tiberkevich}(2009)}]{Slavin2009}%
  \BibitemOpen
  \bibfield  {author} {\bibinfo {author} {\bibfnamefont {A.}~\bibnamefont
  {Slavin}}\ and\ \bibinfo {author} {\bibfnamefont {V.}~\bibnamefont
  {Tiberkevich}},\ }\href@noop {} {\bibfield  {journal} {\bibinfo  {journal}
  {Magnetics, IEEE Transactions on}\ }\textbf {\bibinfo {volume} {45}},\
  \bibinfo {pages} {1875 } (\bibinfo {year} {2009})}\BibitemShut {NoStop}%
\bibitem [{\citenamefont {Chisnell}\ \emph {et~al.}(2015)\citenamefont
  {Chisnell}, \citenamefont {Helton}, \citenamefont {Freedman}, \citenamefont
  {Singh}, \citenamefont {Bewley}, \citenamefont {Nocera},\ and\ \citenamefont
  {Lee}}]{Chisnell2015}%
  \BibitemOpen
  \bibfield  {author} {\bibinfo {author} {\bibfnamefont {R.}~\bibnamefont
  {Chisnell}}, \bibinfo {author} {\bibfnamefont {J.~S.}\ \bibnamefont
  {Helton}}, \bibinfo {author} {\bibfnamefont {D.~E.}\ \bibnamefont
  {Freedman}}, \bibinfo {author} {\bibfnamefont {D.~K.}\ \bibnamefont {Singh}},
  \bibinfo {author} {\bibfnamefont {R.~I.}\ \bibnamefont {Bewley}}, \bibinfo
  {author} {\bibfnamefont {D.~G.}\ \bibnamefont {Nocera}}, \ and\ \bibinfo
  {author} {\bibfnamefont {Y.~S.}\ \bibnamefont {Lee}},\ }\href@noop {}
  {\bibfield  {journal} {\bibinfo  {journal} {Phys. Rev. Lett.}\ }\textbf
  {\bibinfo {volume} {115}},\ \bibinfo {pages} {147201} (\bibinfo {year}
  {2015})}\BibitemShut {NoStop}%
\bibitem [{\citenamefont {Wang}\ \emph
  {et~al.}(2018{\natexlab{b}})\citenamefont {Wang}, \citenamefont {Zhang},\
  and\ \citenamefont {Wang}}]{Wang2018_pra}%
  \BibitemOpen
  \bibfield  {author} {\bibinfo {author} {\bibfnamefont {X.~S.}\ \bibnamefont
  {Wang}}, \bibinfo {author} {\bibfnamefont {H.~W.}\ \bibnamefont {Zhang}}, \
  and\ \bibinfo {author} {\bibfnamefont {X.~R.}\ \bibnamefont {Wang}},\
  }\href@noop {} {\bibfield  {journal} {\bibinfo  {journal} {Phys. Rev.
  Applied}\ }\textbf {\bibinfo {volume} {9}},\ \bibinfo {pages} {024029}
  (\bibinfo {year} {2018}{\natexlab{b}})}\BibitemShut {NoStop}%
\bibitem [{\citenamefont {Owerre}(2017)}]{Owerre2017}%
  \BibitemOpen
  \bibfield  {author} {\bibinfo {author} {\bibfnamefont {S.~A.}\ \bibnamefont
  {Owerre}},\ }\href@noop {} {\bibfield  {journal} {\bibinfo  {journal} {J.
  Phys. Commun.}\ }\textbf {\bibinfo {volume} {1}},\ \bibinfo {pages} {021002}
  (\bibinfo {year} {2017})}\BibitemShut {NoStop}%
\bibitem [{\citenamefont {Rych\l{}y}\ \emph
  {et~al.}(2015{\natexlab{b}})\citenamefont {Rych\l{}y}, \citenamefont
  {K\l{}os}, \citenamefont {Mruczkiewicz},\ and\ \citenamefont
  {Krawczyk}}]{Rychly2015}%
  \BibitemOpen
  \bibfield  {author} {\bibinfo {author} {\bibfnamefont {J.}~\bibnamefont
  {Rych\l{}y}}, \bibinfo {author} {\bibfnamefont {J.~W.}\ \bibnamefont
  {K\l{}os}}, \bibinfo {author} {\bibfnamefont {M.}~\bibnamefont
  {Mruczkiewicz}}, \ and\ \bibinfo {author} {\bibfnamefont {M.}~\bibnamefont
  {Krawczyk}},\ }\href@noop {} {\bibfield  {journal} {\bibinfo  {journal}
  {Phys. Rev. B}\ }\textbf {\bibinfo {volume} {92}},\ \bibinfo {pages} {054414}
  (\bibinfo {year} {2015}{\natexlab{b}})}\BibitemShut {NoStop}%
\bibitem [{\citenamefont {Lisiecki}\ \emph {et~al.}(2019)\citenamefont
  {Lisiecki}, \citenamefont {Rych\l{}y}, \citenamefont
  {Ku\ifmmode~\acute{s}\else \'{s}\fi{}wik}, \citenamefont
  {G\l{}owi\ifmmode~\acute{n}\else \'{n}\fi{}ski}, \citenamefont {K\l{}os},
  \citenamefont {Gro\ss{}}, \citenamefont {Tr\"ager}, \citenamefont {Bykova},
  \citenamefont {Weigand}, \citenamefont {Zelent}, \citenamefont {Goering},
  \citenamefont {Sch\"utz}, \citenamefont {Krawczyk}, \citenamefont
  {Stobiecki}, \citenamefont {Dubowik},\ and\ \citenamefont
  {Gr\"afe}}]{Lisiecki2019}%
  \BibitemOpen
  \bibfield  {author} {\bibinfo {author} {\bibfnamefont {F.}~\bibnamefont
  {Lisiecki}}, \bibinfo {author} {\bibfnamefont {J.}~\bibnamefont {Rych\l{}y}},
  \bibinfo {author} {\bibfnamefont {P.}~\bibnamefont {Ku\ifmmode~\acute{s}\else
  \'{s}\fi{}wik}}, \bibinfo {author} {\bibfnamefont {H.}~\bibnamefont
  {G\l{}owi\ifmmode~\acute{n}\else \'{n}\fi{}ski}}, \bibinfo {author}
  {\bibfnamefont {J.~W.}\ \bibnamefont {K\l{}os}}, \bibinfo {author}
  {\bibfnamefont {F.}~\bibnamefont {Gro\ss{}}}, \bibinfo {author}
  {\bibfnamefont {N.}~\bibnamefont {Tr\"ager}}, \bibinfo {author}
  {\bibfnamefont {I.}~\bibnamefont {Bykova}}, \bibinfo {author} {\bibfnamefont
  {M.}~\bibnamefont {Weigand}}, \bibinfo {author} {\bibfnamefont
  {M.}~\bibnamefont {Zelent}}, \bibinfo {author} {\bibfnamefont {E.~J.}\
  \bibnamefont {Goering}}, \bibinfo {author} {\bibfnamefont {G.}~\bibnamefont
  {Sch\"utz}}, \bibinfo {author} {\bibfnamefont {M.}~\bibnamefont {Krawczyk}},
  \bibinfo {author} {\bibfnamefont {F.}~\bibnamefont {Stobiecki}}, \bibinfo
  {author} {\bibfnamefont {J.}~\bibnamefont {Dubowik}}, \ and\ \bibinfo
  {author} {\bibfnamefont {J.}~\bibnamefont {Gr\"afe}},\ }\href@noop {}
  {\bibfield  {journal} {\bibinfo  {journal} {Phys. Rev. Applied}\ }\textbf
  {\bibinfo {volume} {11}},\ \bibinfo {pages} {054061} (\bibinfo {year}
  {2019})}\BibitemShut {NoStop}%
\bibitem [{\citenamefont {Bhat}\ \emph {et~al.}(2013)\citenamefont {Bhat},
  \citenamefont {Sklenar}, \citenamefont {Farmer}, \citenamefont {Woods},
  \citenamefont {Hastings}, \citenamefont {Lee}, \citenamefont {Ketterson},\
  and\ \citenamefont {De~Long}}]{Bhat2013}%
  \BibitemOpen
  \bibfield  {author} {\bibinfo {author} {\bibfnamefont {V.~S.}\ \bibnamefont
  {Bhat}}, \bibinfo {author} {\bibfnamefont {J.}~\bibnamefont {Sklenar}},
  \bibinfo {author} {\bibfnamefont {B.}~\bibnamefont {Farmer}}, \bibinfo
  {author} {\bibfnamefont {J.}~\bibnamefont {Woods}}, \bibinfo {author}
  {\bibfnamefont {J.~T.}\ \bibnamefont {Hastings}}, \bibinfo {author}
  {\bibfnamefont {S.~J.}\ \bibnamefont {Lee}}, \bibinfo {author} {\bibfnamefont
  {J.~B.}\ \bibnamefont {Ketterson}}, \ and\ \bibinfo {author} {\bibfnamefont
  {L.~E.}\ \bibnamefont {De~Long}},\ }\href@noop {} {\bibfield  {journal}
  {\bibinfo  {journal} {Phys. Rev. Lett.}\ }\textbf {\bibinfo {volume} {111}},\
  \bibinfo {pages} {077201} (\bibinfo {year} {2013})}\BibitemShut {NoStop}%
\bibitem [{\citenamefont {Shi}\ \emph {et~al.}(2018)\citenamefont {Shi},
  \citenamefont {Budrikis}, \citenamefont {Stein}, \citenamefont {Morley},
  \citenamefont {Olmsted}, \citenamefont {Burnell},\ and\ \citenamefont
  {Marrows}}]{Shi2018}%
  \BibitemOpen
  \bibfield  {author} {\bibinfo {author} {\bibfnamefont {D.}~\bibnamefont
  {Shi}}, \bibinfo {author} {\bibfnamefont {Z.}~\bibnamefont {Budrikis}},
  \bibinfo {author} {\bibfnamefont {A.}~\bibnamefont {Stein}}, \bibinfo
  {author} {\bibfnamefont {S.~A.}\ \bibnamefont {Morley}}, \bibinfo {author}
  {\bibfnamefont {P.~D.}\ \bibnamefont {Olmsted}}, \bibinfo {author}
  {\bibfnamefont {G.}~\bibnamefont {Burnell}}, \ and\ \bibinfo {author}
  {\bibfnamefont {C.~H.}\ \bibnamefont {Marrows}},\ }\href@noop {} {\bibfield
  {journal} {\bibinfo  {journal} {Nature Physics}\ }\textbf {\bibinfo {volume}
  {14}},\ \bibinfo {pages} {309 } (\bibinfo {year} {2018})}\BibitemShut
  {NoStop}%
\bibitem [{\citenamefont {Hertel}\ \emph {et~al.}(2004)\citenamefont {Hertel},
  \citenamefont {Wulfhekel},\ and\ \citenamefont {Kirschner}}]{Hertel2004}%
  \BibitemOpen
  \bibfield  {author} {\bibinfo {author} {\bibfnamefont {R.}~\bibnamefont
  {Hertel}}, \bibinfo {author} {\bibfnamefont {W.}~\bibnamefont {Wulfhekel}}, \
  and\ \bibinfo {author} {\bibfnamefont {J.}~\bibnamefont {Kirschner}},\ }\href
  {\doibase 10.1103/PhysRevLett.93.257202} {\bibfield  {journal} {\bibinfo
  {journal} {Phys. Rev. Lett.}\ }\textbf {\bibinfo {volume} {93}},\ \bibinfo
  {pages} {257202} (\bibinfo {year} {2004})}\BibitemShut {NoStop}%
\bibitem [{\citenamefont {Schneider}\ \emph {et~al.}(2008)\citenamefont
  {Schneider}, \citenamefont {Serga}, \citenamefont {Leven}, \citenamefont
  {Hillebrands}, \citenamefont {Stamps},\ and\ \citenamefont
  {Kostylev}}]{schneider2008realization}%
  \BibitemOpen
  \bibfield  {author} {\bibinfo {author} {\bibfnamefont {T.}~\bibnamefont
  {Schneider}}, \bibinfo {author} {\bibfnamefont {A.~A.}\ \bibnamefont
  {Serga}}, \bibinfo {author} {\bibfnamefont {B.}~\bibnamefont {Leven}},
  \bibinfo {author} {\bibfnamefont {B.}~\bibnamefont {Hillebrands}}, \bibinfo
  {author} {\bibfnamefont {R.~L.}\ \bibnamefont {Stamps}}, \ and\ \bibinfo
  {author} {\bibfnamefont {M.~P.}\ \bibnamefont {Kostylev}},\ }\href@noop {}
  {\bibfield  {journal} {\bibinfo  {journal} {Applied Physics Letters}\
  }\textbf {\bibinfo {volume} {92}},\ \bibinfo {pages} {022505} (\bibinfo
  {year} {2008})}\BibitemShut {NoStop}%
\bibitem [{\citenamefont {Vogt}\ \emph {et~al.}(2014)\citenamefont {Vogt},
  \citenamefont {Fradin}, \citenamefont {Pearson}, \citenamefont {Sebastian},
  \citenamefont {Bader}, \citenamefont {Hillebrands}, \citenamefont
  {Hoffmann},\ and\ \citenamefont {Schultheiss}}]{vogt2014realization}%
  \BibitemOpen
  \bibfield  {author} {\bibinfo {author} {\bibfnamefont {K.}~\bibnamefont
  {Vogt}}, \bibinfo {author} {\bibfnamefont {F.~Y.}\ \bibnamefont {Fradin}},
  \bibinfo {author} {\bibfnamefont {J.~E.}\ \bibnamefont {Pearson}}, \bibinfo
  {author} {\bibfnamefont {T.}~\bibnamefont {Sebastian}}, \bibinfo {author}
  {\bibfnamefont {S.~D.}\ \bibnamefont {Bader}}, \bibinfo {author}
  {\bibfnamefont {B.}~\bibnamefont {Hillebrands}}, \bibinfo {author}
  {\bibfnamefont {A.}~\bibnamefont {Hoffmann}}, \ and\ \bibinfo {author}
  {\bibfnamefont {H.}~\bibnamefont {Schultheiss}},\ }\href@noop {} {\bibfield
  {journal} {\bibinfo  {journal} {Nature communications}\ }\textbf {\bibinfo
  {volume} {5}},\ \bibinfo {pages} {1} (\bibinfo {year} {2014})}\BibitemShut
  {NoStop}%
\bibitem [{\citenamefont {Vogt}\ \emph {et~al.}(2012)\citenamefont {Vogt},
  \citenamefont {Schultheiss}, \citenamefont {Jain}, \citenamefont {Pearson},
  \citenamefont {Hoffmann}, \citenamefont {Bader},\ and\ \citenamefont
  {Hillebrands}}]{Vogt2010}%
  \BibitemOpen
  \bibfield  {author} {\bibinfo {author} {\bibfnamefont {K.}~\bibnamefont
  {Vogt}}, \bibinfo {author} {\bibfnamefont {H.}~\bibnamefont {Schultheiss}},
  \bibinfo {author} {\bibfnamefont {S.}~\bibnamefont {Jain}}, \bibinfo {author}
  {\bibfnamefont {J.~E.}\ \bibnamefont {Pearson}}, \bibinfo {author}
  {\bibfnamefont {A.}~\bibnamefont {Hoffmann}}, \bibinfo {author}
  {\bibfnamefont {S.~D.}\ \bibnamefont {Bader}}, \ and\ \bibinfo {author}
  {\bibfnamefont {B.}~\bibnamefont {Hillebrands}},\ }\href {\doibase
  10.1063/1.4738887} {\bibfield  {journal} {\bibinfo  {journal} {Applied
  Physics Letters}\ }\textbf {\bibinfo {volume} {101}},\ \bibinfo {pages}
  {042410} (\bibinfo {year} {2012})},\ \Eprint
  {http://arxiv.org/abs/https://doi.org/10.1063/1.4738887}
  {https://doi.org/10.1063/1.4738887} \BibitemShut {NoStop}%
\bibitem [{\citenamefont {Zakeri}\ \emph {et~al.}(2010)\citenamefont {Zakeri},
  \citenamefont {Zhang}, \citenamefont {Prokop}, \citenamefont {Chuang},
  \citenamefont {Sakr}, \citenamefont {Tang},\ and\ \citenamefont
  {Kirschner}}]{Zakeri2010}%
  \BibitemOpen
  \bibfield  {author} {\bibinfo {author} {\bibfnamefont {K.}~\bibnamefont
  {Zakeri}}, \bibinfo {author} {\bibfnamefont {Y.}~\bibnamefont {Zhang}},
  \bibinfo {author} {\bibfnamefont {J.}~\bibnamefont {Prokop}}, \bibinfo
  {author} {\bibfnamefont {T.-H.}\ \bibnamefont {Chuang}}, \bibinfo {author}
  {\bibfnamefont {N.}~\bibnamefont {Sakr}}, \bibinfo {author} {\bibfnamefont
  {W.~X.}\ \bibnamefont {Tang}}, \ and\ \bibinfo {author} {\bibfnamefont
  {J.}~\bibnamefont {Kirschner}},\ }\href {\doibase
  10.1103/PhysRevLett.104.137203} {\bibfield  {journal} {\bibinfo  {journal}
  {Phys. Rev. Lett.}\ }\textbf {\bibinfo {volume} {104}},\ \bibinfo {pages}
  {137203} (\bibinfo {year} {2010})}\BibitemShut {NoStop}%
\bibitem [{\citenamefont {Jamali}\ \emph {et~al.}(2013)\citenamefont {Jamali},
  \citenamefont {Kwon}, \citenamefont {Seo}, \citenamefont {Lee},\ and\
  \citenamefont {Yang}}]{Jamali2013}%
  \BibitemOpen
  \bibfield  {author} {\bibinfo {author} {\bibfnamefont {M.}~\bibnamefont
  {Jamali}}, \bibinfo {author} {\bibfnamefont {J.~H.}\ \bibnamefont {Kwon}},
  \bibinfo {author} {\bibfnamefont {S.-M.}\ \bibnamefont {Seo}}, \bibinfo
  {author} {\bibfnamefont {K.-J.}\ \bibnamefont {Lee}}, \ and\ \bibinfo
  {author} {\bibfnamefont {H.}~\bibnamefont {Yang}},\ }\href {\doibase
  10.1038/srep03160} {\bibfield  {journal} {\bibinfo  {journal} {Sci. Rep.}\
  }\textbf {\bibinfo {volume} {3}},\ \bibinfo {pages} {3160} (\bibinfo {year}
  {2013})}\BibitemShut {NoStop}%
\bibitem [{\citenamefont {Ot\'alora}\ \emph {et~al.}(2016)\citenamefont
  {Ot\'alora}, \citenamefont {Yan}, \citenamefont {Schultheiss}, \citenamefont
  {Hertel},\ and\ \citenamefont {K\'akay}}]{Otalora2016}%
  \BibitemOpen
  \bibfield  {author} {\bibinfo {author} {\bibfnamefont {J.~A.}\ \bibnamefont
  {Ot\'alora}}, \bibinfo {author} {\bibfnamefont {M.}~\bibnamefont {Yan}},
  \bibinfo {author} {\bibfnamefont {H.}~\bibnamefont {Schultheiss}}, \bibinfo
  {author} {\bibfnamefont {R.}~\bibnamefont {Hertel}}, \ and\ \bibinfo {author}
  {\bibfnamefont {A.}~\bibnamefont {K\'akay}},\ }\href {\doibase
  10.1103/PhysRevLett.117.227203} {\bibfield  {journal} {\bibinfo  {journal}
  {Phys. Rev. Lett.}\ }\textbf {\bibinfo {volume} {117}},\ \bibinfo {pages}
  {227203} (\bibinfo {year} {2016})}\BibitemShut {NoStop}%
\bibitem [{\citenamefont {Chumak}\ \emph {et~al.}(2014)\citenamefont {Chumak},
  \citenamefont {Serga},\ and\ \citenamefont {Hillebrands}}]{chumak2014magnon}%
  \BibitemOpen
  \bibfield  {author} {\bibinfo {author} {\bibfnamefont {A.~V.}\ \bibnamefont
  {Chumak}}, \bibinfo {author} {\bibfnamefont {A.~A.}\ \bibnamefont {Serga}}, \
  and\ \bibinfo {author} {\bibfnamefont {B.}~\bibnamefont {Hillebrands}},\
  }\href@noop {} {\bibfield  {journal} {\bibinfo  {journal} {Nature
  communications}\ }\textbf {\bibinfo {volume} {5}},\ \bibinfo {pages} {1}
  (\bibinfo {year} {2014})}\BibitemShut {NoStop}%
\bibitem [{\citenamefont {Melkov}\ \emph {et~al.}(2006)\citenamefont {Melkov},
  \citenamefont {Vasyuchka}, \citenamefont {Chumak}, \citenamefont
  {Tiberkevich},\ and\ \citenamefont {Slavin}}]{Melkov2006}%
  \BibitemOpen
  \bibfield  {author} {\bibinfo {author} {\bibfnamefont {G.~A.}\ \bibnamefont
  {Melkov}}, \bibinfo {author} {\bibfnamefont {V.~I.}\ \bibnamefont
  {Vasyuchka}}, \bibinfo {author} {\bibfnamefont {A.~V.}\ \bibnamefont
  {Chumak}}, \bibinfo {author} {\bibfnamefont {V.~S.}\ \bibnamefont
  {Tiberkevich}}, \ and\ \bibinfo {author} {\bibfnamefont {A.~N.}\ \bibnamefont
  {Slavin}},\ }\href {\doibase 10.1063/1.2172184} {\bibfield  {journal}
  {\bibinfo  {journal} {Journal of Applied Physics}\ }\textbf {\bibinfo
  {volume} {99}},\ \bibinfo {pages} {08P513} (\bibinfo {year} {2006})},\
  \Eprint {http://arxiv.org/abs/https://doi.org/10.1063/1.2172184}
  {https://doi.org/10.1063/1.2172184} \BibitemShut {NoStop}%
\bibitem [{\citenamefont {Lee}\ \emph {et~al.}(2009)\citenamefont {Lee},
  \citenamefont {Han},\ and\ \citenamefont {Kim}}]{Lee2009WG}%
  \BibitemOpen
  \bibfield  {author} {\bibinfo {author} {\bibfnamefont {K.-S.}\ \bibnamefont
  {Lee}}, \bibinfo {author} {\bibfnamefont {D.-S.}\ \bibnamefont {Han}}, \ and\
  \bibinfo {author} {\bibfnamefont {S.-K.}\ \bibnamefont {Kim}},\ }\href
  {\doibase 10.1103/PhysRevLett.102.127202} {\bibfield  {journal} {\bibinfo
  {journal} {Phys. Rev. Lett.}\ }\textbf {\bibinfo {volume} {102}},\ \bibinfo
  {pages} {127202} (\bibinfo {year} {2009})}\BibitemShut {NoStop}%
\bibitem [{\citenamefont {Chumak}\ \emph {et~al.}(2009)\citenamefont {Chumak},
  \citenamefont {Neumann}, \citenamefont {Serga}, \citenamefont {Hillebrands},\
  and\ \citenamefont {Kostylev}}]{Chumak2009}%
  \BibitemOpen
  \bibfield  {author} {\bibinfo {author} {\bibfnamefont {A.~V.}\ \bibnamefont
  {Chumak}}, \bibinfo {author} {\bibfnamefont {T.}~\bibnamefont {Neumann}},
  \bibinfo {author} {\bibfnamefont {A.~A.}\ \bibnamefont {Serga}}, \bibinfo
  {author} {\bibfnamefont {B.}~\bibnamefont {Hillebrands}}, \ and\ \bibinfo
  {author} {\bibfnamefont {M.~P.}\ \bibnamefont {Kostylev}},\ }\href {\doibase
  10.1088/0022-3727/42/20/205005} {\bibfield  {journal} {\bibinfo  {journal}
  {Journal of Physics D: Applied Physics}\ }\textbf {\bibinfo {volume} {42}},\
  \bibinfo {pages} {205005} (\bibinfo {year} {2009})}\BibitemShut {NoStop}%
\bibitem [{\citenamefont {Topp}\ \emph {et~al.}(2010)\citenamefont {Topp},
  \citenamefont {Heitmann}, \citenamefont {Kostylev},\ and\ \citenamefont
  {Grundler}}]{Topp2010}%
  \BibitemOpen
  \bibfield  {author} {\bibinfo {author} {\bibfnamefont {J.}~\bibnamefont
  {Topp}}, \bibinfo {author} {\bibfnamefont {D.}~\bibnamefont {Heitmann}},
  \bibinfo {author} {\bibfnamefont {M.~P.}\ \bibnamefont {Kostylev}}, \ and\
  \bibinfo {author} {\bibfnamefont {D.}~\bibnamefont {Grundler}},\ }\href
  {\doibase 10.1103/PhysRevLett.104.207205} {\bibfield  {journal} {\bibinfo
  {journal} {Phys. Rev. Lett.}\ }\textbf {\bibinfo {volume} {104}},\ \bibinfo
  {pages} {207205} (\bibinfo {year} {2010})}\BibitemShut {NoStop}%
\bibitem [{\citenamefont {Tacchi}\ \emph {et~al.}(2010)\citenamefont {Tacchi},
  \citenamefont {Madami}, \citenamefont {Gubbiotti}, \citenamefont {Carlotti},
  \citenamefont {Goolaup}, \citenamefont {Adeyeye}, \citenamefont {Singh},\
  and\ \citenamefont {Kostylev}}]{Tacchi2010}%
  \BibitemOpen
  \bibfield  {author} {\bibinfo {author} {\bibfnamefont {S.}~\bibnamefont
  {Tacchi}}, \bibinfo {author} {\bibfnamefont {M.}~\bibnamefont {Madami}},
  \bibinfo {author} {\bibfnamefont {G.}~\bibnamefont {Gubbiotti}}, \bibinfo
  {author} {\bibfnamefont {G.}~\bibnamefont {Carlotti}}, \bibinfo {author}
  {\bibfnamefont {S.}~\bibnamefont {Goolaup}}, \bibinfo {author} {\bibfnamefont
  {A.~O.}\ \bibnamefont {Adeyeye}}, \bibinfo {author} {\bibfnamefont
  {N.}~\bibnamefont {Singh}}, \ and\ \bibinfo {author} {\bibfnamefont {M.~P.}\
  \bibnamefont {Kostylev}},\ }\href {\doibase 10.1103/PhysRevB.82.184408}
  {\bibfield  {journal} {\bibinfo  {journal} {Phys. Rev. B}\ }\textbf {\bibinfo
  {volume} {82}},\ \bibinfo {pages} {184408} (\bibinfo {year}
  {2010})}\BibitemShut {NoStop}%
\bibitem [{\citenamefont {Maruyama}\ \emph {et~al.}(2009)\citenamefont
  {Maruyama}, \citenamefont {Shiota}, \citenamefont {Nozaki}, \citenamefont
  {Ohta}, \citenamefont {Toda}, \citenamefont {Mizuguchi}, \citenamefont
  {Tulapurkar}, \citenamefont {Shinjo}, \citenamefont {Shiraishi},
  \citenamefont {Mizukami} \emph {et~al.}}]{maruyama2009}%
  \BibitemOpen
  \bibfield  {author} {\bibinfo {author} {\bibfnamefont {T.}~\bibnamefont
  {Maruyama}}, \bibinfo {author} {\bibfnamefont {Y.}~\bibnamefont {Shiota}},
  \bibinfo {author} {\bibfnamefont {T.}~\bibnamefont {Nozaki}}, \bibinfo
  {author} {\bibfnamefont {K.}~\bibnamefont {Ohta}}, \bibinfo {author}
  {\bibfnamefont {N.}~\bibnamefont {Toda}}, \bibinfo {author} {\bibfnamefont
  {M.}~\bibnamefont {Mizuguchi}}, \bibinfo {author} {\bibfnamefont
  {A.}~\bibnamefont {Tulapurkar}}, \bibinfo {author} {\bibfnamefont
  {T.}~\bibnamefont {Shinjo}}, \bibinfo {author} {\bibfnamefont
  {M.}~\bibnamefont {Shiraishi}}, \bibinfo {author} {\bibfnamefont
  {S.}~\bibnamefont {Mizukami}},  \emph {et~al.},\ }\href@noop {} {\bibfield
  {journal} {\bibinfo  {journal} {Nature nanotechnology}\ }\textbf {\bibinfo
  {volume} {4}},\ \bibinfo {pages} {158} (\bibinfo {year} {2009})}\BibitemShut
  {NoStop}%
\bibitem [{\citenamefont {Amiri}\ and\ \citenamefont {Wang}(2012)}]{amiri2012}%
  \BibitemOpen
  \bibfield  {author} {\bibinfo {author} {\bibfnamefont {P.~K.}\ \bibnamefont
  {Amiri}}\ and\ \bibinfo {author} {\bibfnamefont {K.~L.}\ \bibnamefont
  {Wang}},\ }in\ \href@noop {} {\emph {\bibinfo {booktitle} {Spin}}},\
  Vol.~\bibinfo {volume} {2}\ (\bibinfo {organization} {World Scientific},\
  \bibinfo {year} {2012})\ p.\ \bibinfo {pages} {1240002}\BibitemShut {NoStop}%
\bibitem [{\citenamefont {Hanyu}\ \emph {et~al.}(2019)\citenamefont {Hanyu},
  \citenamefont {Endoh}, \citenamefont {Ando}, \citenamefont {Ikeda},
  \citenamefont {Fukami}, \citenamefont {Sato}, \citenamefont {Koike},
  \citenamefont {Ma}, \citenamefont {Suzuki},\ and\ \citenamefont
  {Ohno}}]{hanyu2019spin}%
  \BibitemOpen
  \bibfield  {author} {\bibinfo {author} {\bibfnamefont {T.}~\bibnamefont
  {Hanyu}}, \bibinfo {author} {\bibfnamefont {T.}~\bibnamefont {Endoh}},
  \bibinfo {author} {\bibfnamefont {Y.}~\bibnamefont {Ando}}, \bibinfo {author}
  {\bibfnamefont {S.}~\bibnamefont {Ikeda}}, \bibinfo {author} {\bibfnamefont
  {S.}~\bibnamefont {Fukami}}, \bibinfo {author} {\bibfnamefont
  {H.}~\bibnamefont {Sato}}, \bibinfo {author} {\bibfnamefont {H.}~\bibnamefont
  {Koike}}, \bibinfo {author} {\bibfnamefont {Y.}~\bibnamefont {Ma}}, \bibinfo
  {author} {\bibfnamefont {D.}~\bibnamefont {Suzuki}}, \ and\ \bibinfo {author}
  {\bibfnamefont {H.}~\bibnamefont {Ohno}},\ }in\ \href@noop {} {\emph
  {\bibinfo {booktitle} {Advances in non-volatile memory and storage
  technology}}}\ (\bibinfo  {publisher} {Elsevier},\ \bibinfo {year} {2019})\
  pp.\ \bibinfo {pages} {237--281}\BibitemShut {NoStop}%
\end{thebibliography}%

\end{document}